\documentclass[epj]{webofc}
\usepackage[varg]{txfonts}   

\newcommand{\be}{\begin{equation}}  
\newcommand{\ee}{\end{equation}}  
\newcommand{\beq}{\begin{eqnarray}}  
\newcommand{\eeq}{\end{eqnarray}}

\newcommand{\Dlr}{\buildrel \leftrightarrow \over D\raise-1pt\hbox{}}
\newcommand{\<}{\langle}
\renewcommand{\>}{\rangle}
\newcommand{\mc}{\mathcal}
\wocname{EPJ Web of Conferences}
\woctitle{CONF12}
\begin{document}
\selectlanguage{english}
\title{ Novel applications of Lattice QCD: Parton distribution functions, proton charge radius and neutron electric dipole moment}
%
%

\author{Constantia Alexandrou\inst{1,2}\fnsep\thanks{\email{alexand@ucy.ac.cy}} 
}

\institute{Department of Physics, University of Cyprus, P.O. Box 20537, CY-1678 Nicosia, Cyprus \and
Computation-based Science and Technology Research Center, The Cyprus Institute, 20, K. Kavafi Str., 2121  Nicosia, Cyprus
}

\abstract{%
 We briefly discuss the current status of lattice QCD simulations and review selective results on nucleon observables focusing on recent developments in the
lattice QCD evaluation of 
the  nucleon form factors and radii, parton distribution functions and their moments,  and the  neutron electric dipole moment.  Nucleon charges and moments of parton distribution functions are presented using simulations generated at physical values of the quark masses, while exploratory studies are performed for the parton distribution functions and the neutron electric dipole moment at heavier than physical value of the pion mass.
}
\maketitle
\section{Introduction}
\label{intro}
Lattice Quantum Chromodynamics (QCD) provides a non-perturbative approach 
for the evaluation of properties of strongly interacting systems starting directly from the QCD Langragian of the theory,  

\be {\cal L}_{QCD}=-\frac{1}{4} F_{\mu\nu}^aF^{a\, \mu\nu} +\sum_{f=u,d,s,c,b,t} \bar{\psi}_f\left(i\gamma^\mu D_\mu -m_f \right)\psi_f, \,\,
{\rm where} \,\,
 D_\mu=\partial_\mu-ig \frac{\lambda^a}{2} A^a_\mu 
\ee
and using the same degrees of freedom, namely the quarks and the gluons. Lattice QCD is defined on a 4-dimensional hyper cubic lattice with lattice spacing $a$.
There are several ${\cal O}(a)$ improved  fermion discretisation schemes
 each of which has its advantages. Wilson-type actions are pursued by a number of collaborations using mainly the  clover and twisted mass (TM) discretization schemes. They are computationally fast but break chiral symmetry. The clover formulation needs operator improvement, while the TM formulation provides automatic ${\cal O}(a)$ but breaks flavor symmetry, which is only recovered in the continuum limit~\cite{Frezzotti:2003ni,Frezzotti:2003xj}. Staggered fermions are fast to simulate, preserve chiral symmetry but have four doublets and complicated contractions. 
The domain wall (DW) formulation  has improved chiral symmetry and overlap fermions are chiral preserving but both are  computationally expensive as compared to Wilson-type and staggered fermions. Results emerging from the different
discretization schemes must agree in the continuum limit,  $a \rightarrow 0$.
Furthermore, observables need to be  extrapolated to the infinite volume limit, $L\rightarrow \infty$ where $L$ is the spatial lattice length  or at the least one needs to show that the lattice volume is large enough so that finite volume effects are negligible.

\begin{figure}[h]
\begin{minipage}{0.56\linewidth}
   \includegraphics[width=\linewidth]{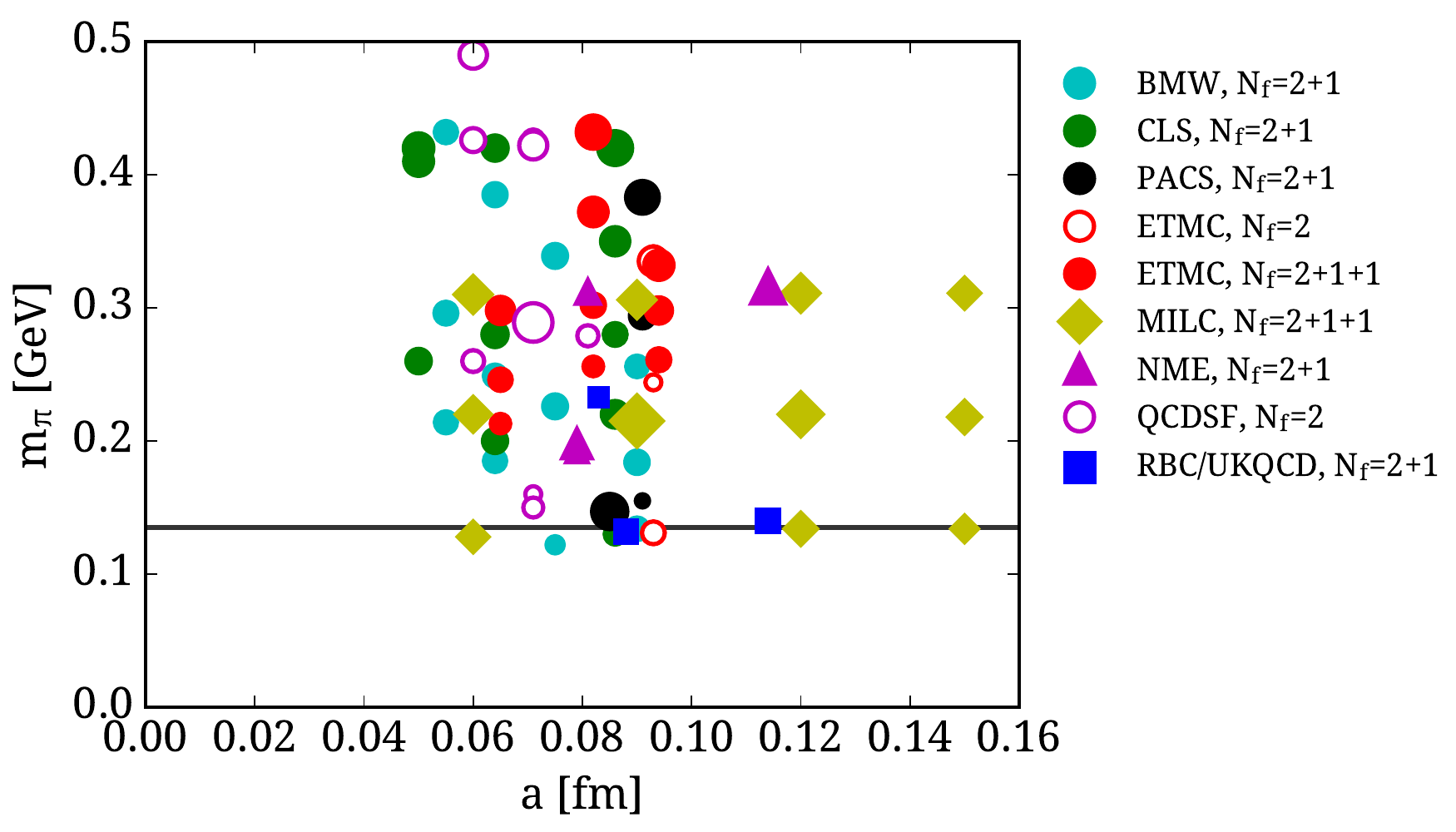}
 \end{minipage}\hfill
\begin{minipage}{0.43\linewidth}
   \includegraphics[width=\linewidth]{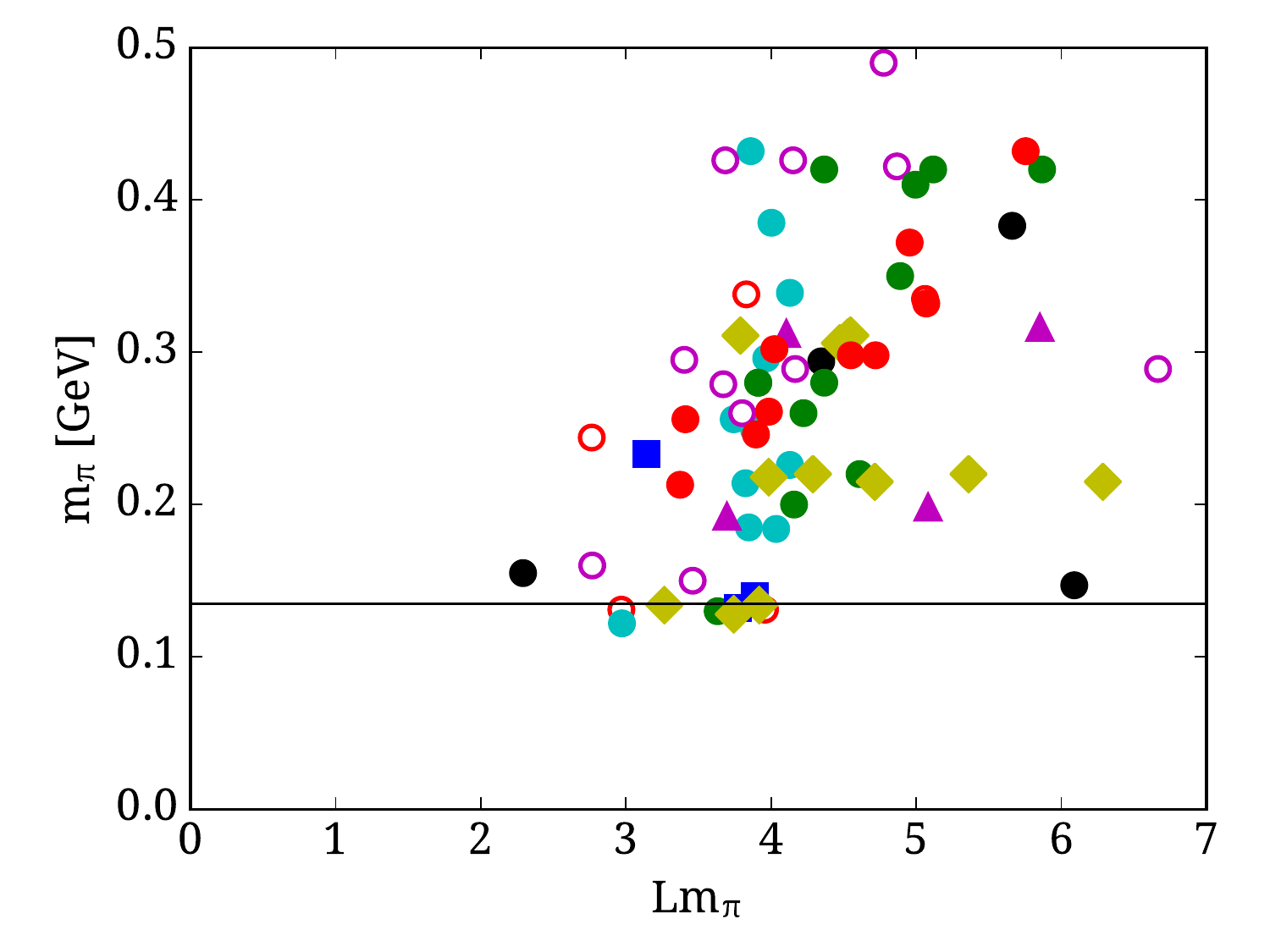}
 \end{minipage}
\caption{A summary of recent simulations showing the value of the pion mass and the lattice spacing (left) and $m_\pi L$ (right): Black filled circles are from PACS using $N_f=2+1$ clover-improved fermions~\cite{Aoki:2009ix,Ishikawa:2015rho}, light blue filled circles from BMW for $N_f=2+1$ clover-improved fermions with HEX smearing~\cite{Durr:2010aw}, yellow filled diamonds from MILC using $N_f=2+1+1$ staggered fermions~\cite{Bazavov:2012xda}, magenta open circles from QCDSF using $N_f=2$ clover-improved fermions~\cite{Horsley:2013ayv}, green filled circles from CLS using $N_f=2+1$ clover-improved fermions~\cite{Bruno:2014jqa}, blue filled squares from RBC-UKQCD using domain wall fermions~\cite{Boyle:2015hfa}, red filled (open) circles from ETMC using $N_f=2+1+1$ twisted mass fermions ($N_f=2$ with a clover term)~\cite{Abdel-Rehim:2015pwa}. The size of the symbols in the left panel is according to the value of $m_\pi L$ with the smallest value taken as $m_\pi L\sim 3$ and the largest $m_\pi L\sim 6.7$.}
\label{fig:sim}
\end{figure}

In Fig.~\ref{fig:sim} we show a summary of dynamical simulations characterized by the value of the pion mass, lattice spacing and lattice extent. As can be seen, a number of collaborations have  simulations at physical values of the pion mass on  sufficiently large volumes using  various types of  Wilson ${\cal O}(a)$-improved, domain wall and staggered fermions~\cite{Aoki:2009ix,Ishikawa:2015rho,Durr:2010aw,Bazavov:2012xda,Horsley:2013ayv,Bruno:2014jqa,Boyle:2015hfa,Abdel-Rehim:2015pwa}. 

Analyzing gauge configurations  at the physical value of the pion mass $m_\pi$ (physical point) typically requires large statistics and it has become computationally as demanding as producing the gauge configurations. A plethora of observables is currently under study in lattice QCD. Here we focus on three sets of key observables:
i) Nucleon  form factors and radii; ii) moments of parton distributions  where results are obtained  using simulations with pion mass close to its physical value. An exploratory study of directly evaluating the parton distribution functions (PDFs) is discussed following the method of  Ref.~\cite{Ji:2013dva} where  results are obtained for clover fermions on  $N_f=2+1+1$ HISQ sea~\cite{Lin:2014zya} for $m_\pi=310$~MeV and for an $N_f=2+1+1$ TM fermion ensemble with $m_\pi=373~$ MeV~\cite{Alexandrou:2015rja}; iii)
 the neutron electric dipole moment  using simulations with heavier than physical pions.

The evaluation of hadron matrix elements requires the computation of the
appropriate  Euclidean three-point function given by
\be
G^{\mu\nu}(\Gamma,\vec q,\vec{p}^\prime, t_s, t_{\rm ins}) =\sum_{\vec x_s, {\vec x}_{\rm ins}} \, e^{i{\vec x}_{\rm ins} \cdot \vec q}\,  e^{-i{\vec x}_s\cdot \vec p^\prime}\, 
     {\Gamma_{\beta\alpha}}\, \langle {J^{\alpha}_H(\vec x_s,t_s)} {\cal O}_\Gamma^{\mu\nu}({\vec x}_{\rm ins},t_{\rm ins}) {\overline{J}^{\beta}_H(\vec{x}_0, t_0)} \rangle,
\ee
where $\vec p^\prime$ is the final momentum and $\vec q$ is the momentum trasfer. 
Dividing by an appropriate combination of two-point functions one obtains a ratio that in the large Euclidean time limit converges to the ground state matrix element as 
 \begin{align}
    R(t_s,t_{\rm ins},t_0) \xrightarrow[(t_s-t_{\rm ins})\Delta \gg 1]{(t_{\rm ins}-t_0)\Delta \gg 1} \mathcal{M}[1
      +  \mathcal{T}_1e^{-\Delta({\bf p})(t_{\rm ins}-t_0)} + \mathcal{T}_2 e^{-\Delta({\bf
          p'})(t_s-t_{\rm ins})}+\cdots],
  \end{align}
The matrix element of interest is given by
 $\mathcal{M}$. The ratio depends on $t_s,t_{\rm ins},t_0$, which are the
  sink, insertion and source times, respectively and $\Delta({\bf p})$ the
  energy gap between the ground state and  the first excited state.  To extract  $\mathcal{M}$ one either
takes $t_{\rm ins}-t_0$ and $t_s-t_{\rm ins}$ large enough so excited states give a negligible contribution and fits to a constant (plateau method) or includes the first excited state introducing additional fit parameters (two-state fit). For diagonal matrix elements ${\cal T}_1={\cal T}_2$ and resulting in two additional fit parameters.  Alternatively, summing over $t_{\rm ins}$ (summation method) we obtain
  \be
  \sum_{t_{\rm ins}=t_0}^{t_s} R(t_s,t_{\rm ins},t_0) = {\sf Const.} + \mathcal{M}[(t_s-t_0) + \mathcal{O}(e^{-\Delta({\bf p})(t_s-t_0)})  + \mathcal{O}(e^{-\Delta({\bf p'})(t_s-t_0)})].
  \ee
In the summed ratio,  excited state contributions are suppressed by exponentials decaying with $t_s-t_0$, rather than $t_s-t_{\rm ins}$ and $t_{\rm ins}-t_0$. However, one needs to fit the slope rather than to a constant or take differences and then fit to a constant~\cite{Maiani:1987by} yielding larger errors as compared to the plateau method. If excited states are under control these three methods should yield the same results, and this can be used as a criterion for the onset of ground state dominance. 

\begin{figure}[h]
\begin{minipage}{0.49\linewidth}
  \includegraphics[width=0.8\linewidth]{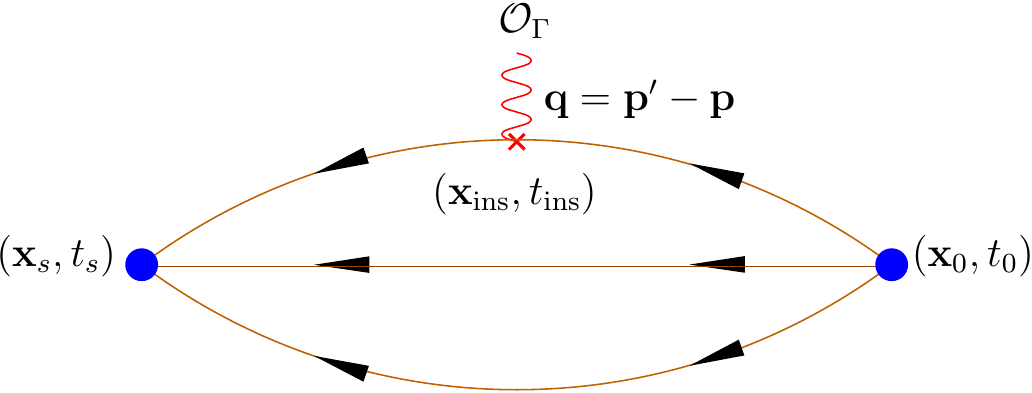}\\[12pt]
\includegraphics[width=0.8\linewidth]{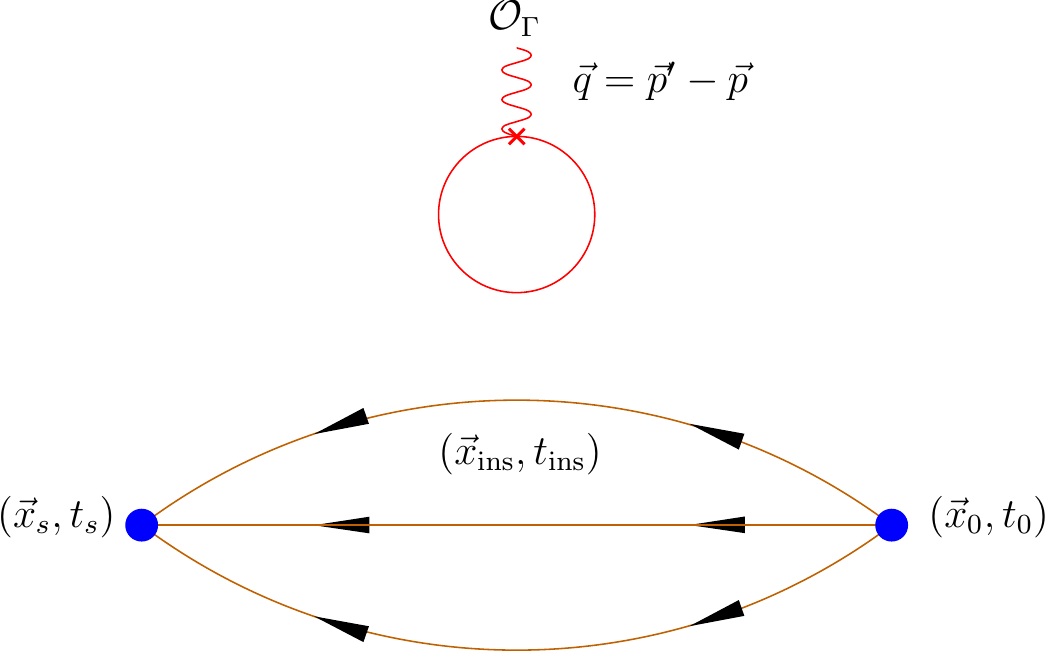}
\caption{Connected (upper) and disconnected (lower) contributions to the three-point function of quark bilinear operators. }
\label{fig:diagrams}
\end{minipage}\hfill
\begin{minipage}{0.49\linewidth}
   \includegraphics[width=\linewidth]{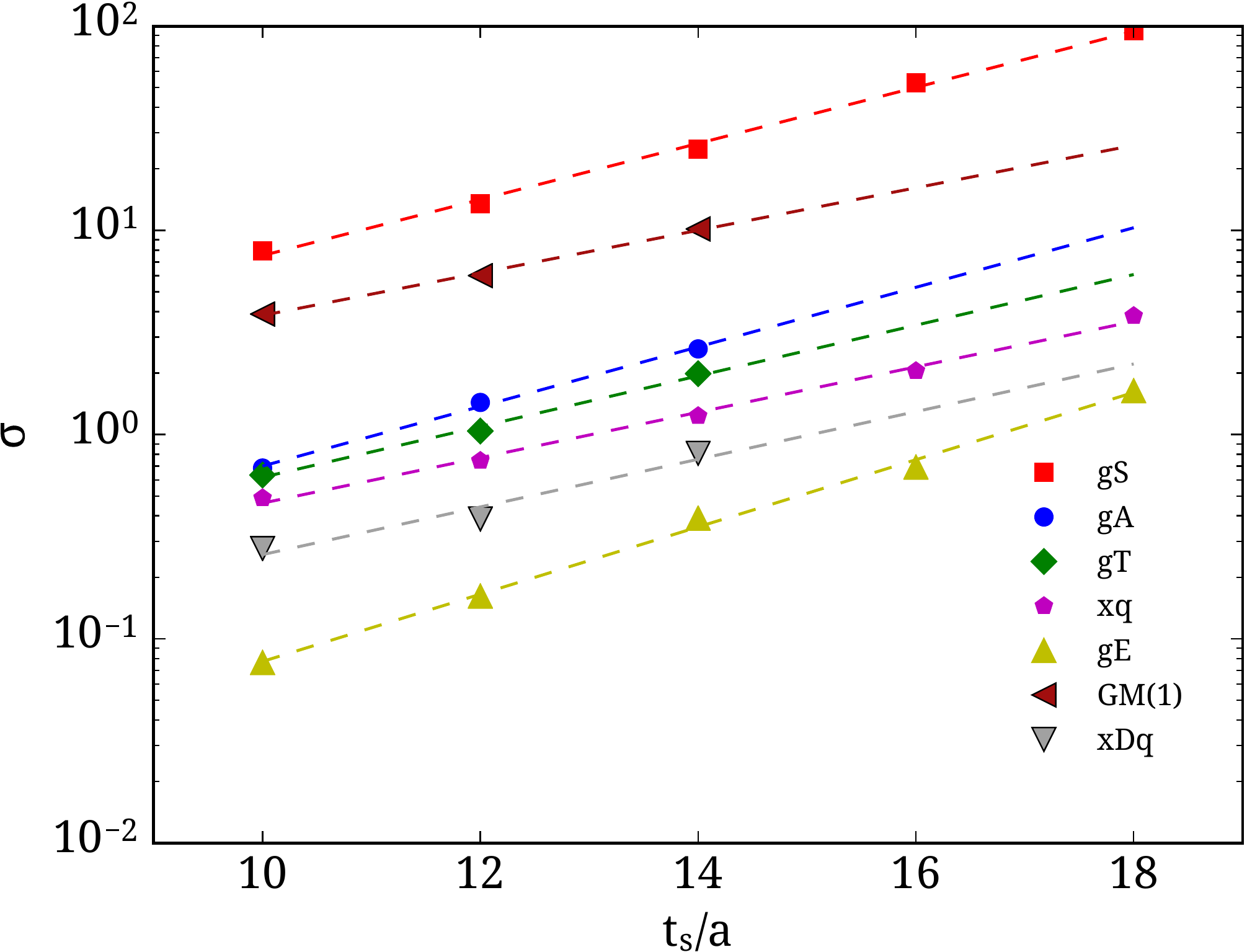}
\caption{The dependence of the statistical errors on the time separation $t_s-t_0$ for $N_f=2$ TM clover-improved fermions at the physical point. }
\label{fig:error}
 \end{minipage}
\end{figure}

A three-point function has, in general, two-type of contributions, one when the current couples to a valence quark (connected) and one when it couples to a sea quark (disconnected) as depicted diagrammatically in Fig.~\ref{fig:diagrams}.  Recent improvements have allowed the computation of the disconnected contribution that was ignored in the past being computationally very expensive. Further developments are needed in oder to reduce the errors  for some of the disconnected contributions in order to enable accurate results using simulations  at the physical point.
Although methods to compute the connected contribution are by now standard, increasing the sink-source time separation  to extract the ground state matrix element as required in particular for the physical point, leading to large  gauge since the noise to signal increases exponentially with the time separation. This is demonstrated in Fig.~\ref{fig:error} for several of the nucleon observables discussed here. These results are obtained using two degenerate flavors ($N_f=2$) of TM clover-improved fermions at the physical point.

\section{Nucleon form factors and radii}

\subsection{Nucleon charges}\label{charges}

Nucleon form factors  at zero momentum transfer yield the nucleon charges. Therefore we first discuss the evaluation of the nucleon matrix element at zero momentum transfer i.e.
 $\langle N(\vec{p^\prime}){\cal O}_X N(\vec{p})\rangle|_{q^2=0}$, and  consider 
the scalar 
${\cal O}_{S}^a=\bar{\psi}(x)\frac{\tau^a}{2}\psi(x)$
the axial-vector 
${\cal O}_{A}^a=\bar{\psi}(x)\gamma^{\mu}\gamma_5\frac{\tau^a}{2}\psi(x)$
and the  tensor 
${\cal O}_{T}^a=\bar{\psi}(x)\sigma^{\mu\nu}\frac{\tau^a}{2}\psi(x)$
operators.
The  nucleon axial charge $g_A$ is measured in neutron $\beta$-decays and it is an isovector quantity. This means that only the connected contribution is non-vanishing in the isospin limit, simplifying its computation within lattice QCD.
Despite the technical simplicity in its evaluation however, the value of $g_A$ has so far eluded lattice QCD evaluations. In the past this was attributed to results using simulations with heavier that physical pion mass. With gauge configurations at the physical point one is examining lattice artifacts such as excited states contribution to the ground state matrix element, finite volume and  lattice spacing effects so as to obtain a value with controlled systematic errors  that can 
be compared with the experimental one.

\begin{figure}[h]
\begin{minipage}{0.49\linewidth}
{\includegraphics[width=\linewidth]{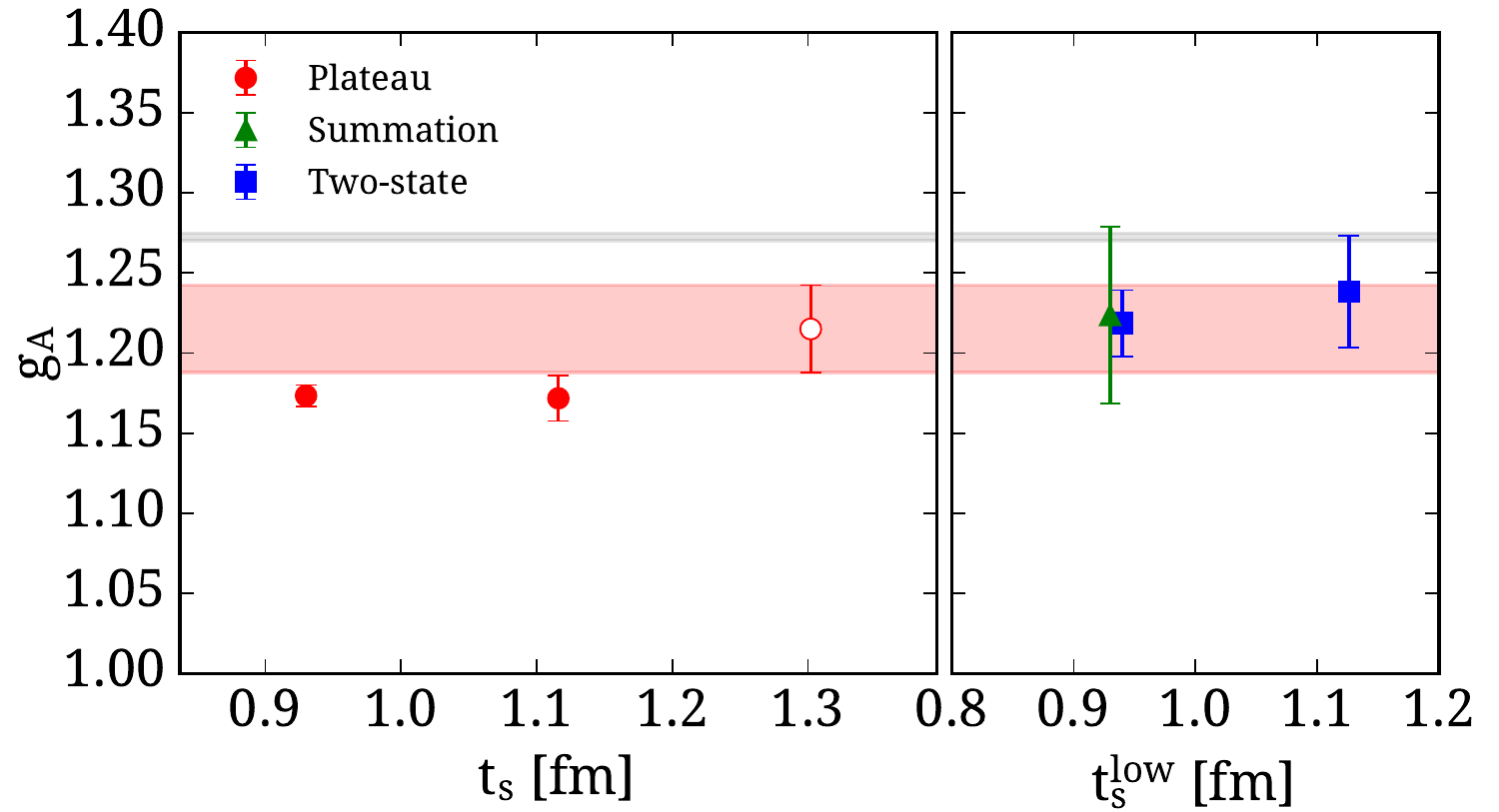}}\\
\end{minipage}
\begin{minipage}{0.46\linewidth}
\hspace*{-0.2cm} {\includegraphics[width=\linewidth]{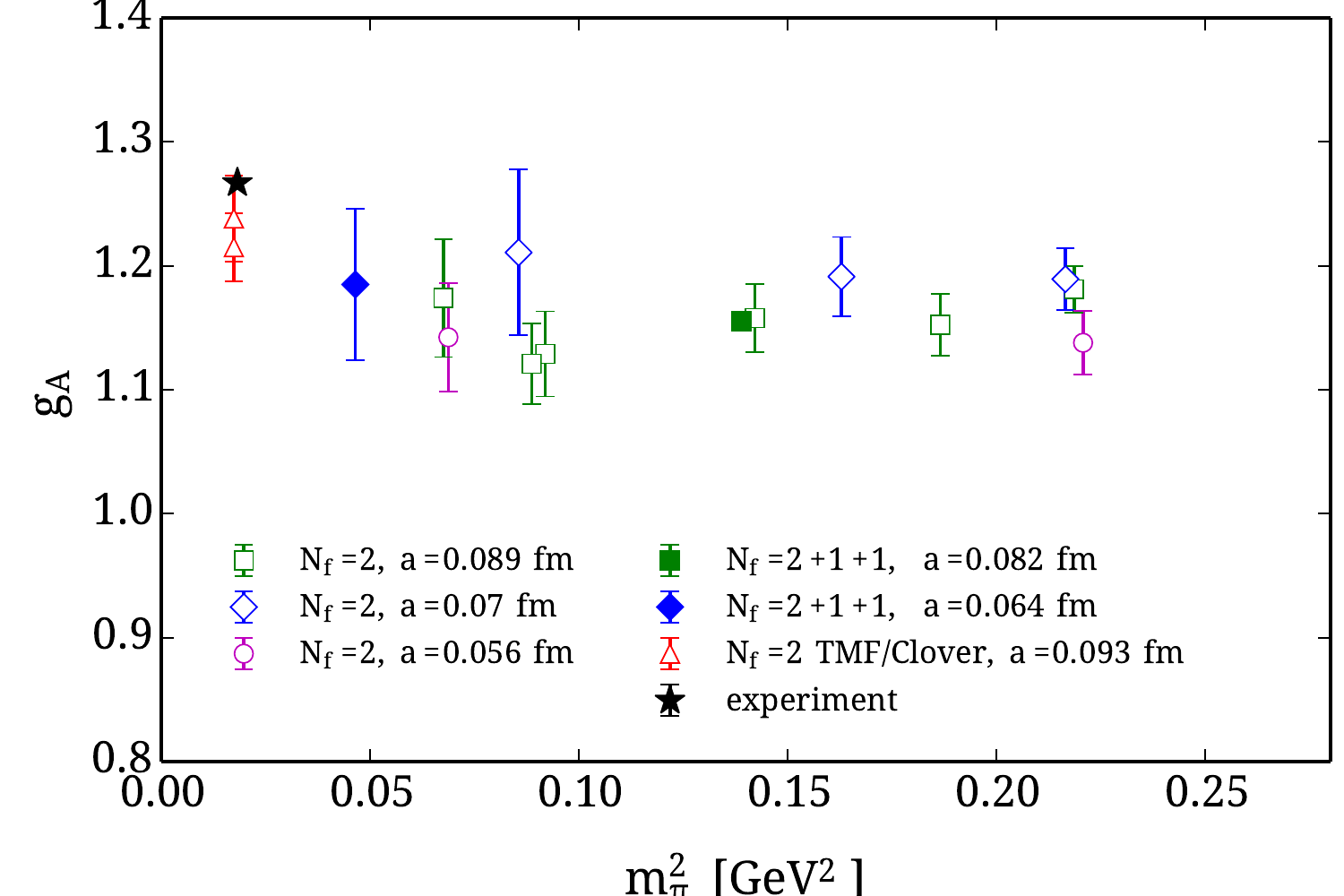}}
\end{minipage}
\caption{Isovector axial charge $g_A$. Left showing the value as we vary the sink-source time separation  $t_s-t_0$. $t_s^{\rm low}$ is the lowest value of $t_s$ used in the fits. Right we show the pion mass dependence of $g_A$ using TM fermion results.}
\label{fig:gA}
\end{figure}
In Fig.~\ref{fig:gA} we examine the extraction of the ground state matrix element as the sink-source separation is increased for an ensemble of $N_f=2$  twisted mass clover-improved fermions on a lattice of size  $48^3\times 96$, $a$ = 0.093(1)~fm, and $m_\pi$ = 131~MeV (referred to as the {\it physical TM ensemble}). At time separation $t_s-t_0=1.3$~fm, which is the largest we currently have, we find from the plateau method $g_A=1.22(3)(2)$, where the first error is statistical and the second systematic determined by the difference between the values from the plateau and two-state fits~\cite{Abdel-Rehim:2015owa}. This is close but still slightly below the measured value. Comparing with TM fermion results at heavier than physical pions masses we find that finite volume and  lattice spacing  effects are small. However, one needs to ensure that these remain small at the physical point. A number of collaborations are investigating lattice artifacts  (see Ref.~\cite{Collins} for a recent review) and the expectation is that soon one will be able to  understand the small
difference currently observed between the lattice QCD value for $g_A$ and the experimental value.

\begin{figure} 
 \begin{minipage}{0.49\linewidth}
\includegraphics[width=\linewidth]{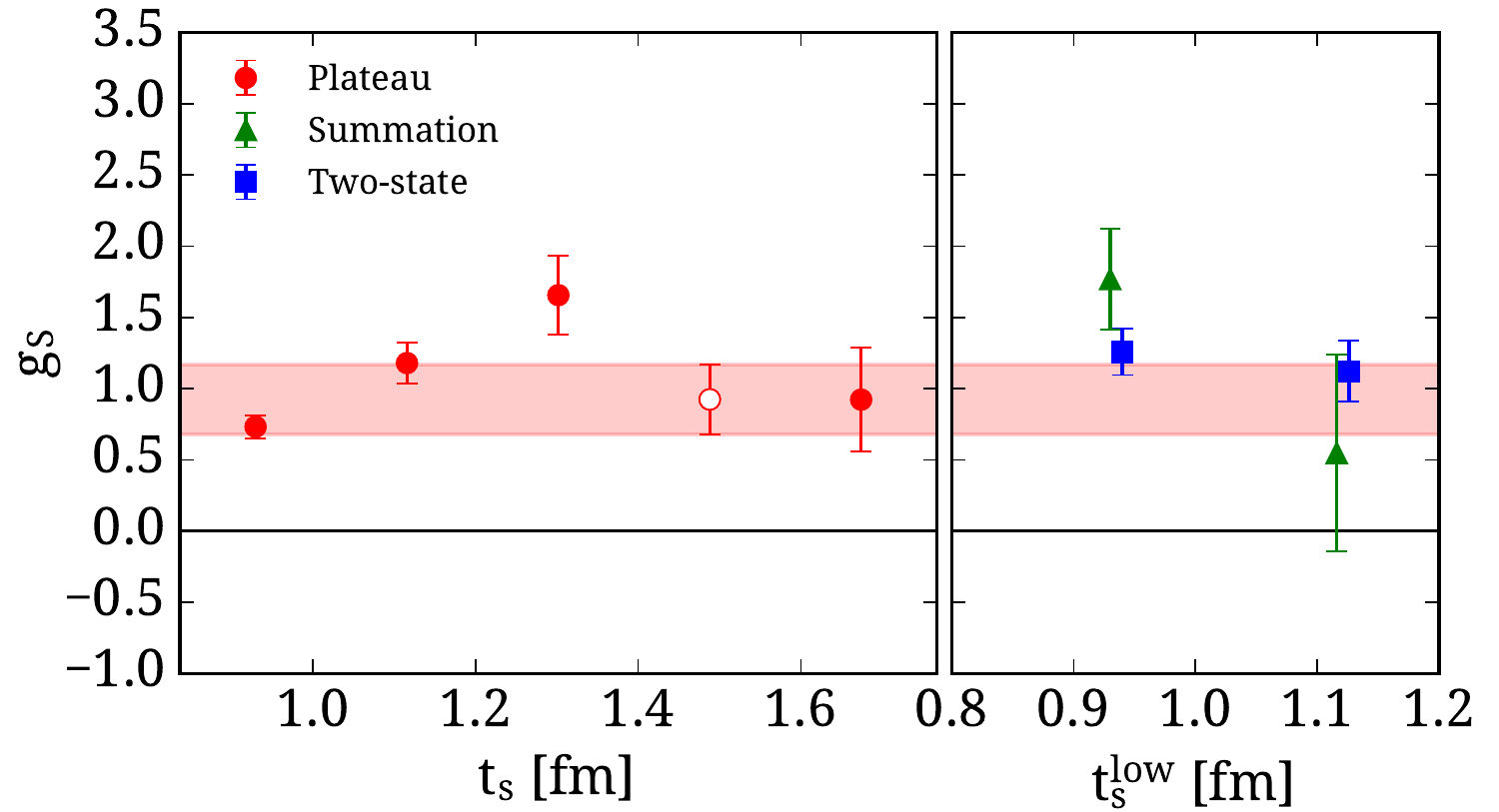}
   \end{minipage}\hfill
 \begin{minipage}{0.49\linewidth}
   \includegraphics[width=\linewidth]{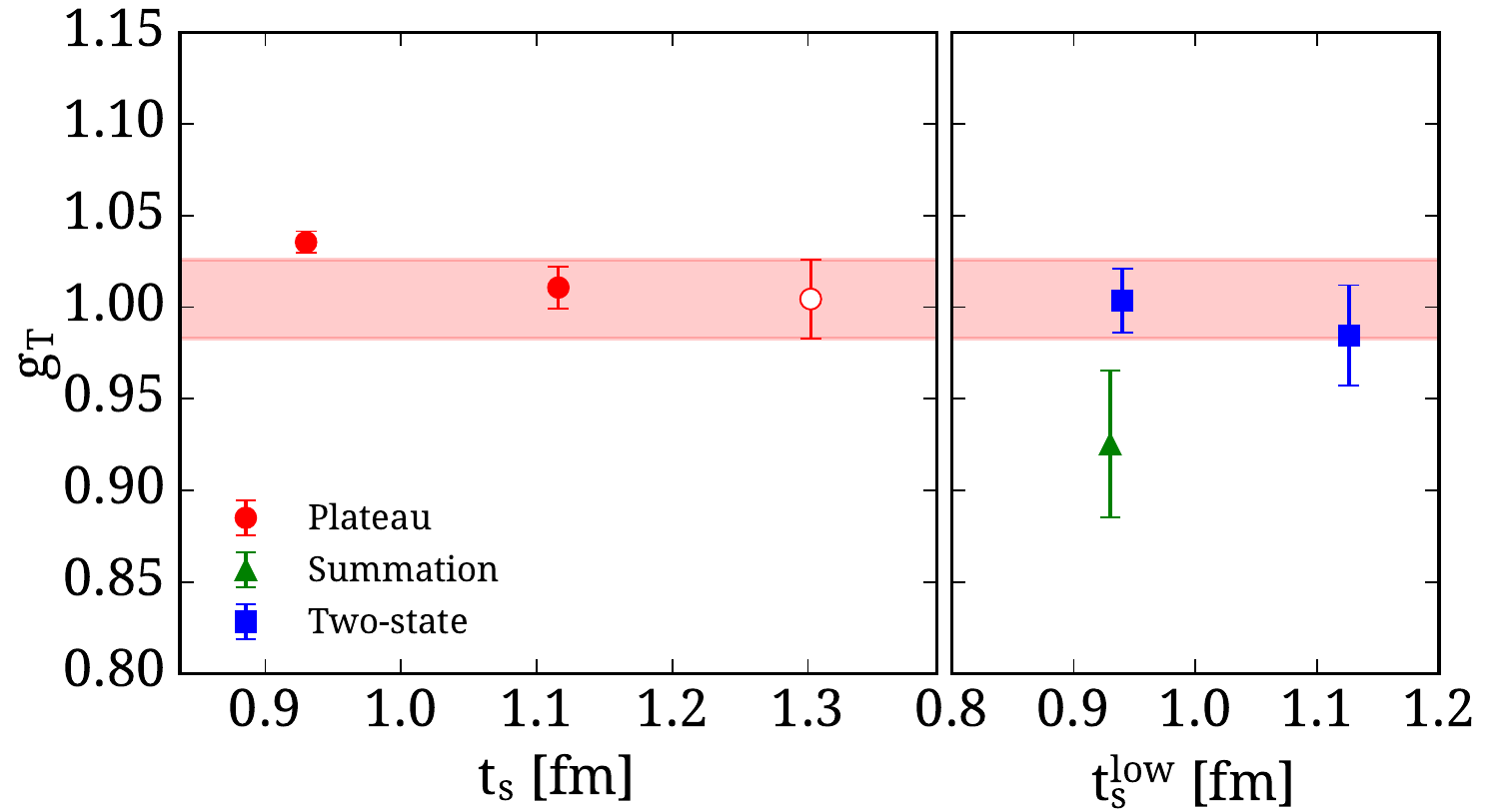}
  \end{minipage}
\caption{Results on the isovector scalar (left) and tensor (right) charges for TM clover-improved fermions at the physical point.}
\label{fig:gSgT}
\end{figure}

\begin{figure}[h]
\begin{minipage}{0.49\linewidth}
\includegraphics[width=\linewidth]{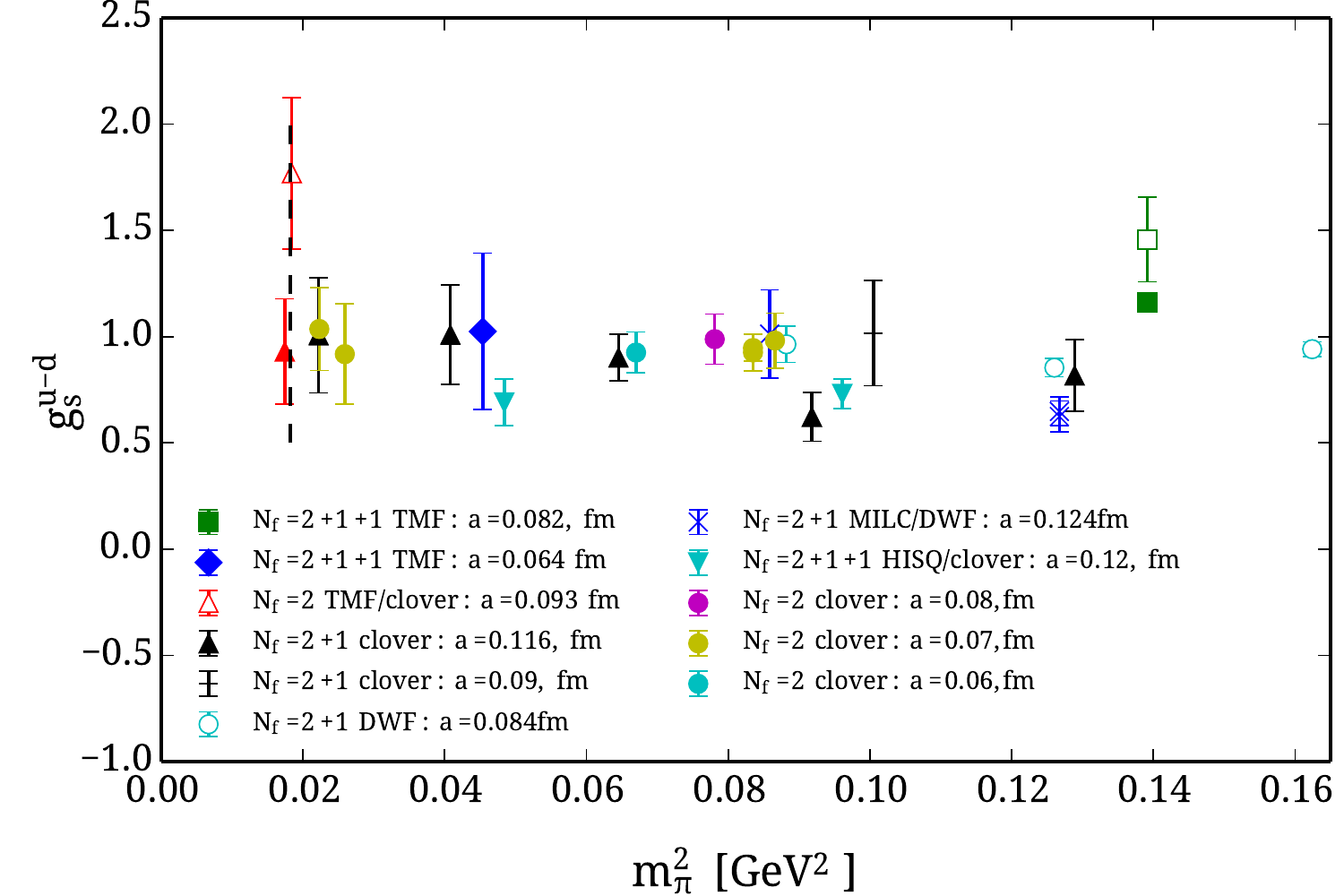}
\end{minipage}\hfill
\begin{minipage}{0.49\linewidth}
\includegraphics[width=\linewidth]{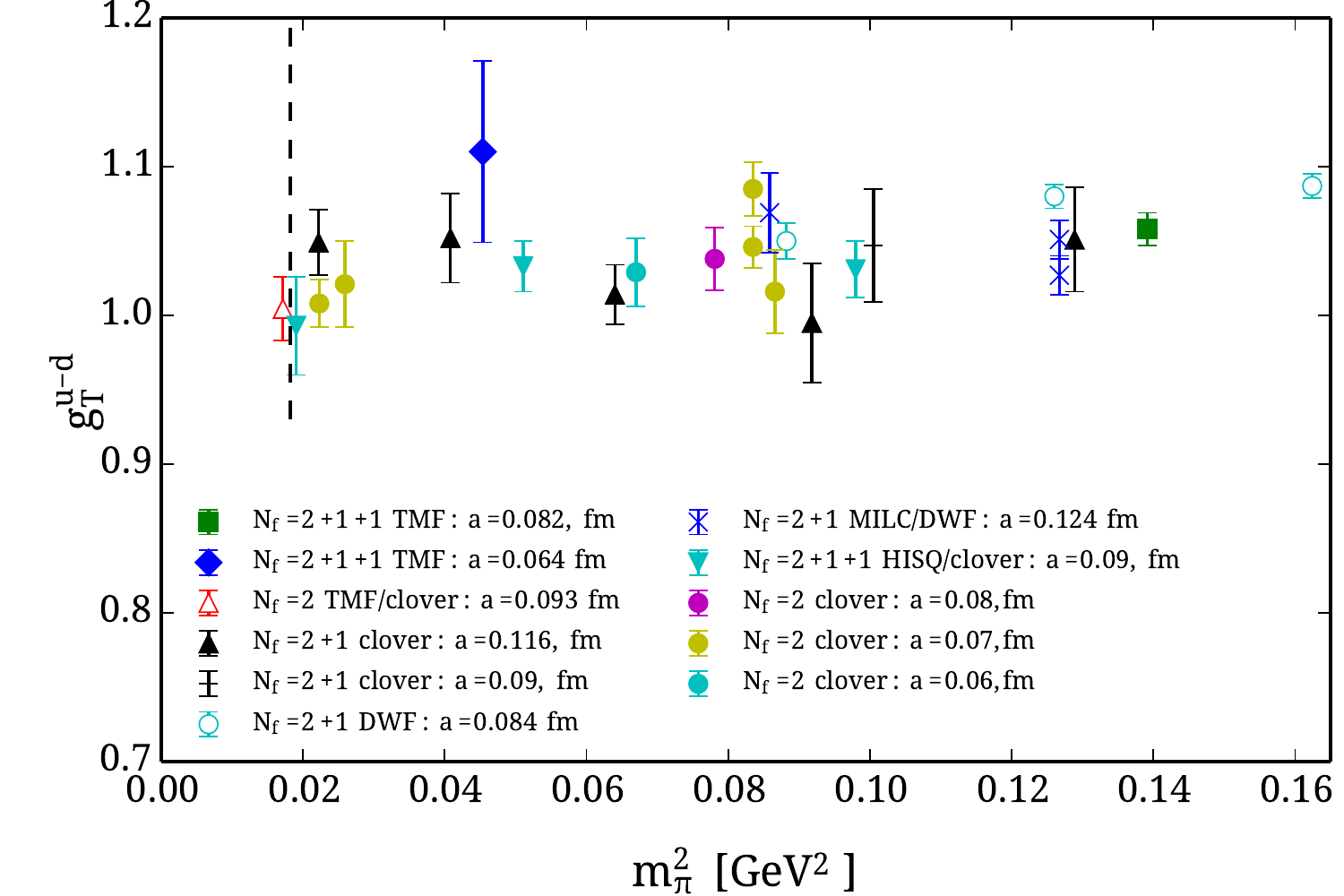}
\end{minipage}
\caption{Comparison of results for the isovector scalar (left) and tensor (right).}
\label{fig:gSgT all}
\end{figure}

The value of the  scalar and tensor charges on the other hand are not well known. These are important quantities for probing scalar and tensor interactions beyond the standard model. For the isovector tensor charge $g_T$  a global analysis of HERMES, COMPASS and Belle $e^+ e^-$ data yields  $g_T^{u-d}\sim 0.54 ^{+0.30}_{-0.13}$~\cite{Anselmino:2013vqa}, while a new analysis of COMPASS and Belle data resulted in $g_T^{u-d}=0.81(44)$~\cite{Radici:2015mwa}. Given this large uncertainty, a lattice QCD determination can provide a valuable input, especially in view of plans to measure $g_T$ in the  SIDIS experiment  on $^3$He/Proton  at JLab.
In Fig.~\ref{fig:gSgT} we show results on the isovector $g_S$ and $g_T$ for the  physical TM ensemble, extracted using $\sim 9260$ statistics for $t_s/a=10,12,14$, $\sim 48 000$ for $t_s/a=16$ and $\sim 70 000$ for $t_s/a=18$. As can be seen, the scalar charge shows large excited state contributions  
 and a larger $t_s-t_0$ is required as compared to e.g. $g_T$ in order to extract the correct matrix element. We find that  $t_s-t_0 \stackrel{\sim}{>} 1.5$~fm at the physical point is needed for convergence.
Using  the plateau method we find
$g_S^{u-d}=0.93(25)(33)$ and $g_T^{u-d}=1.00(2)(1)$, where the second error is estimated from the difference between the plateau value and the two state fit result~\cite{Alexandrou2016}. In Fig.~\ref{fig:gSgT all} we compare lattice results using different discretization schemes, lattice spacings and volumes. The fact that lattice results agree among them  indicates small lattice artifacts for these improved lattice actions.

\begin{figure}[h]
\begin{minipage}{0.49\linewidth}
{\includegraphics[width=\linewidth]{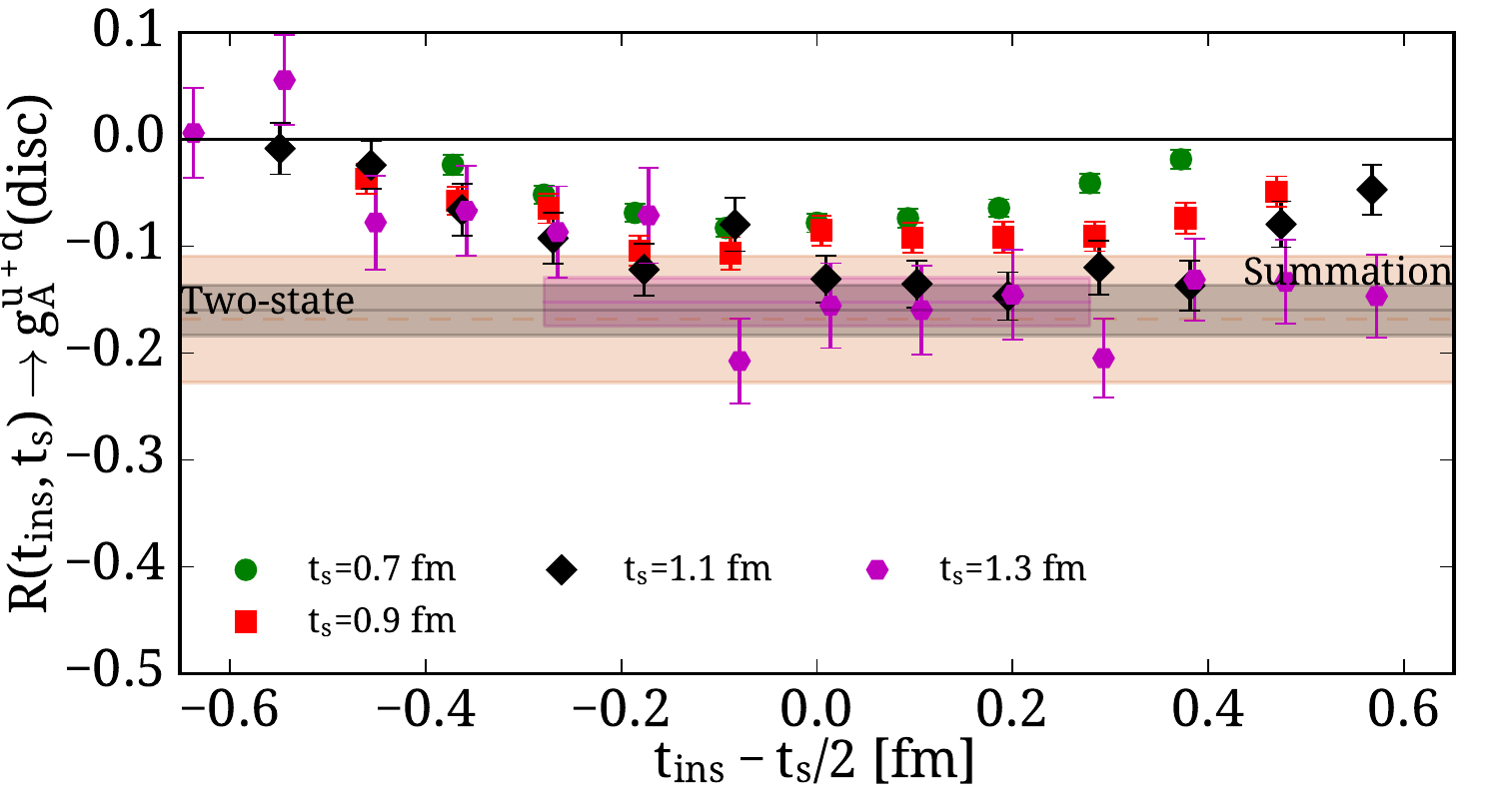}}\\
{\includegraphics[width=\linewidth]{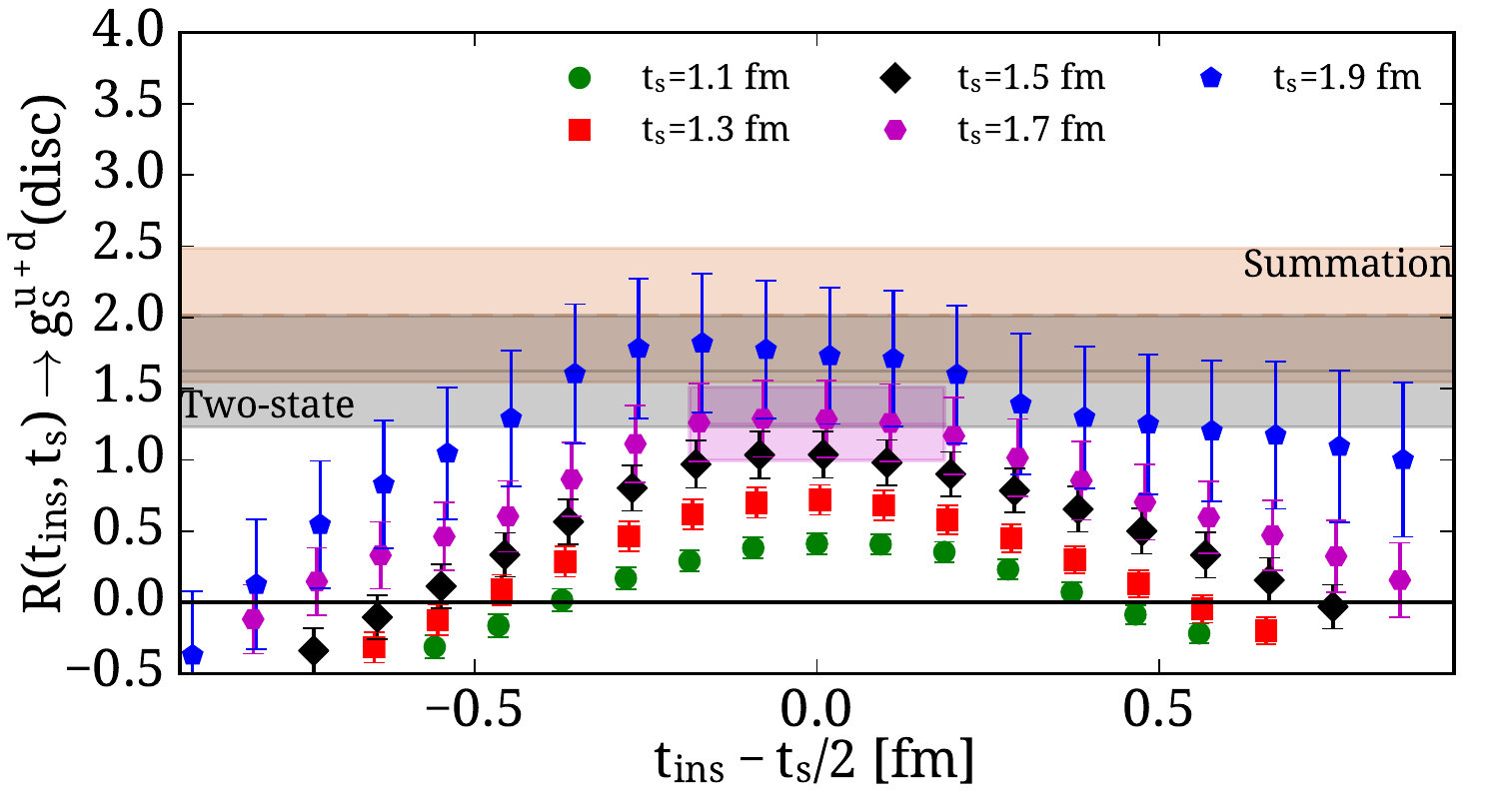}}
\end{minipage}\hfill
\begin{minipage}{0.49\linewidth}
 {\includegraphics[width=\linewidth]{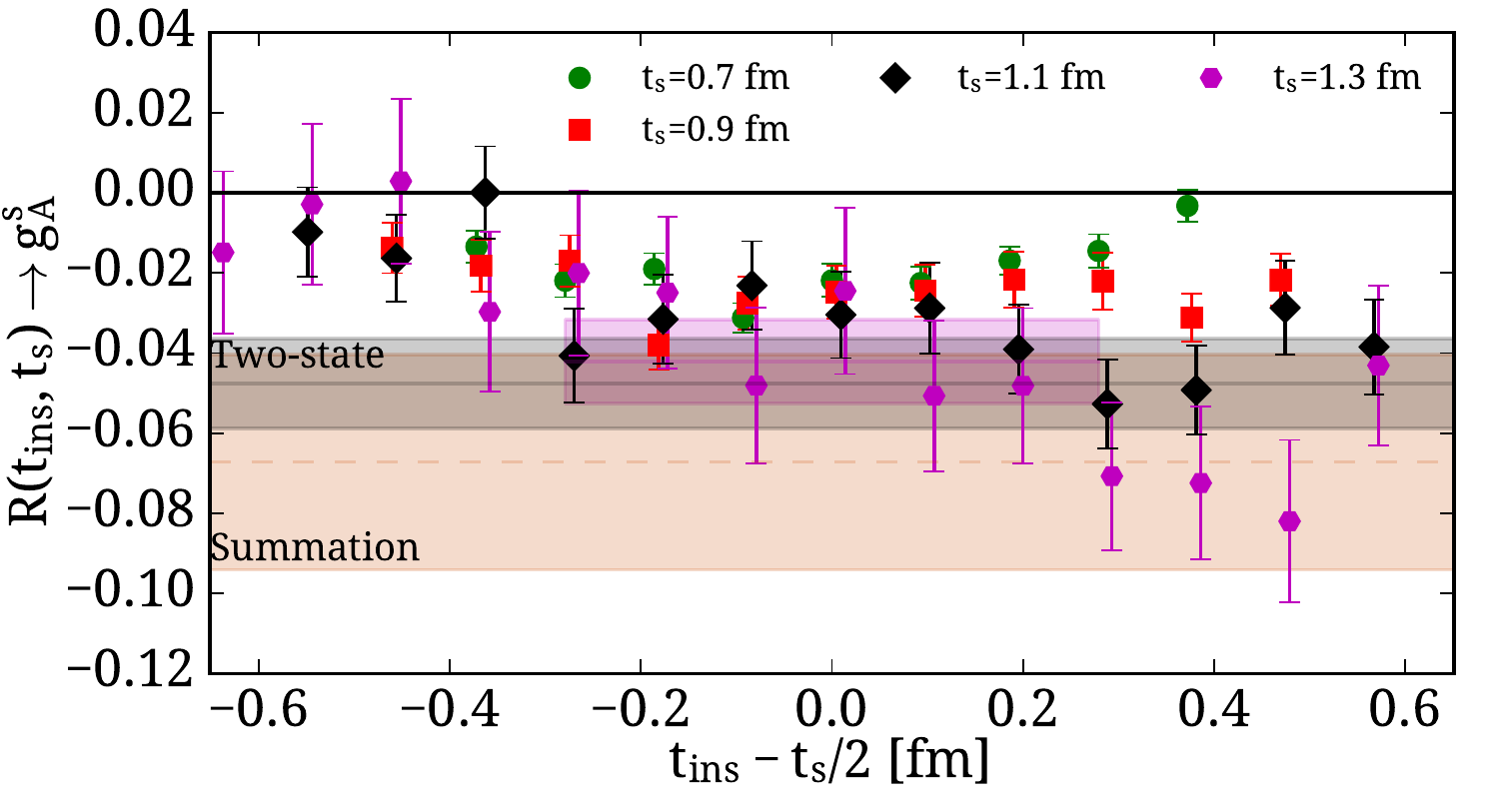}}\\
\includegraphics[width=\linewidth]{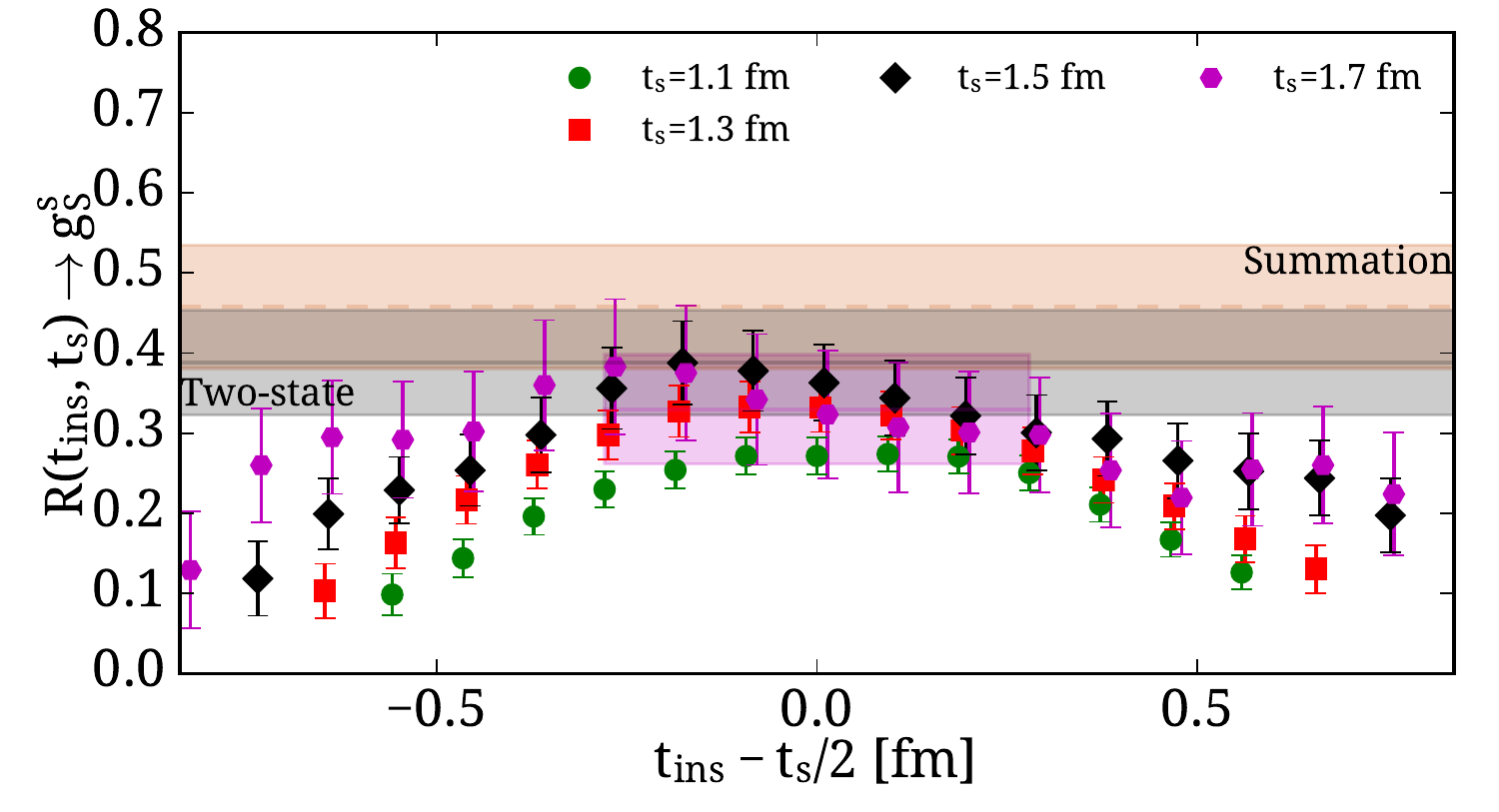}
\end{minipage}
\caption{Disconnected contribution to the isoscalar axial  (upper left) and
strange axial charge (upper right) and isoscalar scalar (lower left) and strange scalar (lower right) charge of the nucleon for the physical TM ensemble as $t_s-t_0$ increases
. The scalar charge is computed in  the $\overline{\rm MS}$ at 2~GeV. }
\label{fig:gA disc}
\end{figure}

In order to compute individual quark contributions one needs to evaluate the disconnected contributions to the matrix element. Using recently developed methods that combine exact deflation and stochastic evaluation of the all-to-all quark propagator on graphics cards (GPUs) we are able to reach enough statistics to extract meaningful values. In Fig.~\ref{fig:gA disc} we show the disconnected contribution  to the isoscalar axial charge $g_A^{u+d}$ and strange $g_A^s$. Both contributions are non-zero and have to be taken into account.  We find for the disconnected contribution $g_A^{u+d}=  -0.15(2)$ using  854,400  statistics 
and combining with the isovector we obtain  $g_A^u = 0.826(26)$ and $g_A^d = -0.386(14)$ while $g_A^s= -0.042(10)$ computed with 861,200 statistics. The corresponding values for the scalar charge are 
$g_S^{u+d}$= 8.25(51)(13) (conn) and 1.25(26) (disconn) yielding $g_S^u = 5.21(31)$ and $g_S^d = 4.28(31)$. For the tensor charge we find $g_T^{u+d}$=0.584(16)(17) (conn) and 0.0007(11) (disconn) yielding  $g_T^u = 0.795(13)$, $g_T^d = -0.210(13)$,
 where the first error is statistical and the second error on the connected is the systematic determined by the difference between the values from the plateau and two-state fits.

Having the scalar charge, we can obtain the quark content of the nucleon, given by
$\sigma_f \equiv m_f\langle N|{\bar q}_f q_f|N\rangle$. Apart from  measuring the explicit breaking of chiral symmetry, these quantities  
represent the largest uncertainty in interpreting experiments for dark matter searches.
With our increased statistics we find  $\sigma_{\pi N}= 36(2)~$MeV, $\sigma_s= 37(8)$~MeV,  $\sigma_c=83(17)$~MeV\cite{Abdel-Rehim:2016won}. These values are in agreement with other recent lattice QCD results but they are in tension with recent phenomenological extractions using Roy-Steiner equations, an issue that needs to be further investigated.

\subsection{Electromagnetic form factors}
The electromagnetic form factors $F_1$ and $F_2$ of the nucleon can be extracted from the nucleon matrix element of the electromagnetic current 
\be
\langle N(p',s^\prime) |j^\mu |N (p,s)\rangle  
= \bar u_N (p',s^\prime)  \left[\gamma^\mu { F_1(q^2)}+\frac{i\sigma^{\mu\nu}q_\nu}{2m}{F_2(q^2)} \right]u_N(p,s).
\ee
The electric and magnetic Sachs form factors are given by $G_E(q^2)=F_1(q^2)+\frac{q^2}{4m_N^2}F_2(q^2)$ and $ G_M(q^2)=F_1(q^2)+F_2(q^2)$, respectively.
Having the form factors one can determine the root mean square (r.m.s) radius by finding the slope in the limit of zero momentum transfer squared $q^2$. A ground breaking experiment using muonic hydrogen extracted the proton charge radius ten times more accurately than what was determined previously from electron scattering. The muonic hydrogen value is smaller by about seven standard deviations    from the one extracted from electron scattering~\cite{Pohl:2010zza}.
The Mainz A1 collaboration at MAMI has measured the electric form factor at low $Q^2=-q^2$ and finds $r_p = 0.879(5)_{\rm stat}(4)_{\rm syst}(2)_{\rm model}(4)_{\rm group}$ fm in agreement with the CODATA06 value of 0.8768(69) fm~\cite{Bernauer:2013tpr}. 
Other analyses of electron scattering data that include the Mainz data yield consistency with the muonic hydrogen determination see e.g. Refs.~\cite{Griffioen:2015hta,Lorenz:2012tm,Higinbotham:2015rja}. For a review see Ref.~\cite{Hill}. 

\begin{figure}[h]
\begin{minipage}{0.49\linewidth}
\includegraphics[width=\linewidth]{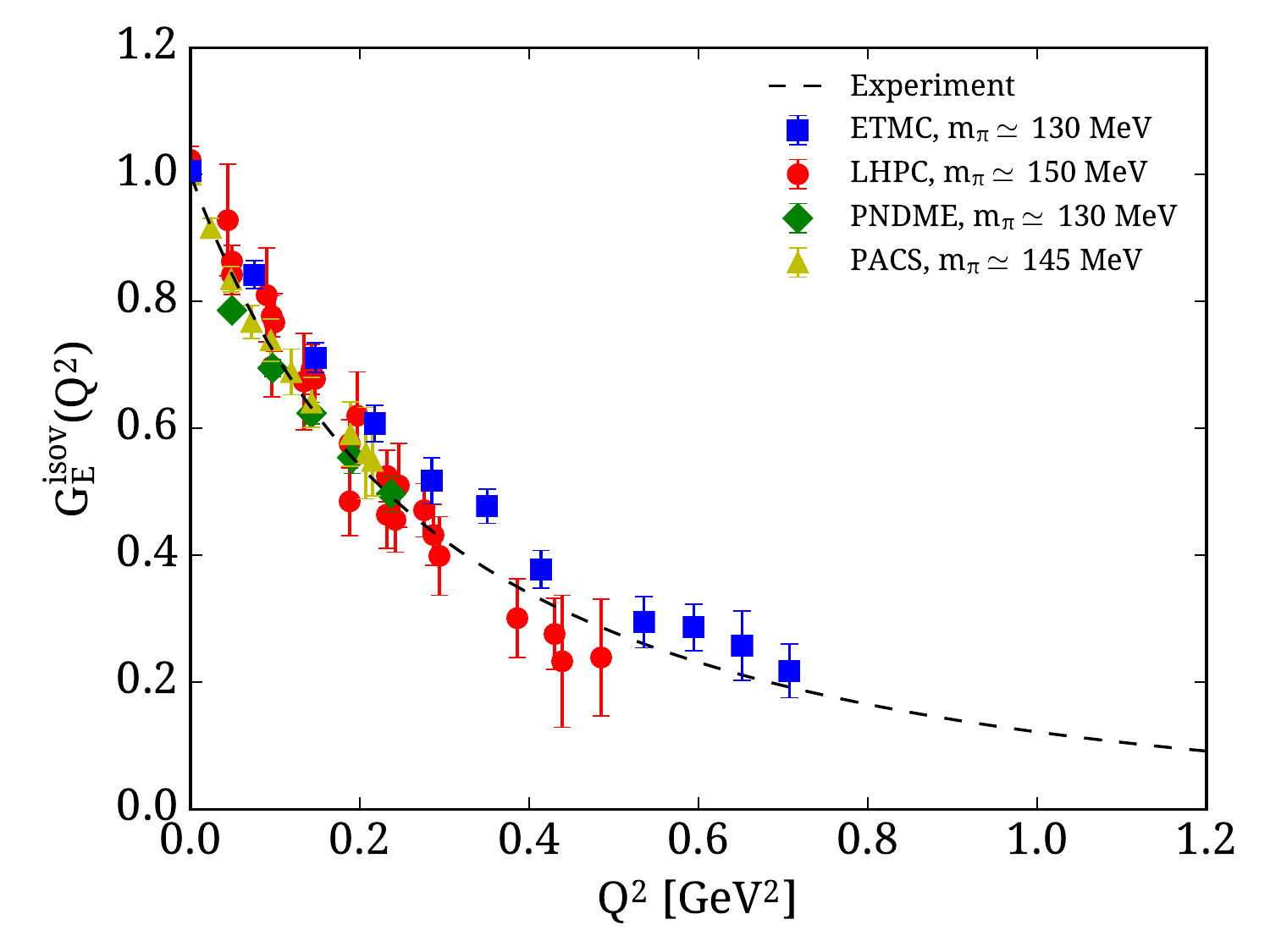}
\end{minipage}\hfill
\begin{minipage}{0.49\linewidth}
{\includegraphics[width=\linewidth]{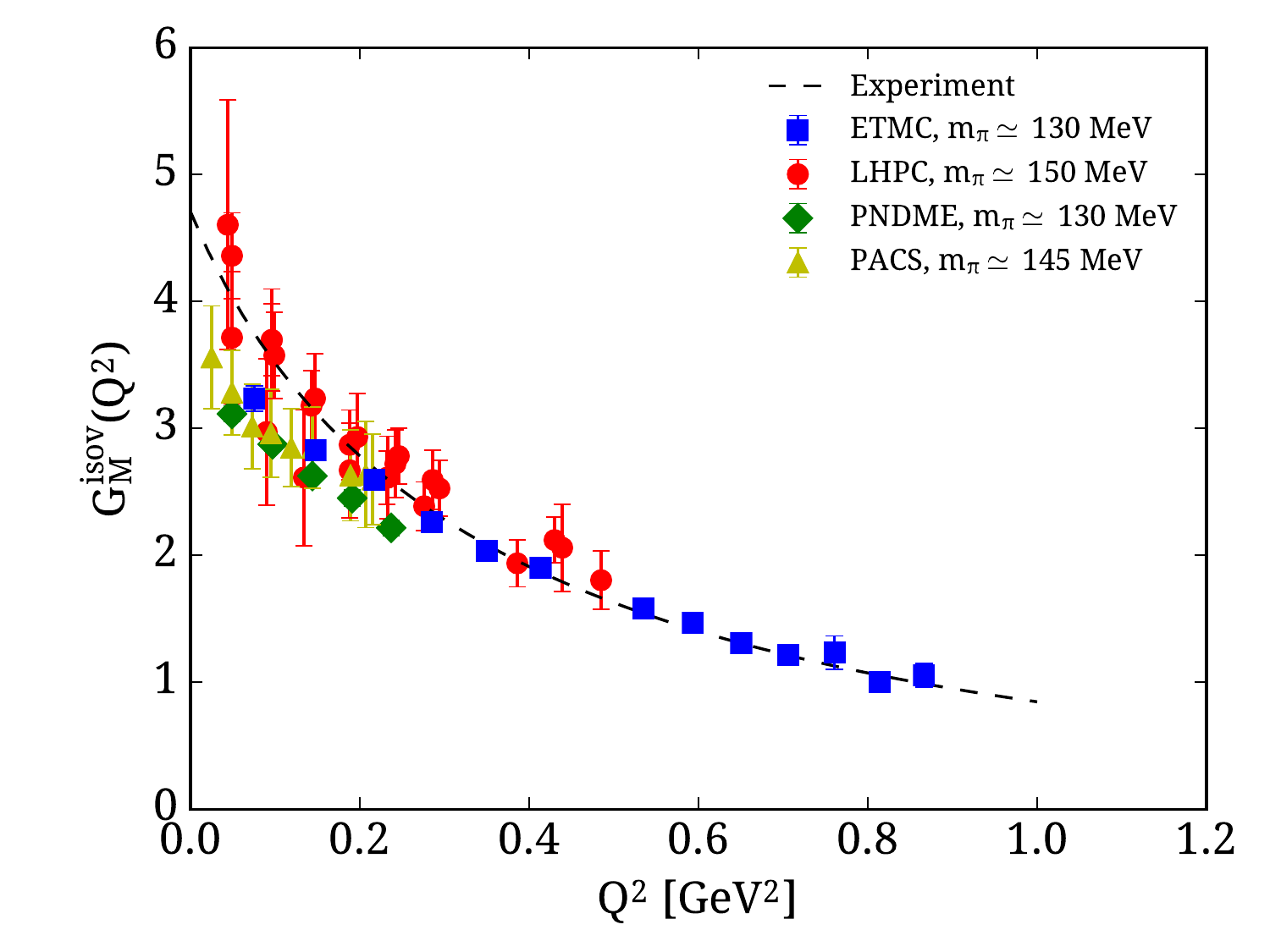}}
\end{minipage}\hfill
\caption{Results on the isovector  $G_E$ (left) and $G_M$ (right) with TF mass fermions for the physical ensemble using   $t_s-t_0=1.7$~fm and 66,000 statistics, $G_M$ with $t_s-t_0=1.3$~fm and 9,300 statistics~\cite{koutsou2016}, from LHPC using $N_f=2+1$ clover fermions, $a=0.116$~fm, $48^4$, summation method with 3 values of $t_s-t_0$ from 0.9 fm to 1.4~fm  and $\sim 7,800$ statistics~\cite{Green:2014xba}, from PNDME mixed action HISQ $N_f=2+1+1$ and  clover valence, $a=0.087$~fm, $64^3\times 96$, summation method with 3 values of $t_s$ from 0.9 fm to 1.4~fm  and $\sim 7,00$ high precision  and $\sim 85,000$ low precision~\cite{Jang}, PACS using $N_f=2+1$ clover fermions, $a=0.085$~fm,  $96^3\times 192$, $t_s=1.3$~fm, 9,300 statistics~\cite{Kuramashi}.}
\label{fig:EM compare}
\end{figure}

An evaluation of the  electromagnetic nucleon form factors  in lattice QCD can be used to determine the charge and magnetic radii. In the past, these quantities were calculated using simulations with pion mass larger than physical yielding radii that were smaller than what one extracted from the measured form factors.
ETMC, LHPC, PACS and PNDME have recently produced results at near physical values of the pion mass. Very recent results, some of which still preliminary,  for the isovector Sachs form factors are shown in Fig.~\ref{fig:EM compare}, which are in good agreement. For the electric form factor lattice results agree well with experimental results, while lattice QCD results tend to underestimate the magnetic form factor $G_M(Q^2)$ at low $Q^2$. Like for $g_A$, lattice systematics need to carefully evaluated.

\begin{figure}
\begin{minipage}{0.7\linewidth}
\includegraphics[width=0.5\linewidth,angle=-90]{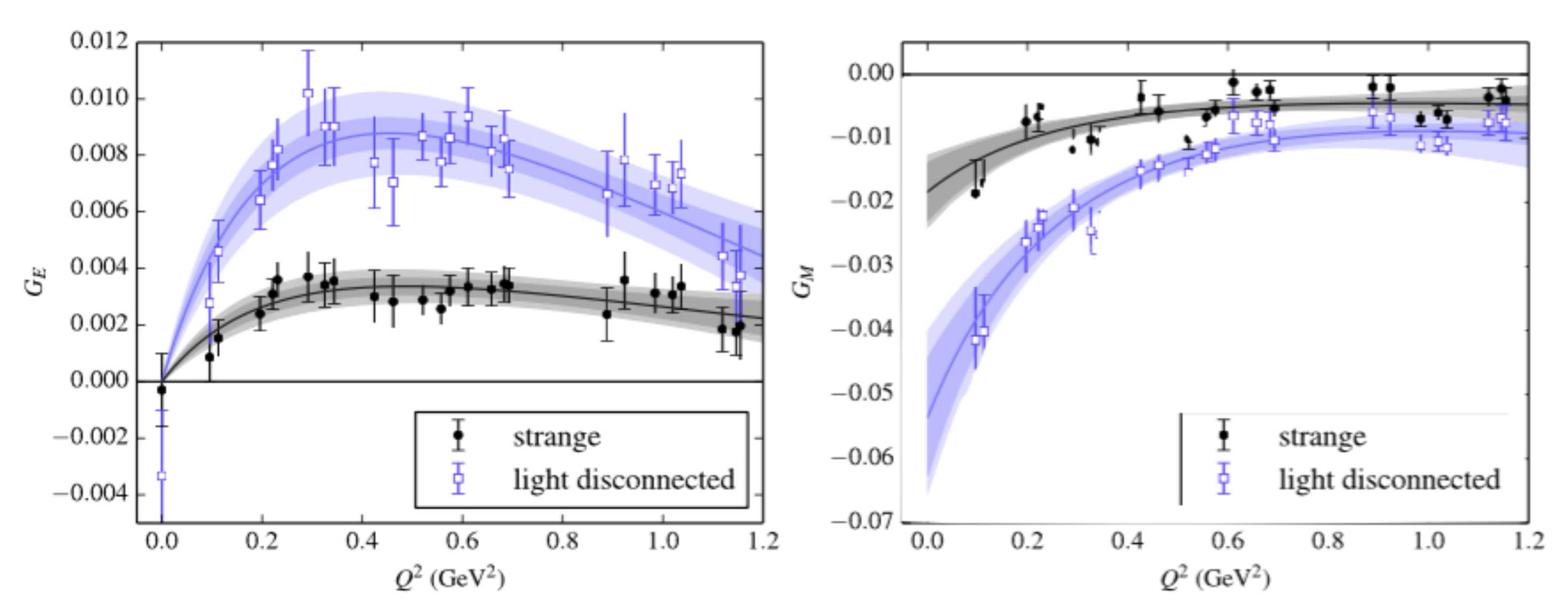}\\
\end{minipage}
\caption{Results on the disconnected contributions to the nucleon electromagnetic form factors with $N_f=2+1$ clover fermions, $m_\pi\sim 320$~MeV taken from Ref.~\cite{Green:2015wqa}.
}
\label{fig:strange FF}
\end{figure}

\begin{figure}
\begin{minipage}{0.49\linewidth}
\includegraphics[width=\linewidth]{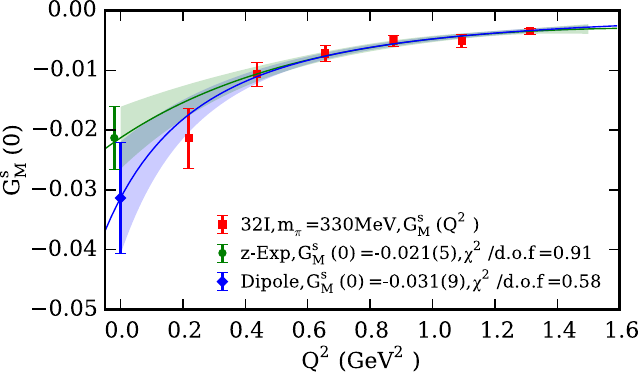}
\end{minipage}
\begin{minipage}{0.49\linewidth}
\includegraphics[width=\linewidth]{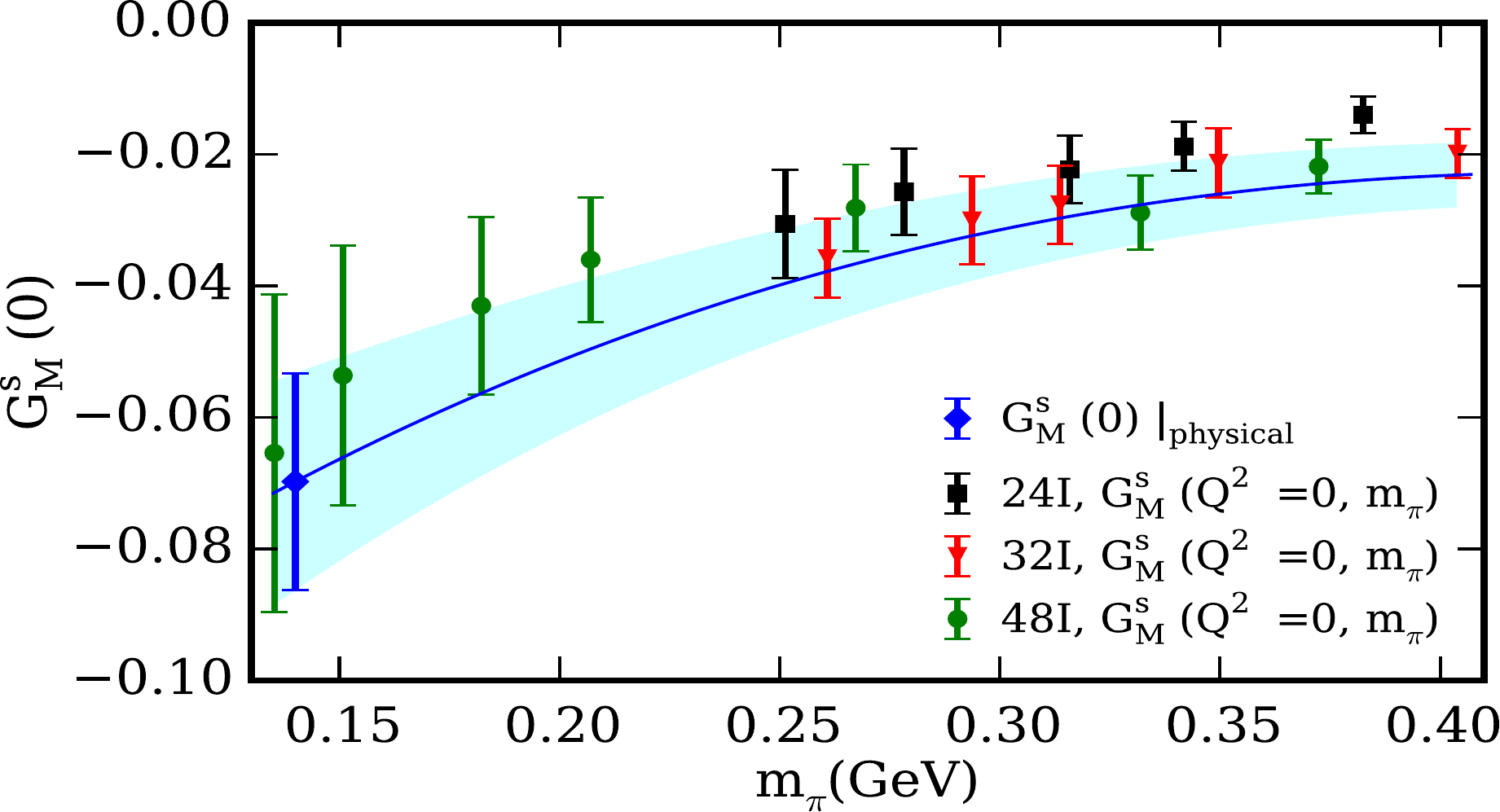}
\end{minipage}
\caption{Overlap valence on $N_f=2+1$ domain wall fermions, $24^3\times 64$, $a=0.11$~fm, $m_\pi=330$~MeV; $32^3\times 64$, $a=0.083$~fm, $m_\pi=300$~MeV and $48^3\time 96$, a=0.11~fm, $m_\pi=139$~MeV~\cite{Sufian:2016pex}.}
\label{fig:chiQCD}
\end{figure}
 
\begin{figure}
\begin{minipage}{0.49\linewidth}
{\includegraphics[width=0.8\linewidth]{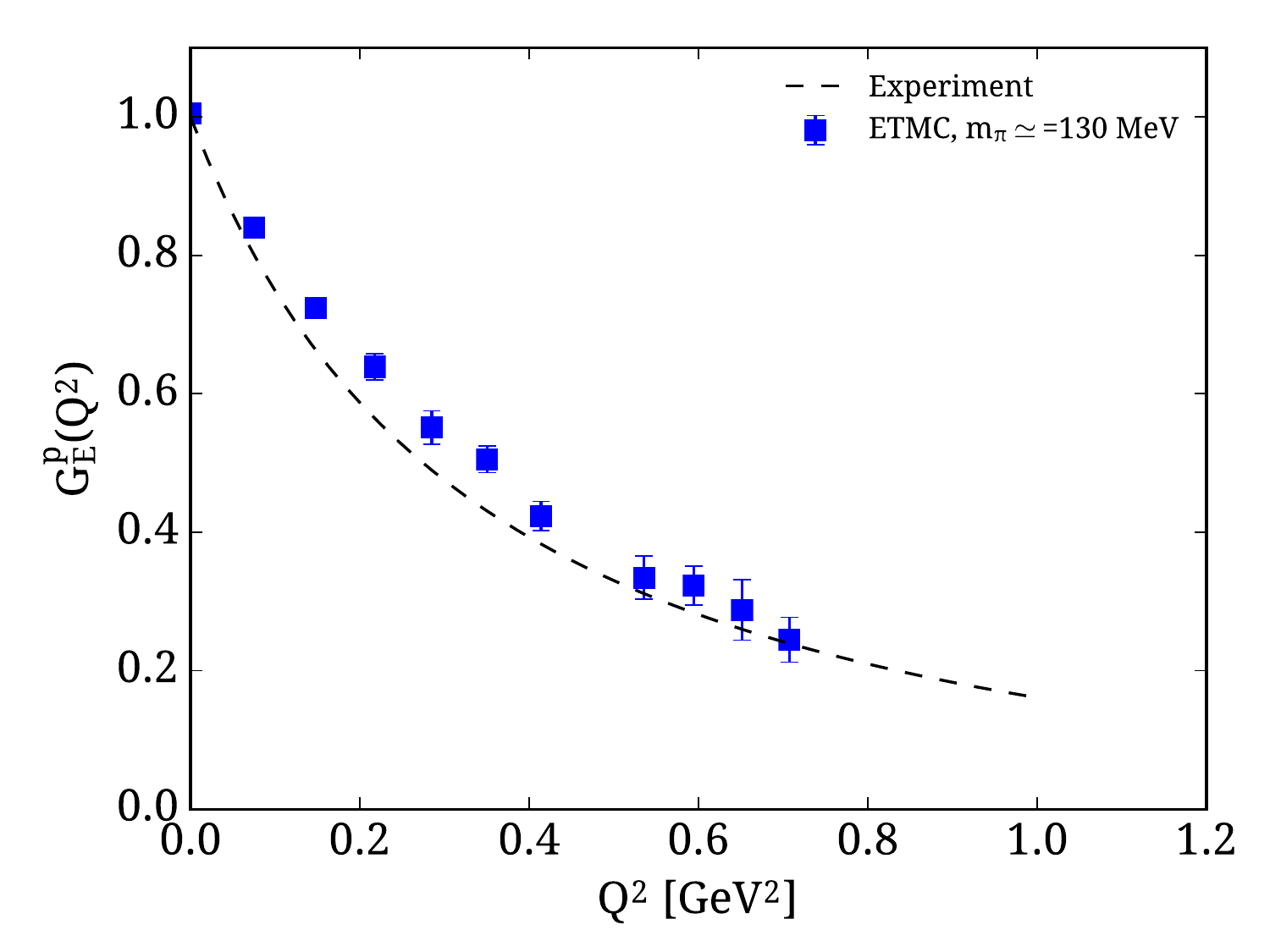}\\
\includegraphics[width=0.8\linewidth]{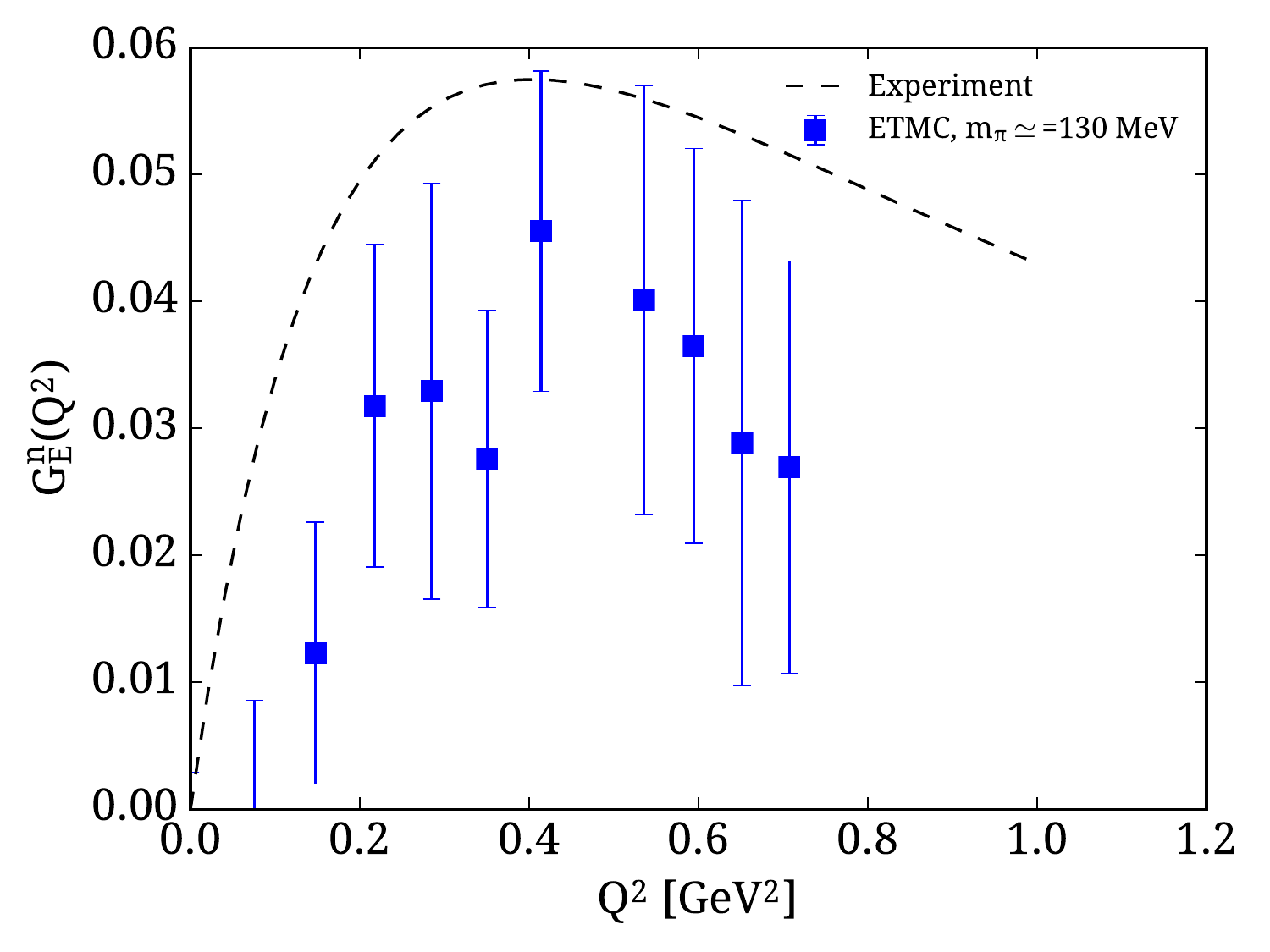}}
\end{minipage}\hfill
\begin{minipage}{0.49\linewidth}
{\includegraphics[width=0.8\linewidth]{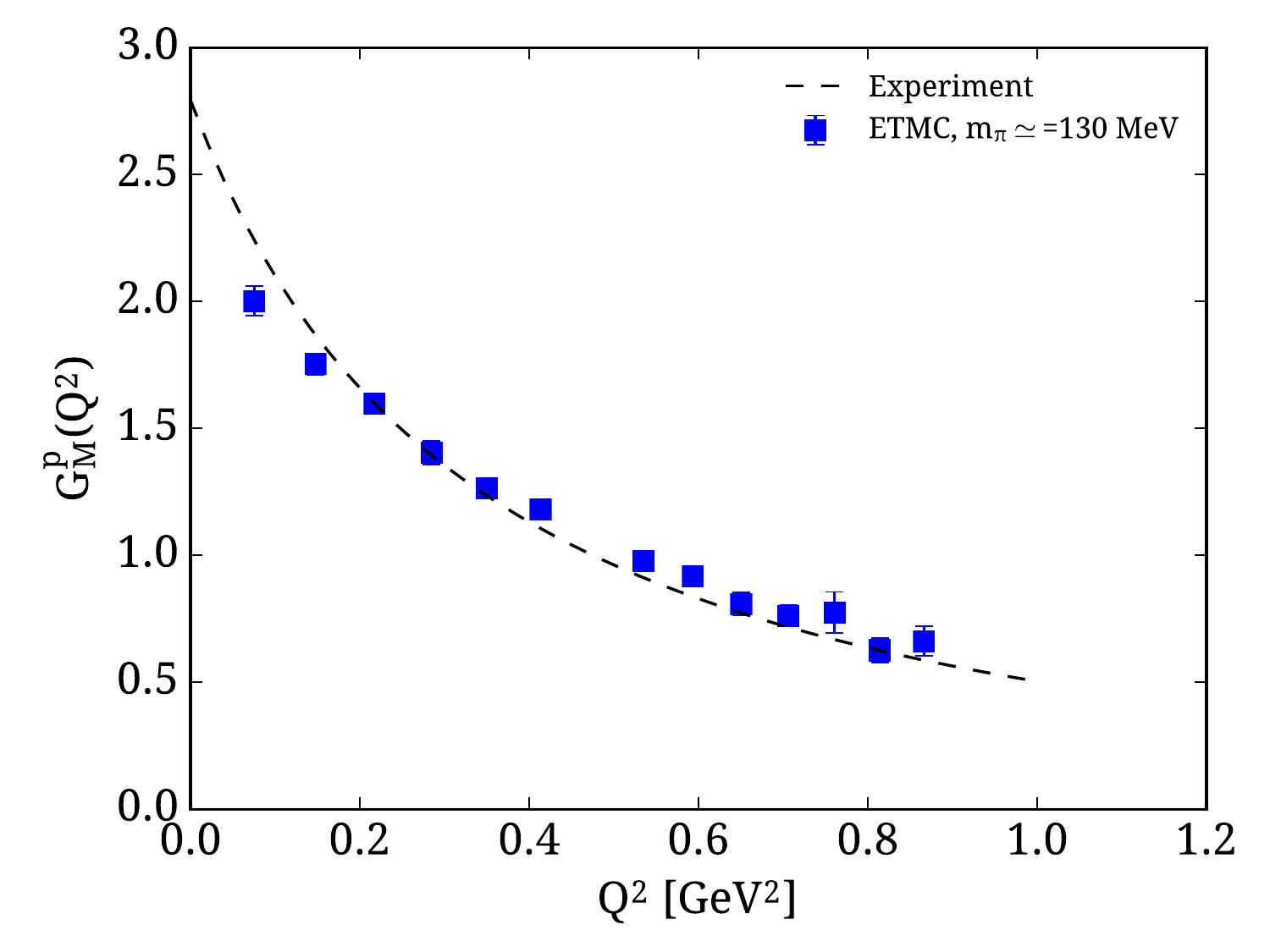}\\
\includegraphics[width=0.8\linewidth]{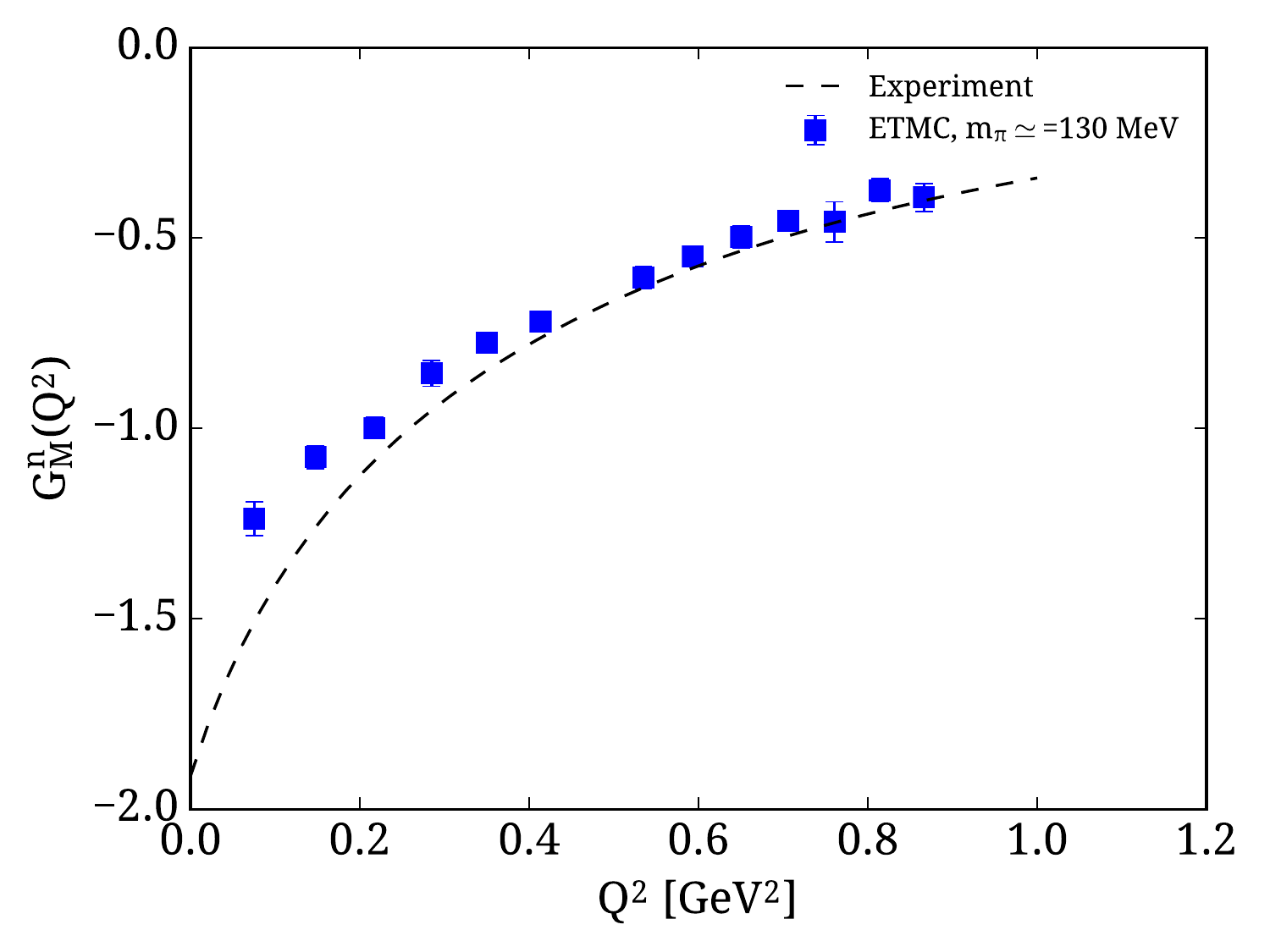}}
\end{minipage}
\caption{Results by ETMC for the physical TM ensemble. $G_E$ is extracted for  $t_s-t_0=1.7$~fm and 66,000 statistics, while for $G_M$ for $t_s-t_)=1.3$~fm and 9,300 statistics~\cite{koutsou2016}.}
\label{fig:GE GM}
\end{figure}

Disconnected contributions to the nucleon  electromagnetic form factors are computed to unprecedented accuracy using hierarchical  probing for an ensemble of clover improved fermions at pion mass of about 320~MeV~\cite{Stathopoulos:2013aci}. In Fig.~\ref{fig:strange FF} we show the disconnected contributions due to the light and strange quark loops. Although for both form factors  they are non-zero for   $G_E(Q^2)$ they are smaller than 1\%. An important conclusion of this study is that the strange form factors are non-zero.
These form factors are determined experimentally  from parity violating $e-N$ scattering. The HAPPEX experiment finds $G_E^s(0.62) = 0.047(34)$ and $G_M^s(0.62)=-0.070(67)$~\cite{Ahmed:2011vp}. 
The $\chi$QCD collaboration has also computed  $G_M^s$ using overlap valence on $N_f=2+1$ DW fermion sea for three ensembles of  $m_\pi=330$~MeV, 300~MeV and 139~MeV. The results for $G_M^s(0)$ are shown in Fig.~\ref{fig:chiQCD} as a function of the pion mass. $G_M^s(0)$ becomes more negative as the pion mass decreases, while the two lattice QCD evaluations are consistent for $m_\pi\sim 320$~MeV.
The predictions from lattice QCD, being very accurate, provide a very valuable input. 

In Fig.~\ref{fig:GE GM} we show the electromagnetic form factor for the proton and neutron for the physical TM ensemble. Only the connected contributions are evaluated. As can be seen, lattice QCD results are consistent with the Kelly's parameterization of the experimental data for the proton electric $G_E^p(Q^2)$ and magnetic $G_M^p(Q^2)$  form factors with small deviations observed for $G_M^p(Q^2)$  at low $Q^2$ and for the slope of $G_E^p(Q^2)$. For the neutron, $G_M^n(Q^2)$ shows larger deviations at low $Q^2$  agreeing with experiment for $Q^2>0.5$~GeV$^2$.  Lattice QCD results on $G_E^n(Q^2)$ yield non-zero values with the same qualitative behaviour as the experimental data, albeit with large statistical errors.  Efforts to compute the disconnected contributions and quantify lattice systematics are ongoing.

\subsection{Electromagnetic radii $\langle r_E^2\rangle$, $\langle r_M^2\rangle$}

\begin{figure}[h]
\begin{minipage}{0.49\linewidth}
{\includegraphics[width=\linewidth,]{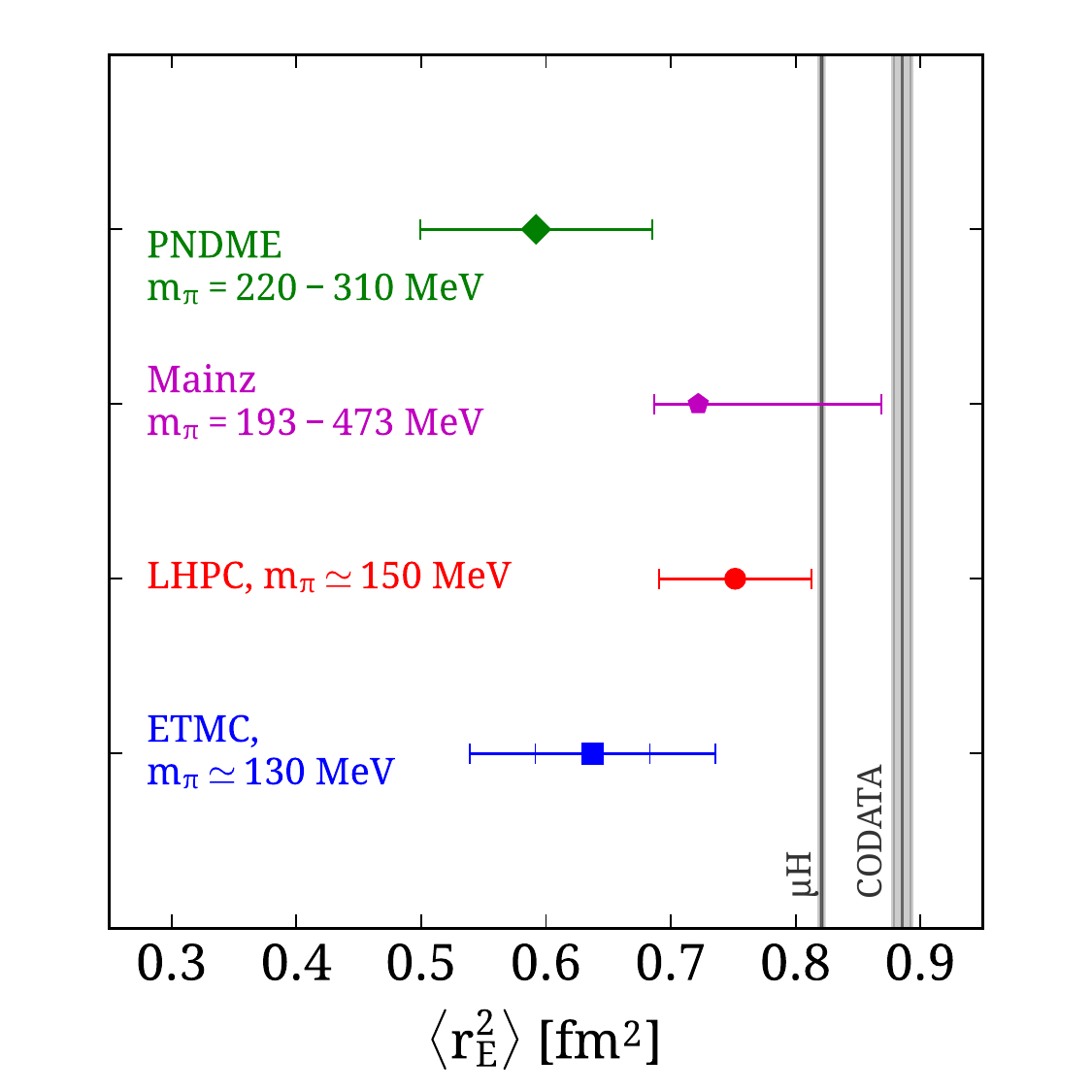}}
\end{minipage}\hfill
\begin{minipage}{0.49\linewidth}
{\includegraphics[width=\linewidth]{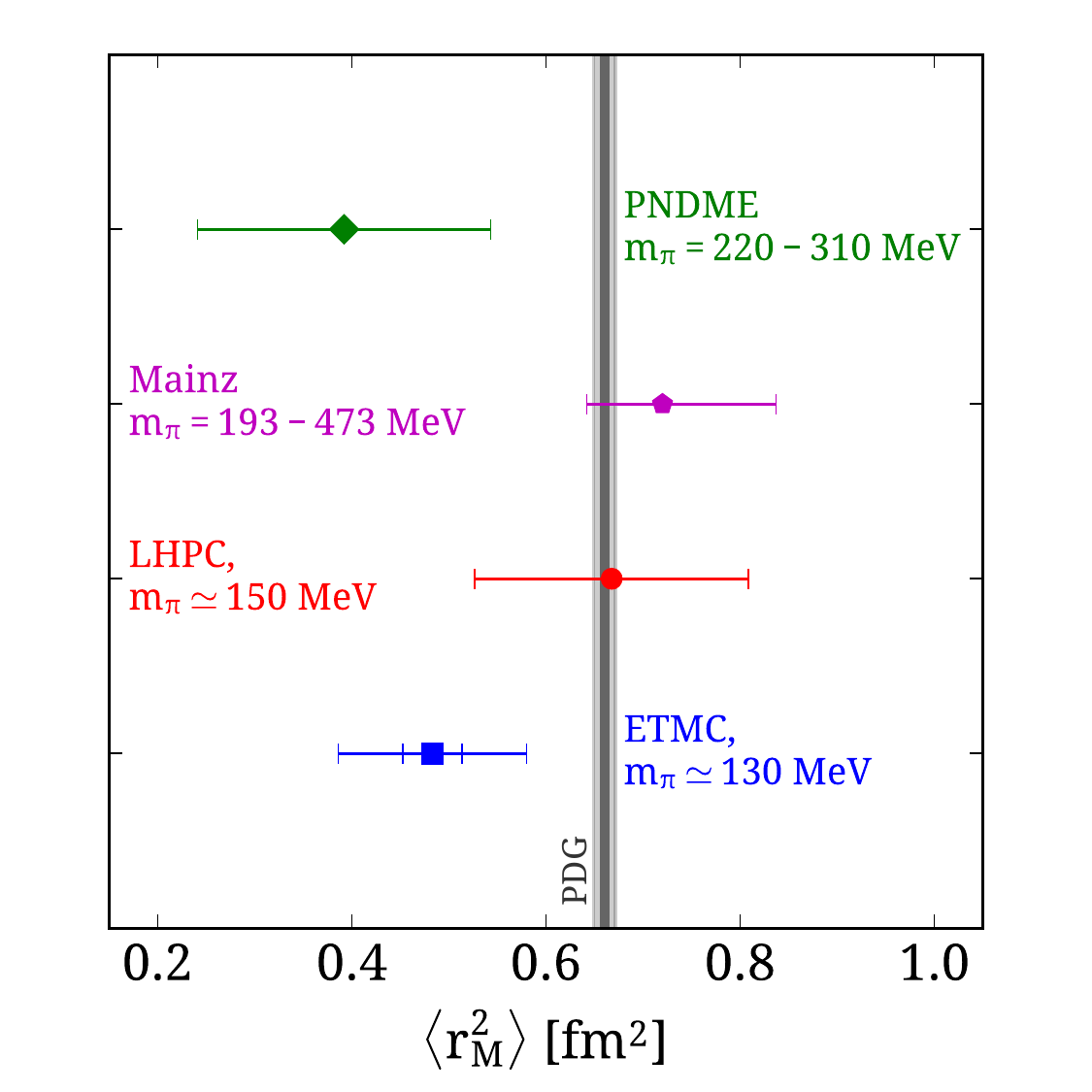}}
\end{minipage}
\caption{Lattice QCD results on the proton charge (left) and magnetic (right) r.m.s. radius. The gray bands are the muonic and electron scattering determination of the proton charge r.m.s radius. The ETMC results are computed using the physical TM ensemble with  $r_E$ (left) extracted from the slope of $G_E$ computed with  $t_s-t_0=1.7$~fm and 66,000 statistics, and $r_M$ (right) from $G_M$ with $t_s-t_0=1.3$~fm and 9,300 statistics. PNDME are taken from Ref.~\cite{Bhattacharya:2013ehc},  LHPC from Ref.~\cite{Green:2014xba} and CLS/Mainz from Ref.~\cite{Capitani:2015sba}.}
\label{fig:radii}
\end{figure}

The slope of the form factors  at $Q^2\rightarrow 0$ yields the radius, namely $\langle r_{EM}^2\rangle =-\frac{6}{G_{EM}(0)}\frac{dG_{EM}(Q^2)}{dQ^2}|_{Q^2=0}$.
Thus to extract the radii we need an Ansatz for the $Q^2$-dependence of the form factors. Usually one considers a dipole fit $\frac{G_0}{(1+Q^2/M^2)^2}$ or the  $z-$expansion. Since on the lattice the momentum is discretized the lowest momentum for periodic boundary conditions is $2\pi/L$, introducing a systematic error from the Ansatz used to fit the $Q^2$-dependence in order to extract the slope at $Q^2=0$. In Fig.~\ref{fig:radii} we show the lattice QCD results for the charge and magnetic radii. As can be seen, the errors on the lattice QCD results are still much larger than the experimental error and we thus cannot at this point discriminate between the muonic and electron determination of the radii.

\begin{figure}[h]
   \begin{minipage}{0.49\linewidth}
    \includegraphics[width=\linewidth]{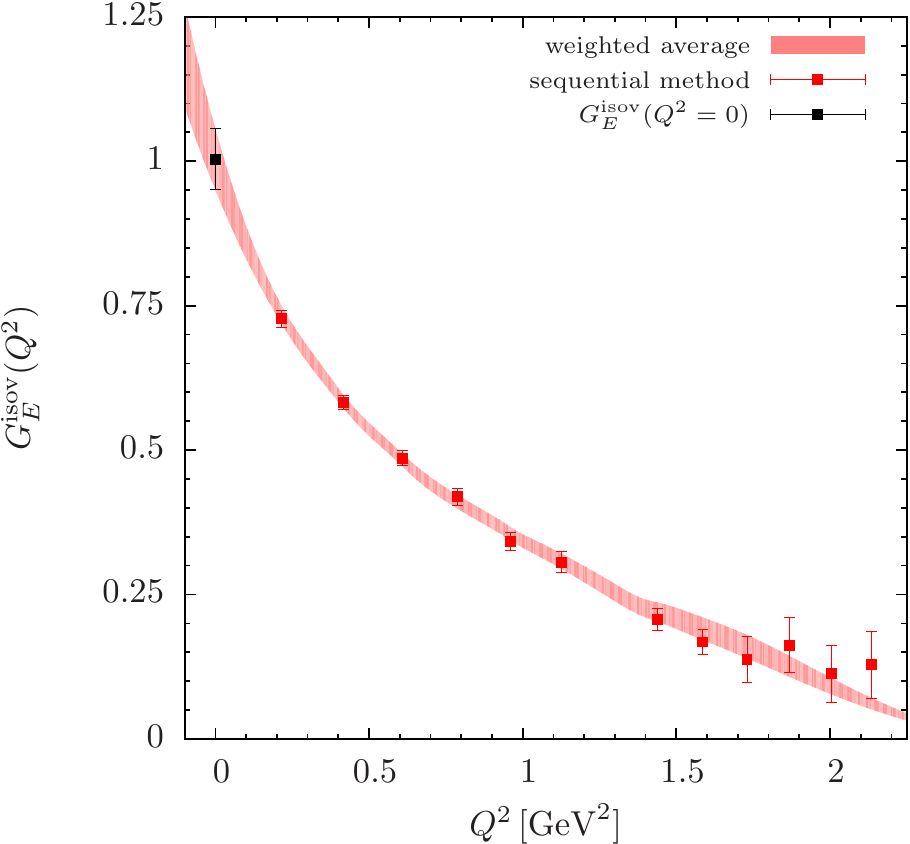}\\
\end{minipage}\hfill
   \begin{minipage}{0.49\linewidth}
 \includegraphics[width=\linewidth]{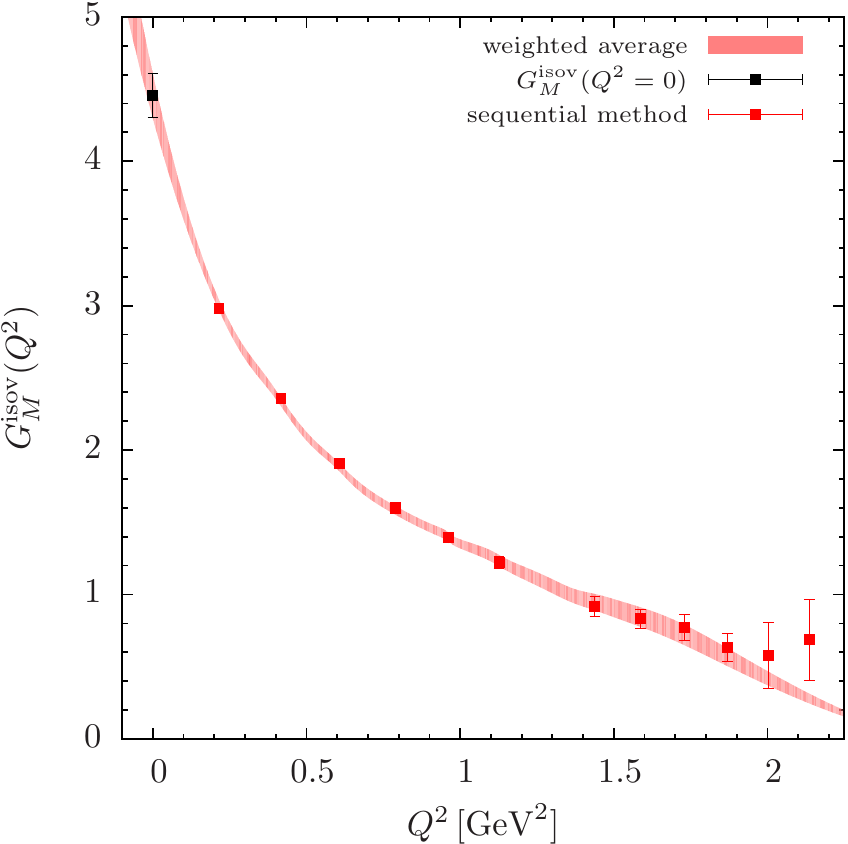}\\
   \end{minipage}
\caption{$G_E^{\rm isov}(0)$ (left) and $G_M^\mathrm{isov}(0)$ (right) from $\mathcal{O}(4700)$ gauge confs of B55 and $(t_\mathrm{s}-t_0)/a=14$ using the momentum elimination method.}
\label{fig:space position method}
\end{figure}

Space position methods can extract form factors at $Q^2=0$ and radii directly avoiding model dependence-fits. We briefly describe this new method using as an example the simpler case of the 
   nucleon isovector magnetic moment $G_M^\mathrm{isov}(0)$. Let us consider the
ratio 
 \be
 \lim\limits_{t\rightarrow\infty} \ \lim\limits_{t_s-t\rightarrow\infty} \frac{G_{3pt}^\mu(\Gamma_\nu,\vec{q},t_s,t_{\rm ins})}{G_{2pt}(t_s)} = {\Pi^\mu\left(\vec{q}, \Gamma_\nu\right)} \,. 
\ee
$G_M$ is then extracted from
 \be
 {{\Pi_i\left(\vec{q},\Gamma_k\right)}} = -C\frac{1}{4m_N} \epsilon_{ijk} {{q_j}} G_M\left(Q^2\right).
\ee
However, due to the momentum factor $q_j$, the magnetic moment $G_M(0)$ cannot be extracted directly. One can eliminate the factor $q_j$ by differentiating w.r.t. $q_j$
\begin{equation}
   {{\lim\limits_{q^2\rightarrow 0} \frac{\partial}{\partial q_j} \Pi_i(t, \vec{q}, \Gamma_k) =  \frac{1}{2m_N} \, \epsilon_{ijk} G_M(0)}} \,.
  \end{equation}
This approach can be tested in the case of  $G_E(0)$ for which we know that the value should be unity. We use the expression
\be
 \Pi_i\left(\vec{q},\Gamma_0\right) = -C\frac{i}{2m_N} {{q_i}} G_E\left(Q^2\right) \ee
 Since the application of a continuous derivative to the three-point  is not strictly correct leading to a time-dependent ratio, we have developed an alternative approach where one takes the derivative in the plateau region~\cite{Alexandrou:2016rbj}. We show the evaluation of 
$G_E(0)$ in Fig.~\ref{fig:space position method}, which indeed yields unity.
In the same figure we also show the corresponding  evaluation of the magnetic moment. These were computed using an ensemble of $N_f=2+1+1$ TM fermions for $m_\pi=373$, volume $32^3\times 64$ and $a=0.082$~fm (refer to as the B55 ensemble). We obtain for ${{G_M^\mathrm{isov}=4.45(15)_\mathrm{stat}}}$ larger than  $3.99(9)_\mathrm{stat}$ extracted from a dipole fit and  closer to the  experimental  value of 4.71.
   Application of this method to extract the radii at the physical point is in progress.

\subsection{Nucleon axial form factors}

\begin{figure}
\begin{minipage}{0.49\linewidth}
\includegraphics[width=\linewidth]{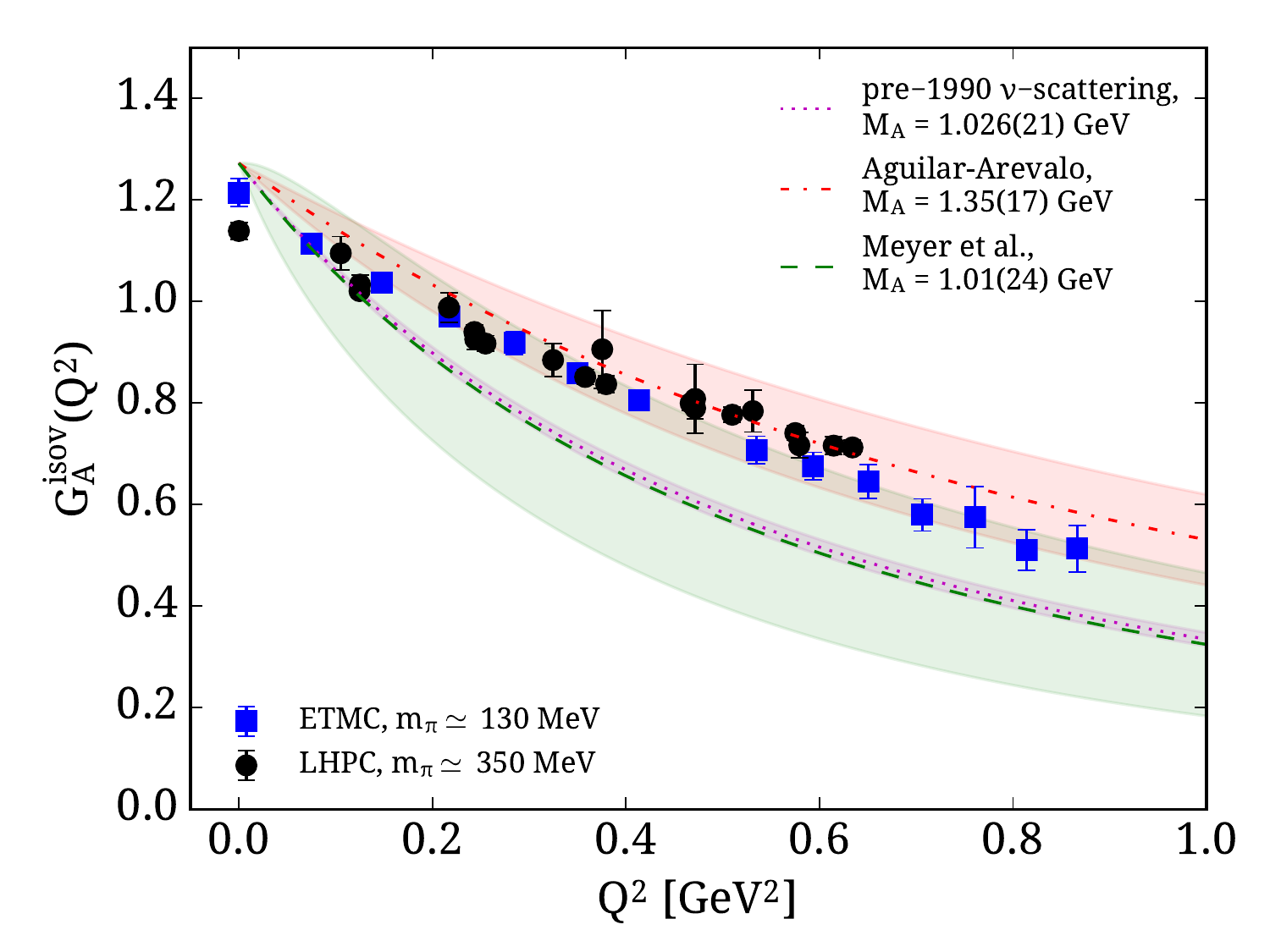}
\end{minipage}\hfill
\begin{minipage}{0.49\linewidth}
\includegraphics[width=\linewidth]{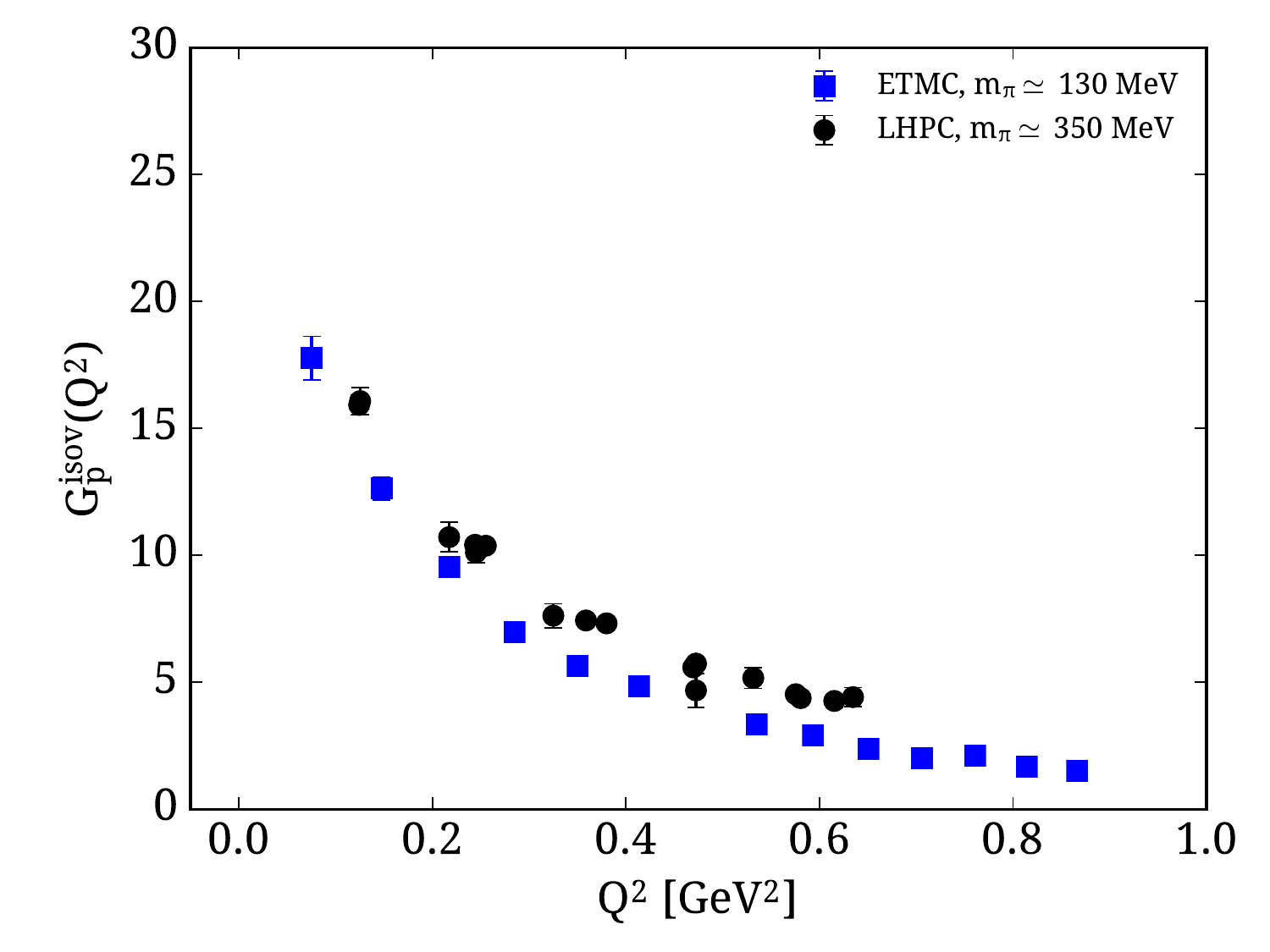}
\end{minipage}\hfill
\caption {The axial (left) and induced pseudoscalar  (right) form factors for the physical TM ensemble where we use  $t_s-t_0=1.3$~fm and 9,300 statistics. The black squares  show results by LHPC using $N_f=2+1$ clover fermions with $m_\pi=317$~MeV~\cite{Green2016}. }
\label{fig:axial ff}
\end{figure}

\begin{figure}[h]
\begin{minipage}{0.33\linewidth}
{\includegraphics[width=\linewidth]{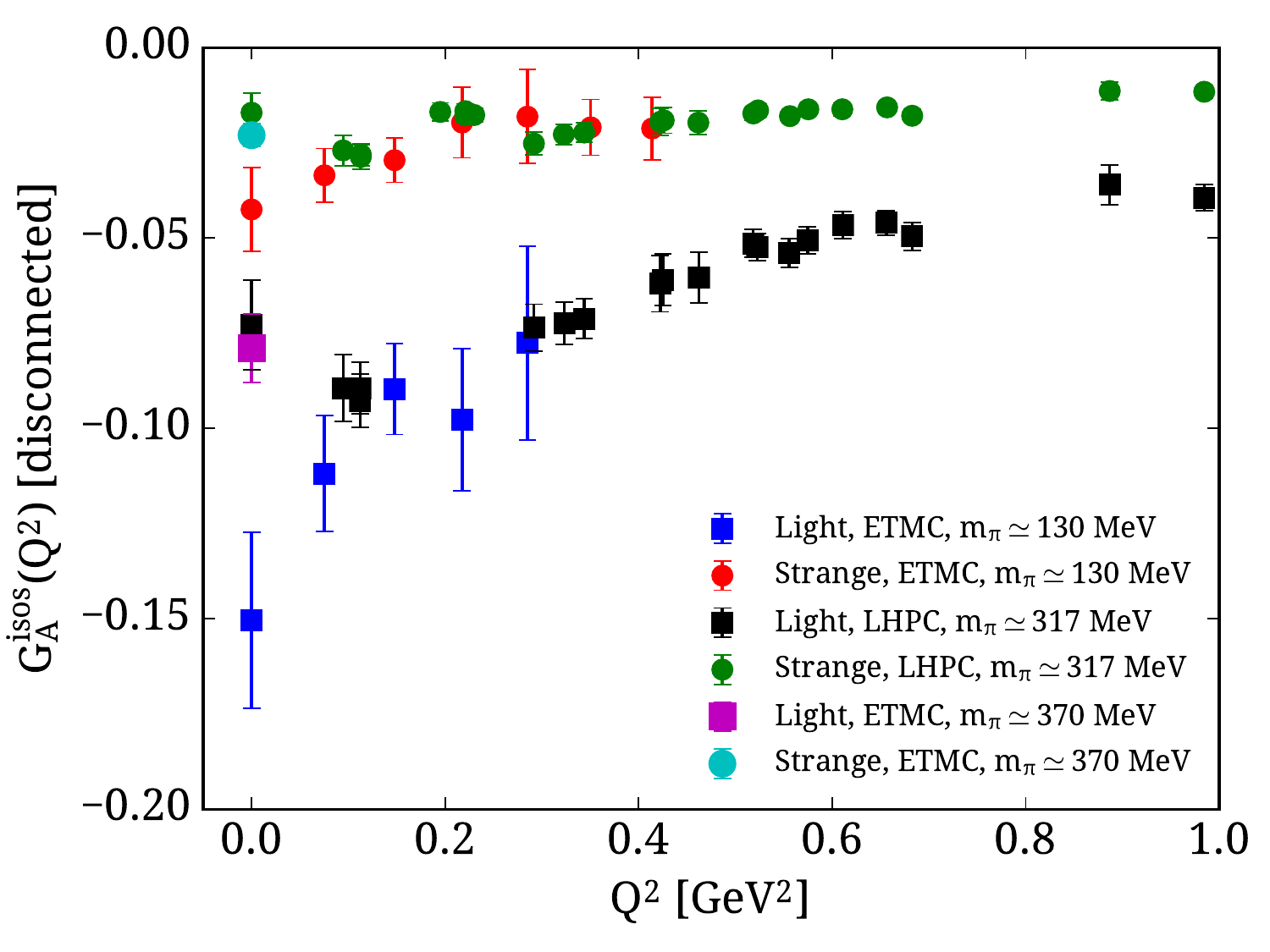}}
\end{minipage}\hfill
\begin{minipage}{0.33\linewidth}
{\includegraphics[width=\linewidth]{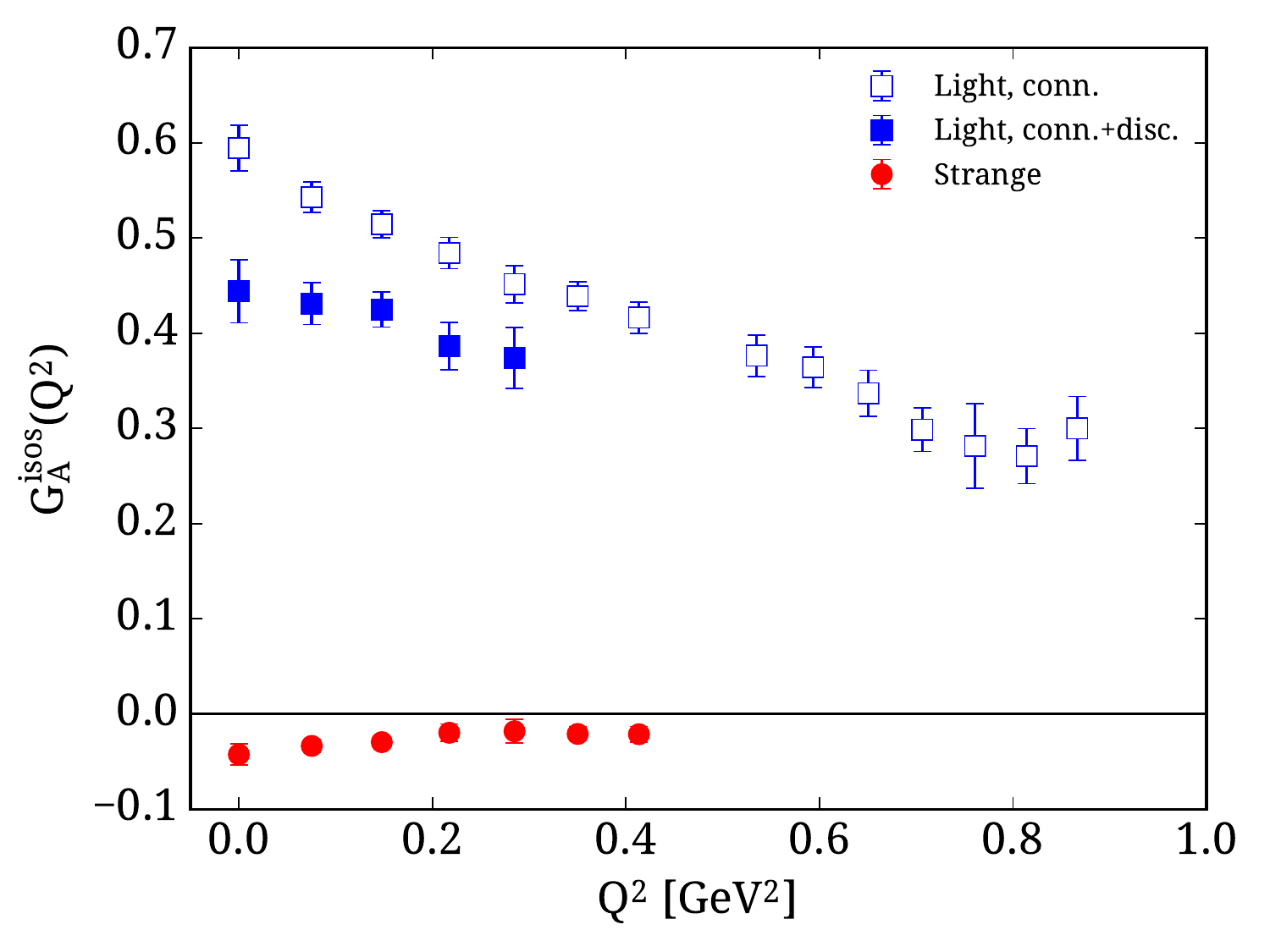}}
\end{minipage}\hfill
\begin{minipage}{0.3\linewidth}
{\includegraphics[width=\linewidth]{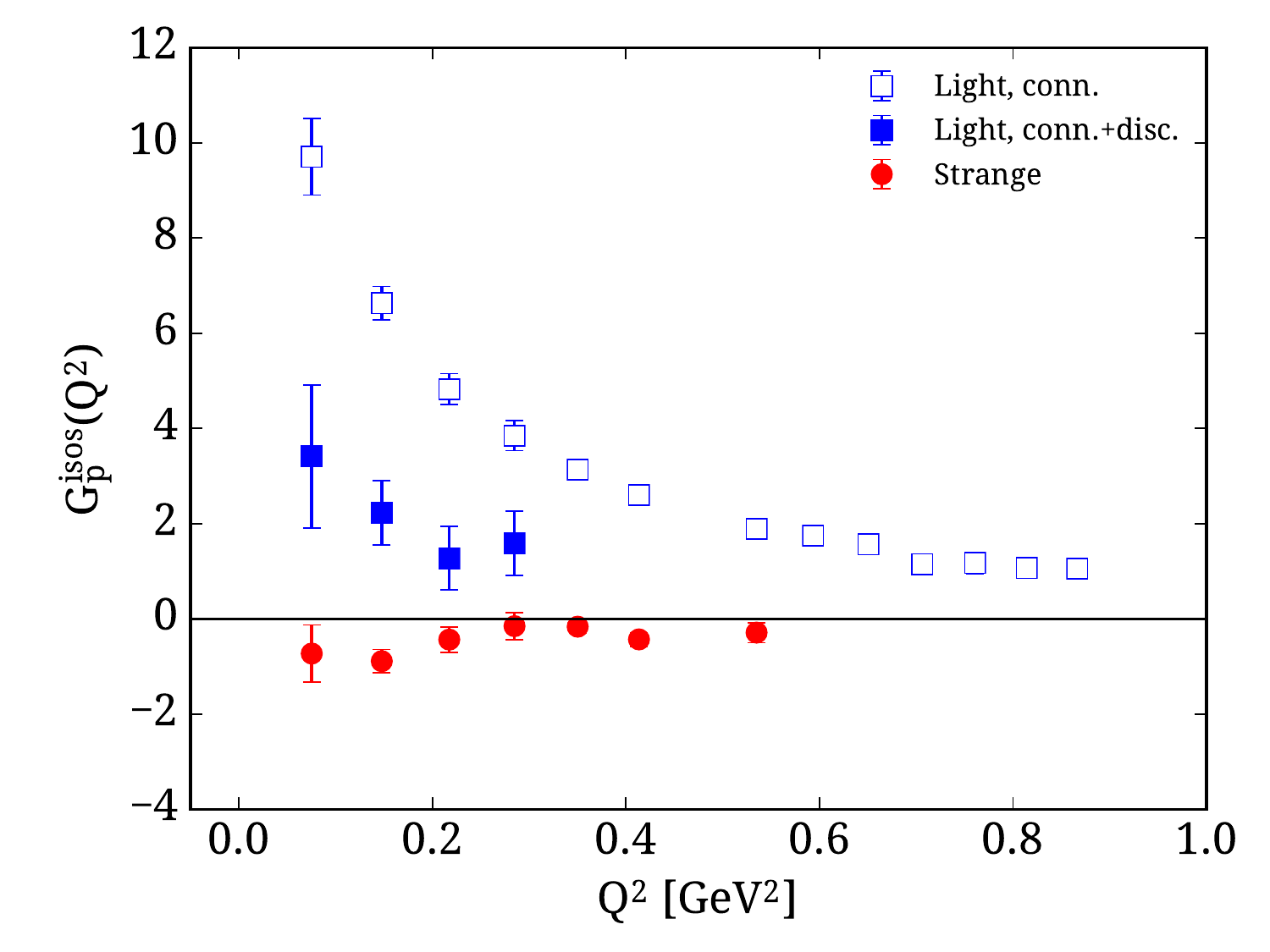}}
\end{minipage}
\caption{Results on the disconnected contribution to the isoscalar axial (left), the axial (middle) and induced pseudoscalar (right) form factors using the physical TM ensemble with 855,000 statistics for the disconnected contributions~\cite{Abdel-Rehim:2016pjw}.  }
\label{fig:axial disc}
\end{figure}

We briefly comment on our recent results on the nucleon axial form factors (see Ref.~\cite{Collins} for a review). In Fig.~\ref{fig:axial ff} we show results for the isovector axial form factors for the physical TM ensemble as well as for $N_f=2+1$ clover fermions with $m_\pi=317$~MeV from LHPC. The disconnected contributions to the isoscalar axial form factors  computed recently  are sizable as shown in Fig.~\ref{fig:axial disc} and need to be taken into account in the discussion of the spin carried by quark in the  nucleon.

\section{Generalized Parton Distributions}

Another set of observables that probes the structure of hadrons are Generalized Parton Distributions (GPDs) measured in deep inelastic scattering.
These are matrix elements in the infinite momentum frame but
factorization leads to a set of three twist-two local operators, namely the vector operator 
${\cal O}_{V^a}^{\mu_1 \cdots \mu_n}=\bar{\psi}(x)\gamma^{\{\mu_1}i\Dlr^{\mu_2}\ldots i\Dlr^{\mu_n \}}\frac{\tau^a}{2}\psi(x)$, the  axial-vector operator
${\cal O}_{A^a}^{\mu_1 \cdots \mu_n}=\bar{\psi}(x)\gamma^{\{\mu_1}i\Dlr^{\mu_2}\ldots i\Dlr^{\mu_n \}}\gamma_5\frac{\tau^a}{2}\psi(x)$ and the  tensor operator
${\cal O}_{T^a}^{\mu_1 \cdots \mu_n}=\bar{\psi}(x)\sigma^{\{\mu_1,\mu_2}i\Dlr^{\mu_3}\ldots i\Dlr^{\mu_n \}}\frac{\tau^a}{2}\psi(x)$.
In the special case where  we have  no derivatives these yield the usual hadron form factors, while for $Q^2=0$ they reduce to the  parton distribution functions (PDFs) yielding for  instance 
the  average momentum fraction or the unpolarized moment $\langle x \rangle$  in the case of  the one-derivative vector operator.

For a spin-1/2 particle, like the nucleon, the decomposition of the matrix element of the one-derivative vector operator is given by

\be 
\langle N(p^\prime,s^\prime) | {\cal O}_{V^3}^{\mu\nu}| N(p,s) \rangle = 
    \bar u_N(p^\prime, s^\prime) 
     \Biggl[  {A_{20}(q^2)} \gamma^{\{\mu}P^{\nu\}}+{B_{20}(q^2)} \frac{i\sigma^{\{\mu \alpha}q_\alpha P^{\nu\}}}{2m}
+C_{20}(q^2) \frac{q^{\{\mu}q^{\nu\}}}{m} \biggr] \frac{1}{2}u_N(p,s) \quad.
\label{vector derivative}
\ee
Extracting $A_{20}$ and $B_{20}$ is particularly relevant for  understanding the nucleon spin $J^q$ carried by a quark since $J^q=\frac{1}{2}\biggl [A^q_{20}(0)+B^q_{20}(0)\biggr]$ as well as the momentum fraction  $\langle x \rangle_q=A^q_{20}(0)$.

\begin{figure}[h]
\begin{minipage}{0.55\linewidth}
 \includegraphics[width=\linewidth,height=\linewidth]{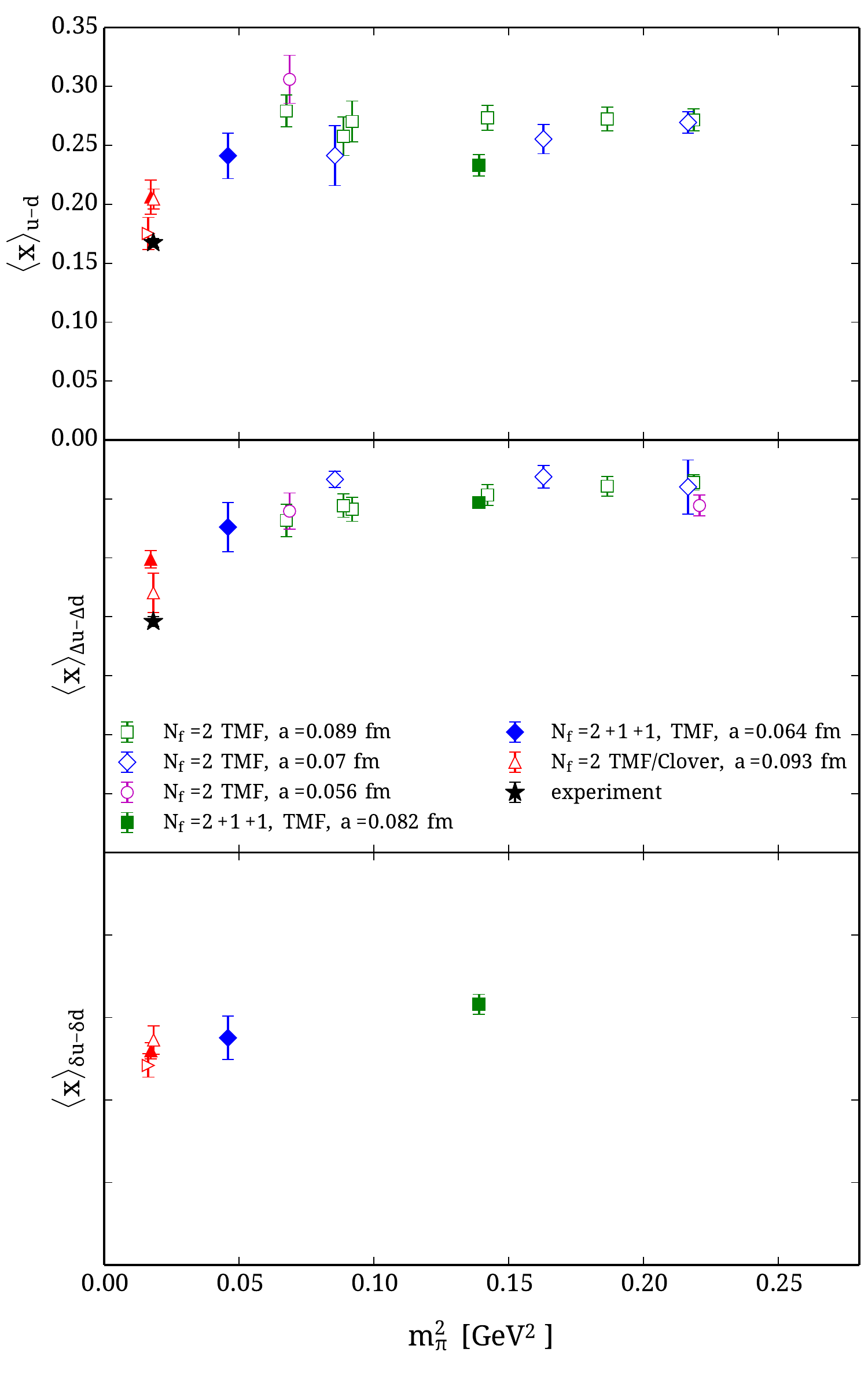}
\caption{The isovector momentum fraction $\langle x\rangle_{u-d}$ (top), the helicity  $\langle x\rangle_{\Delta u-\Delta d}$ (middle) and  transversity $\langle x\rangle_{\delta u-\delta d}$
The lattice QCD results are from Refs.~\cite{Abdel-Rehim:2015pwa}. The experimental values are from Refs.~\cite{Alekhin:2016tkw,Blumlein:2010rn}, respectively.}
\label{fig:moments}\hfill
\end{minipage}\hfill
\begin{minipage}{0.33\linewidth}
{\includegraphics[width=\linewidth]{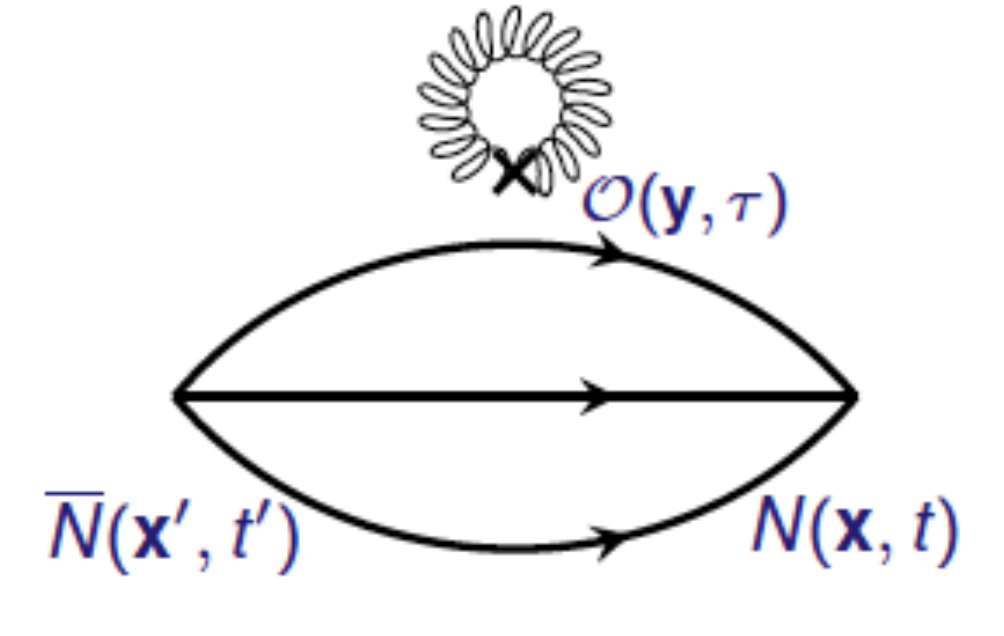}}\\
{\includegraphics[width=0.5\linewidth,angle=-90]{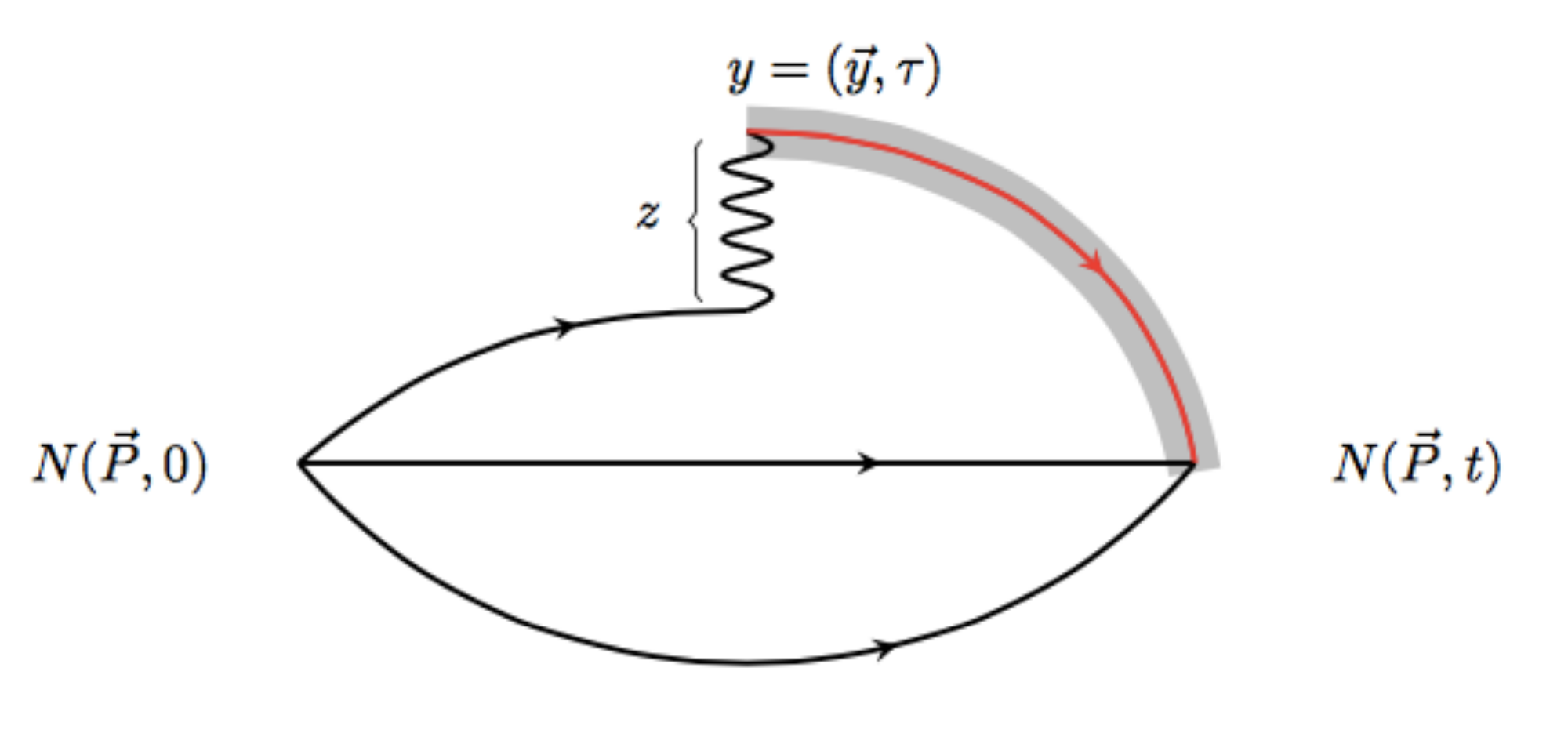}}
\caption{Upper: Gluonic contribution to the nucleon three point function. Lower: Diagrammatic presentation of the three-point function entering the determination of the quasi-parton distribution.}
\label{fig:diagrams2}
\end{minipage}
\end{figure}

A number of collaborations have computed the nucleon matrix elements of the one-derivative vector, axial-vector and tensor operators mostly for pion masses larger than physical. In Fig.`\ref{fig:moments} we show results on the isovector moments using a number of $N_f=2$ and $N_f=2+1+1$ twisted mass gauge configurations including the physical TM ensemble. The sink-source time separation is increased to 1.7~fm and the value of both the unpolarized ($\langle x\rangle_{u-d}\rangle$) and polarized ($\langle x\rangle_{\Delta u-\Delta d}\rangle$) moments decreases approaching their experimental value.  Since both the plateau and two-state fits are consistent while the summation method gives a smaller value increasing the sink-source separation is still necessary to extract the final values. Extracting the value from the plateau method  we find 
 {$\langle x\rangle_{u-d}= 0.194(9)(10)$} and  {$\langle x\rangle_{u+d+s} = 0.74(10)$}, in the $\overline{\rm MS}$ at 2~GeV, where the mixing of 
   $\langle x\rangle_{u+d+s}$ with the gluon operator  is perturbatively computed using one-loop lattice perturbation theory. For the helicity we find  $\langle x\rangle _{\Delta u-\Delta d}$= 0.259(9)(10). 
For the tensor moment the predicted value is $\langle x\rangle_{\delta u- \delta d}$=0.273(17)(18). 
In all cases the first error is statistical and the second systematic determined by the difference between the values from the plateau and two-state fits.

 Gluonic contributions to the  momentum and spin
  in the nucleon can be evaluated in an analogous manner by computing the nucleon matrix element of the gluon operator as shown diagrammatically in Fig.~\ref{fig:diagrams2} and extracting the generalized form factors 
$A_{20}^g(0)$ and $B_{20}^g(0)$ to obtain
 $\langle x \rangle_g= A_{20}^g$
and $J_g = \frac{1}{2} (A_{20}^g + B_{20}^g)$, respectively.
The momentum fraction is extracted from the matrix element  $\langle N|O_{44}-\frac{1}{3}O_{jj}|N\rangle$ at zero momentum, which yields directly $\langle x \rangle_g$, where we considered
the gluon operator $O_{\mu\nu}=-{\rm Tr}[ G_{\mu\rho} G_{\nu\rho}]$. The three-point function  depicted in Fig.~\ref{fig:diagrams2} is a disconnected correlation function and known
  to be very noisy. We employ several steps of
 stout smearing in order to remove gauge ultra-violet fluctuations.
 Results are obtained for the B55 ensemble~\cite{Alexandrou:2013tfa} as well as for the physical TM  ensemble~\cite{Alexandrou:2016ekb}.
 2094 gauge configurations with 100 different source
  positions resulting to more than 200\,000
  measurements are utilized to extract the matrix element at the physical point.
The bare value is  $\langle x \rangle_{g,\mbox{bare}} = 0.318(24)$
and the renormalized
$\langle x\rangle ^R_g = Z_{gg} \langle x\rangle_g + Z_{gq} \langle x\rangle_{u+d+s}$= 0.273(23)(24) where the 
mixing of the gluon operator  with $\langle x \rangle _{u+d+s}$ is taken into account perturbatively~\cite{Alexandrou:2016ekb}. The systematic error is the difference between using one- and two-levels of stout smearing. 
Having both quark and gluon  momentum fractions we can check the momentum sum $\sum_q\langle x \rangle_q+\langle x \rangle_g=\langle x \rangle_{u+d}^{CI}+\langle x \rangle_{u+d+s}^{DI}+\langle x \rangle_g=1.01(10)(2)$, which is very well satisfied.

\subsection{Nucleon spin}

The contributions to the total spin of the nucleon should add up to one-half. Both quarks and gluons  contribution  to the spin sum:  
 $ \frac{1}{2}=\sum_{q}\underbrace{\left(\frac{1}{2}\Delta \Sigma^q +L^q\right)}_{J^q} +J^g $, where
$J^q=\frac{1}{2}\left(A_{20}^q(0)+B_{20}^q(0)\right)$ and $\Delta \Sigma^q=g_A^q$.

\begin{figure}[h]
\begin{minipage}{0.49\linewidth}
\includegraphics[width=0.98\linewidth]{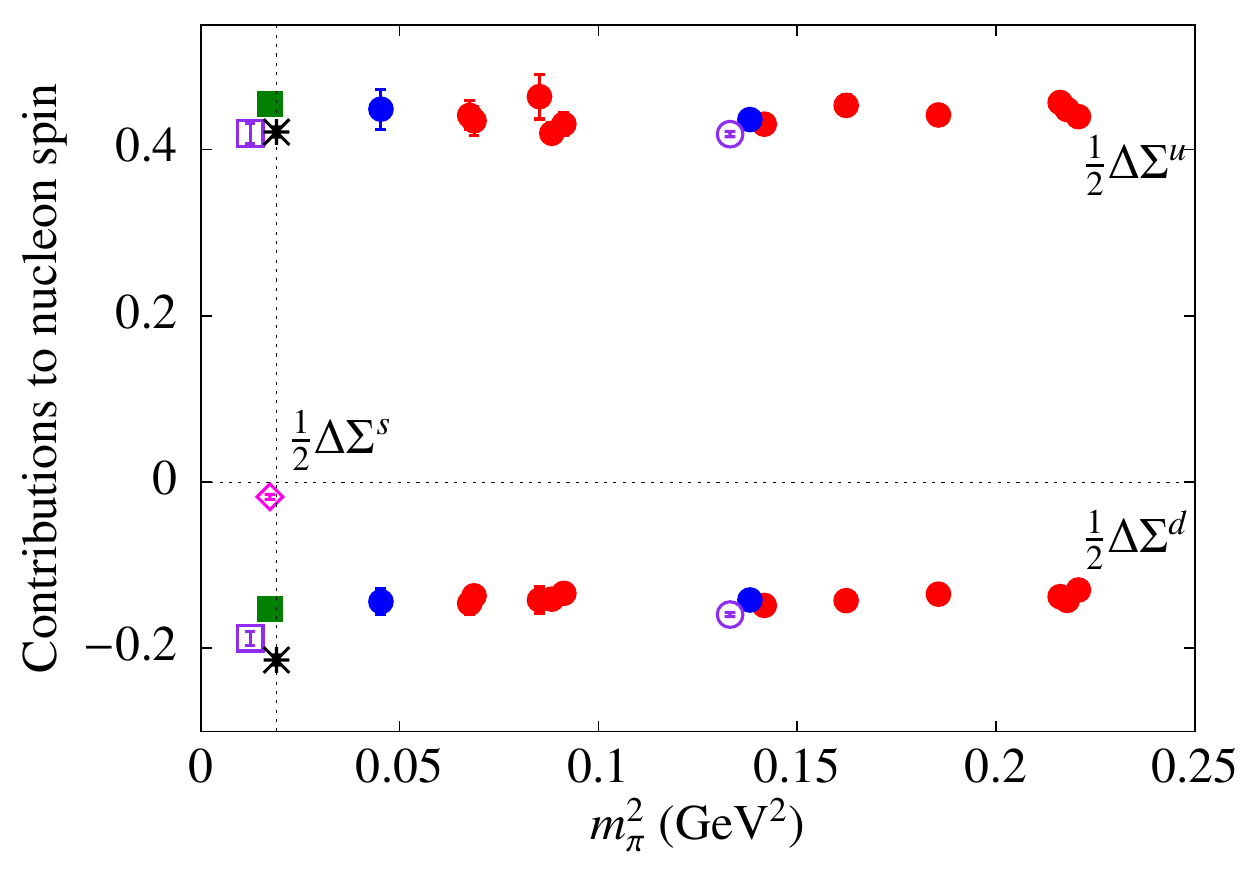}\\
\includegraphics[width=\linewidth]{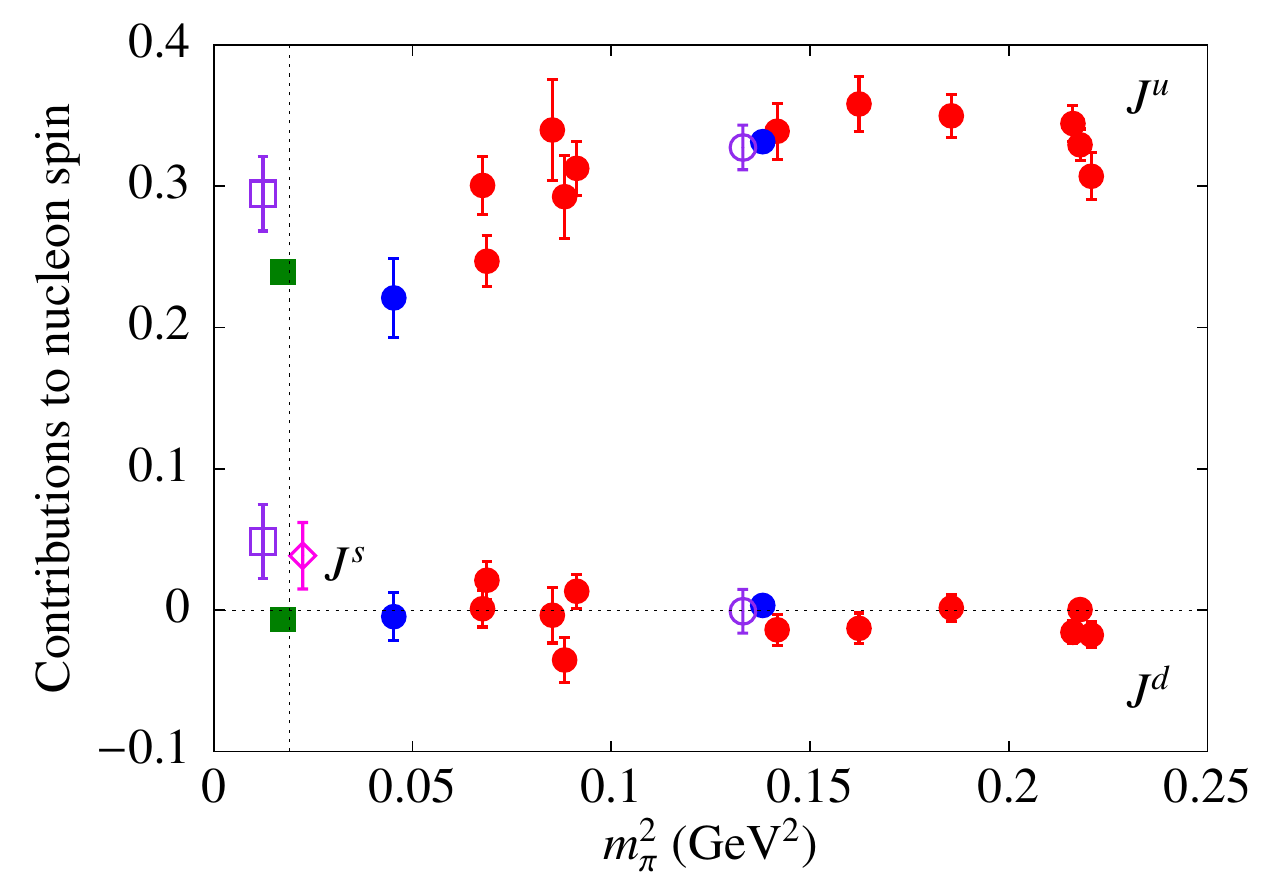}
\end{minipage}
\begin{minipage}{0.49\linewidth}
{\includegraphics[width=\linewidth]{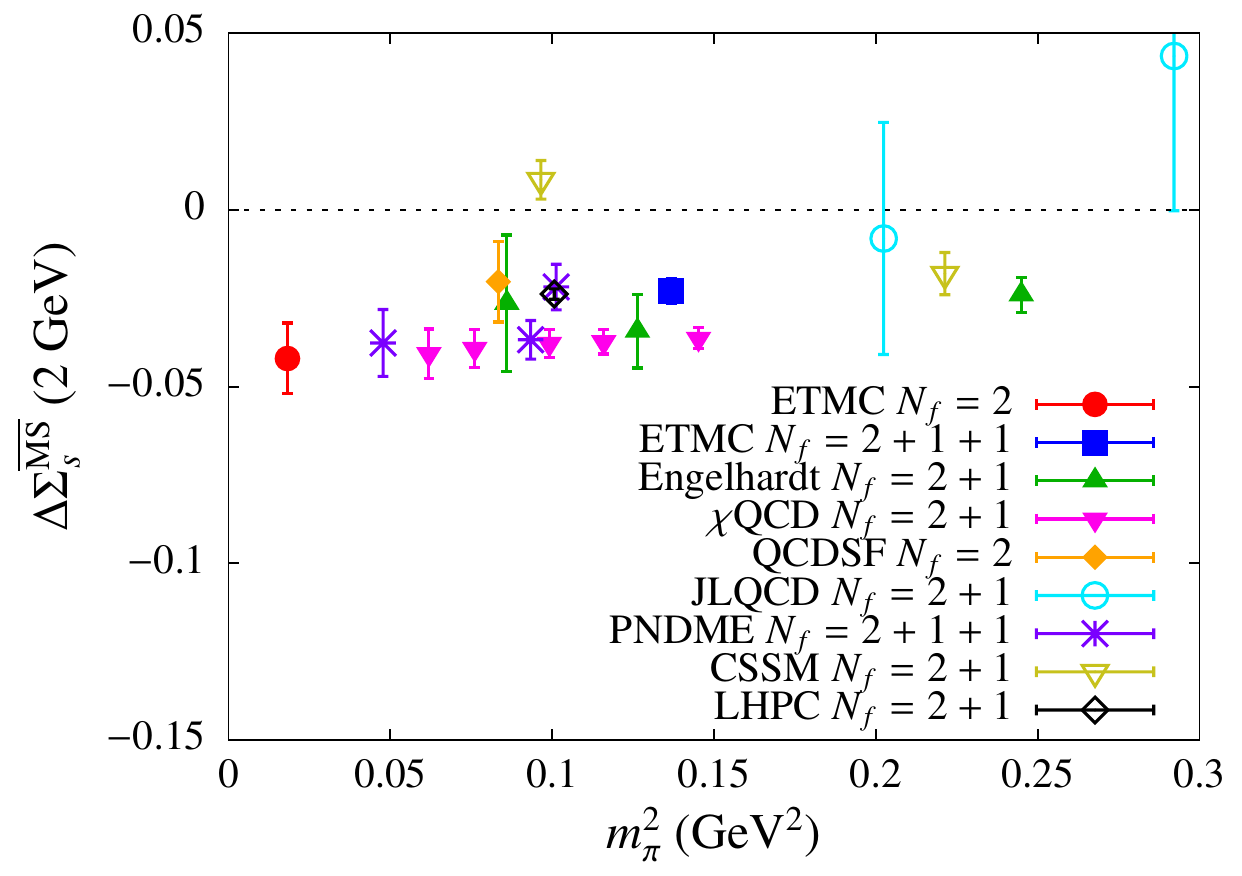}}\\
{\includegraphics[width=\linewidth]{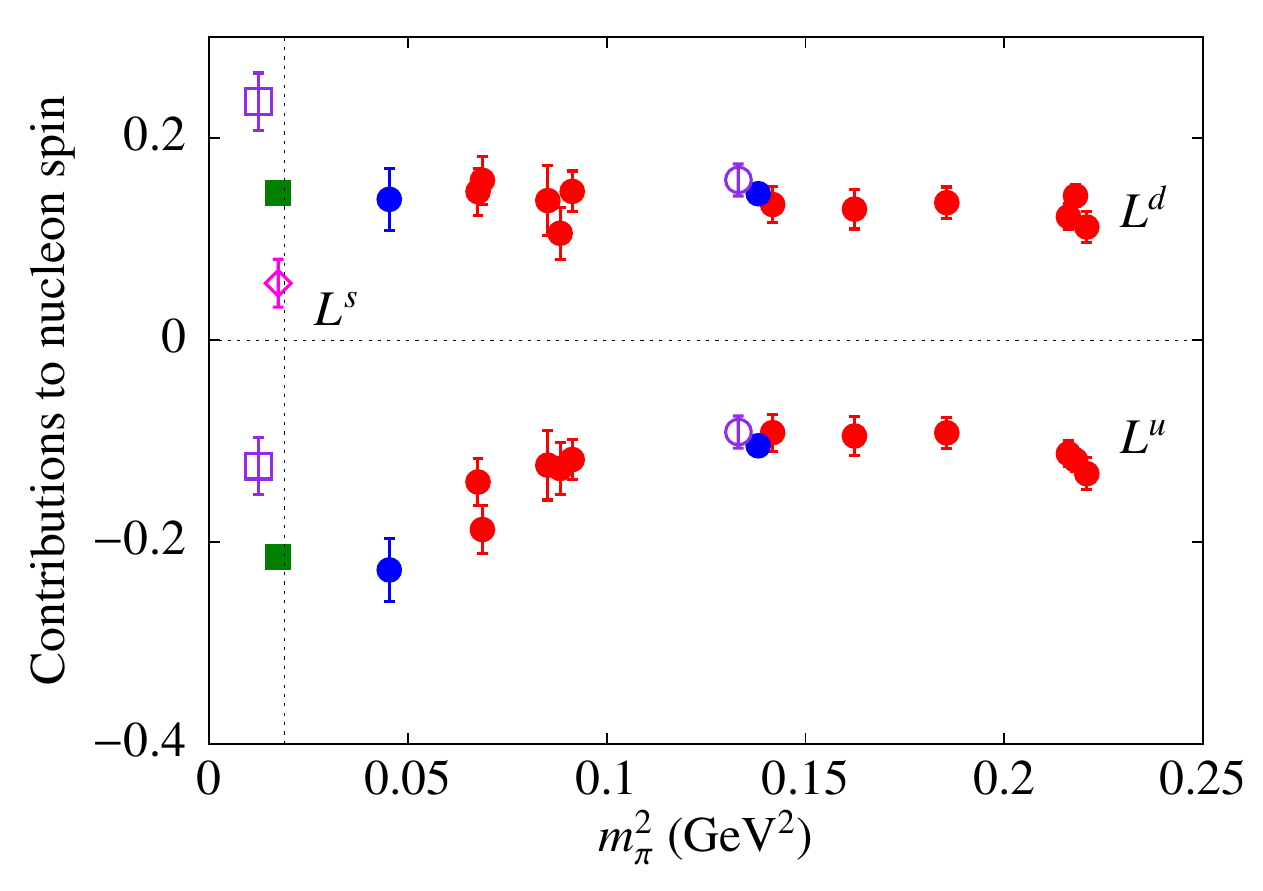}}
\end{minipage}
\caption{Intrinsic spin, total spin and orbital angular momentum in the $\overline{MS}$ at 2 GeV.}
\label{fig:spin}
\end{figure}

As can be seen in Fig.~\ref{fig:spin} $\Delta \Sigma^{u,d}$ at the physical point is consistent with the experimental values after disconnected contributions are included. Also disconnected contributions affect the value of $L^q$ where we find 
 that both $L^u$ and $L^{d}$ increase if disconnected are included. The total 
quark contribution to the spin is $J^{u+d+s}=0.374(51)(42)$ where mixing with the gluon operator is taken into account.  The second error is the systematic determined as the difference between this value and $J^{u+d+s}=0.412(55)$ using non-perturbative renormalization and ignoring the mixing. For the gluon spin  we only have $A_{20}^g(0)$. If we neglect $B_{20}^g(0)$ then we find for the nucleon spin $J_N=0.510(50)(42)$ consistent with the spin sum.

\subsection{Direct evaluation of parton distribution functions - an exploratory study}
\begin{figure}[h]
\begin{minipage}{0.58\linewidth}
{\includegraphics[width=\linewidth]{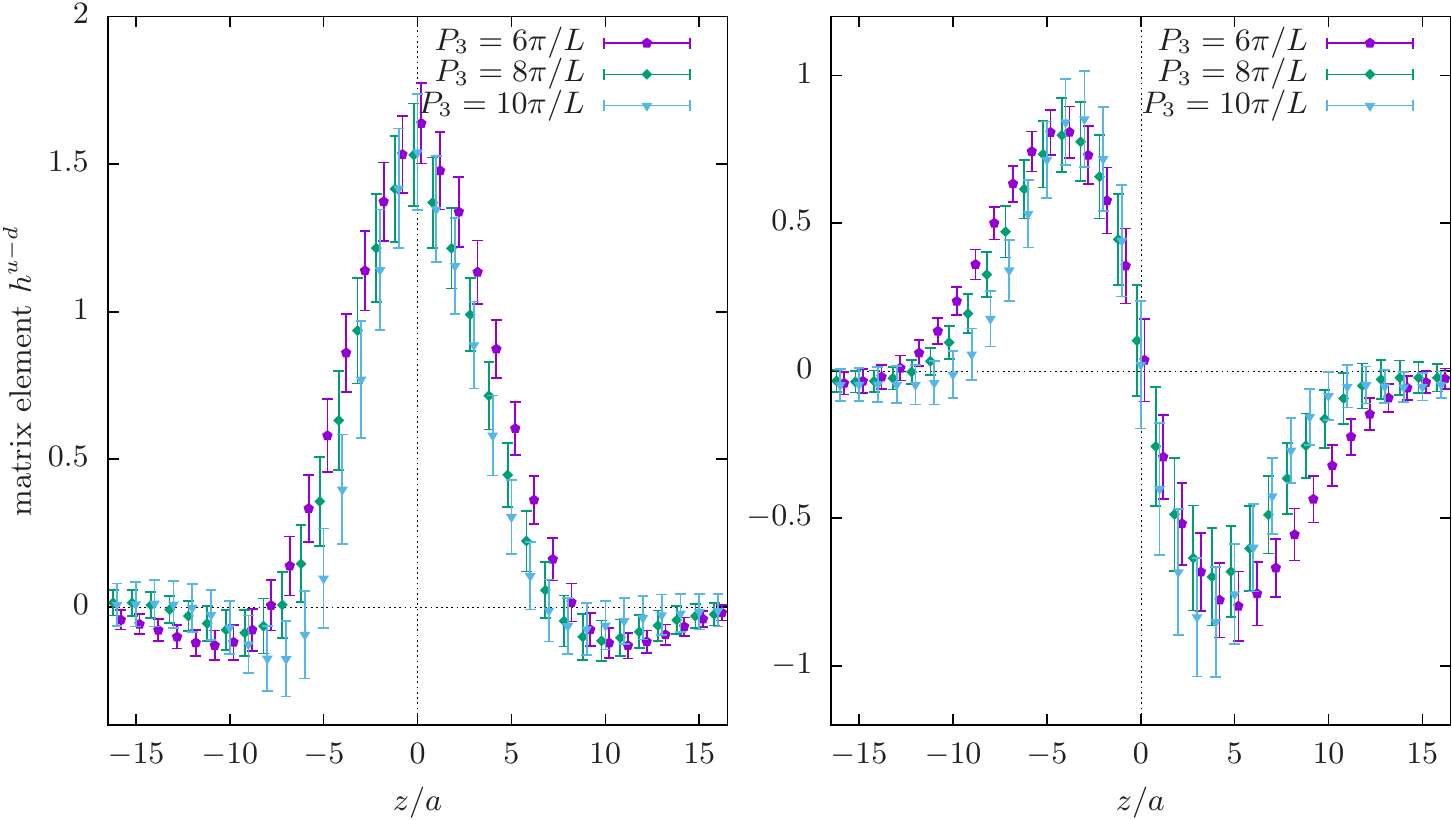}}
\end{minipage}
\begin{minipage}{0.41\linewidth}
\includegraphics[width=\linewidth]{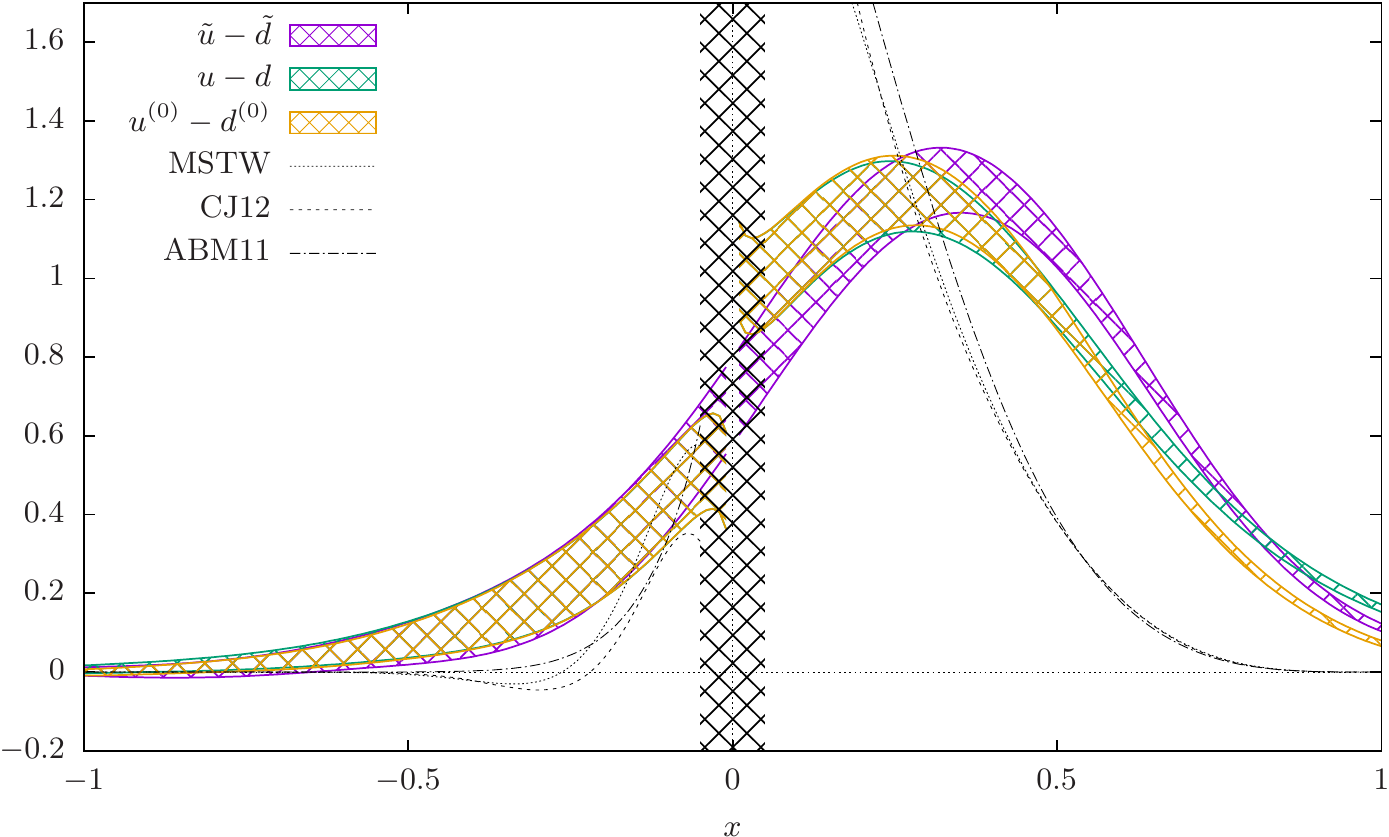}
\end{minipage}
\caption{Results on the unrenormalized unpolarized isovector quasi-distribution for an ensemble of $N_f=2+1+1$ TM fermions, lattice size $32^3 \times 64$, $m_{\pi}= 373\,\text{MeV}$ and $a\approx 0.082\,\text{fm}$ (referred to as B55 ensemble). It uses 5-steps of HYP smearing and  momentum smeared sources~\cite{Bali:2016lva} and 150 measurements for $6\pi/L$ (left) and 300 for $8\pi/L$ and $10\pi/L$~\cite{Alexandrou:2016jqi} (middle). Right: The isovector unpolarized parton distribution for B55.}
\label{fig:PDF}
\end{figure}

We discuss an exploratory study  to extract  directly in lattice QCD
the PDFs following the proposal of  Ref.~\cite{Ji:2013dva}. Consider the matrix element 
\be
\tilde{q}(x,\Lambda,P_3)=\int_{-\infty}^{+\infty}  \frac{dz}{4\pi} e^{-izxP_3}{\langle P|\bar{\psi}(z,0)\,\gamma_3 \,W(z)\psi(0,0)|P\rangle}_{h(P_3,z)},
\ee
where $\tilde{q}(x)$ is the quasi-distribution to be related to the PDFs. First results are obtained for $N_f=2+1+1$ clover fermions on HISQ sea~\cite{Lin:2014zya,Chen:2016utp} and for the B55 ensemble~\cite{Alexandrou:2015rja,Alexandrou:2016jqi} for which we show results in  Fig.~\ref{fig:PDF} on the isovector distribution 
$q^{u-d}(x)$ for 5 steps of HYP smearing. 
The matching to the PDF $q(x)$ is done using 
\be
q(x,\mu)=\tilde{q}(x,\Lambda,P_3)-\frac{\alpha_s}{2\pi}\tilde{q}(x, \Lambda,P_3)\delta Z_F^{(1)}\left(\frac{\mu}{P_3},\frac{\Lambda}{P_3}\right)-\frac{\alpha_s}{2\pi}\int_{-1}^1  \frac{dy}{y} Z^{(1)}\left(\frac{x}{y},\frac{\mu}{P_3},\frac{\Lambda}{P_3}\right)\tilde{q}(y,\Lambda,P_3) +{\cal O}(\alpha_s^2)
\label{PDF}
\ee

Subtracting the target mass corrections, which are small for the larger momenta considered and doing the matching one converts the quasi-distribution measured on the lattice into the distribution shown in Fig.~\ref{fig:PDF}. It shows  good qualitative agreement with the phenomenologically extracted PDF. We observe in particular, that it exhibits an asymmetry between the quark and the anti-quark distributions, which is a highly non-trivial outcome. The polarized and transversity PDFs have been similarly computed in Ref.~\cite{Alexandrou:2016jqi}. For the latter, the uncertainties from the phenomenological analyses are large and thus the lattice QCD calculation, after a suitable renormalization that still needs to be carried out, has the potential  of providing  a prediction based only on QCD.

\section{Neutron Electric Dipole Moment}

Possible experimental observation of a non-zero neutron dipole moment (nEDM)  will probe  beyond the standard model physics since it will indicate violation of $P$ and $T$ symmetries.  The   best experimental bound currently is~\cite{Baker:2006ts}
      \be
      \vert \vec{d}_N \vert  < 3.0 \times 10^{-13} e \cdot {\rm fm}\,\nonumber
      \label{eq:nEDM_smaller_experimental_upperbound}
      \ee
 A neutron dipole moment  is induced by  the $CP$-violating Cherns-Simons (CS) term, 
\beq
{\cal L}_{\rm CS} \left( x \right) \equiv - i \theta \frac{1}{ 32 \pi^2} \epsilon^{\mu \nu \rho \sigma} {\rm Tr} \left[ G_{\mu \nu} \left( x \right) G_{\rho \sigma} \left( x \right) \right]\equiv -i\theta q(x) \,.\nonumber
\label{eq:QCD_Chern_Simons}
\eeq
which we add to the QCD Lagrangian resulting in the Lagrangian density
\beq
{\cal L} \left( x \right) = {\cal L}_{\rm QCD} \left( x \right) + {\cal L}_{\rm CS}\left( x \right)\,.\nonumber
\label{eq:Lagrangian_Density}
\eeq
 Model dependent studies as well as chiral perturbation theory (ChPT) predictions find
$
\vert d_N \vert \sim \theta \cdot {\cal O} \left( 10^{-2} \sim 10^{-3} \right) e \cdot {\rm fm}\,$,
yielding $\theta \lesssim {\cal O} \left( 10^{-10} \sim  10^{-11}   \right)$.

 In order to compute the nEDM moment in lattice QCD we need to compute expectations values  with ${\cal L}(x)$ (in Euclidean time):
\beq
    \< \mc{O}(x_1,...,x_n) \>_{\theta} = \frac{1}{Z_{\theta}} \int d[U] d[\psi_f] 
    d[\bar{\psi}_f] ~ \mc{O}(x_1,...,x_n) ~ e^{-S_{\rm QCD}+i{\theta} \int q(x) d^4x}\,.\nonumber
    \label{eq:vev}
\eeq
However, the $\theta$-term leads to a complex action prohibiting  simulations. Several methods have been developed to enable the computation of the neutron  electric dipole moment, namely i) Measure the neutron energy in an external electric field; ii)  Simulate with imaginary $\theta$, see e.g. QCDSF~\cite{Guo:2015tla}; iii) Assume $\theta$ is small and expand to first order and compute the CP-violating form factor $F_3(0)$ yielding $|d_n|=\lim_{q^2\rightarrow 0}\frac{F_3(q^2)}{2m_N}$. If one adopts the latter method one needs  $F_3(0)$, which  cannot be determined directly requiring either  to fit the $Q^2$-dependence
or use position space methods described in Section~3.2.
Here we  assume that $\theta$ is small and expand to first order
\beq
  \< \mc{O}(x_1,...,x_n) \>_{\theta}= \left\< \mc{O}(x_1,...,x_n) \right\>_{\theta=0}+ \left\< \mc{O}(x_1,...,x_n) \left(i{\theta} \int  d^4x q(x) \right)\right\>_{\theta=0}+O(\theta^2)\,.
\eeq
We then compute the neutron CP-violating electromagnetic form factor 
$F_3(Q^2)$ 
and extract  $F_3(0)$ either by fitting  the $Q^2$-dependence to a dipole form
or use our newly developed  position space methods to extract it directly at $Q^2=0$. 
With the CP-violating $\theta$-term the form factor decomposition of the nucleon electromagnetic form factor reads
\be
	  \langle N^\theta (\vec{p_f},s)\vert J_\mu^{\rm EM}\vert N^\theta (\vec{p_i},s^\prime)\rangle \sim \bar u_N^\theta (\vec{p_f},s) \Lambda_\mu^\theta (q) u_N^\theta (\vec{p_i},s^\prime),
\ee
where $\Lambda_\mu^\theta (q) =   \Lambda_\mu^{\rm even}(q) + i\theta\ \Lambda_\mu^{\rm odd}(q) + \mathcal{O}(\theta^2)$ contains the (standard) $CP$-even and the $CP$-odd part, given by
\be
\Lambda_\mu^{\rm even}(q)= \gamma_\mu
F_1(q^2)+\frac{F_2(q^2)}{2m_N}q_\nu\sigma_{\mu\nu}\,, 
\Lambda_\mu^{\rm odd}(q)= {{\frac{F_3(q^2)}{2m_N}q_\nu\sigma_{\mu\nu}\gamma_5}}
\ee
 
\begin{figure}[h]
\begin{minipage}{0.49\linewidth}
   \includegraphics[width=0.9\linewidth]{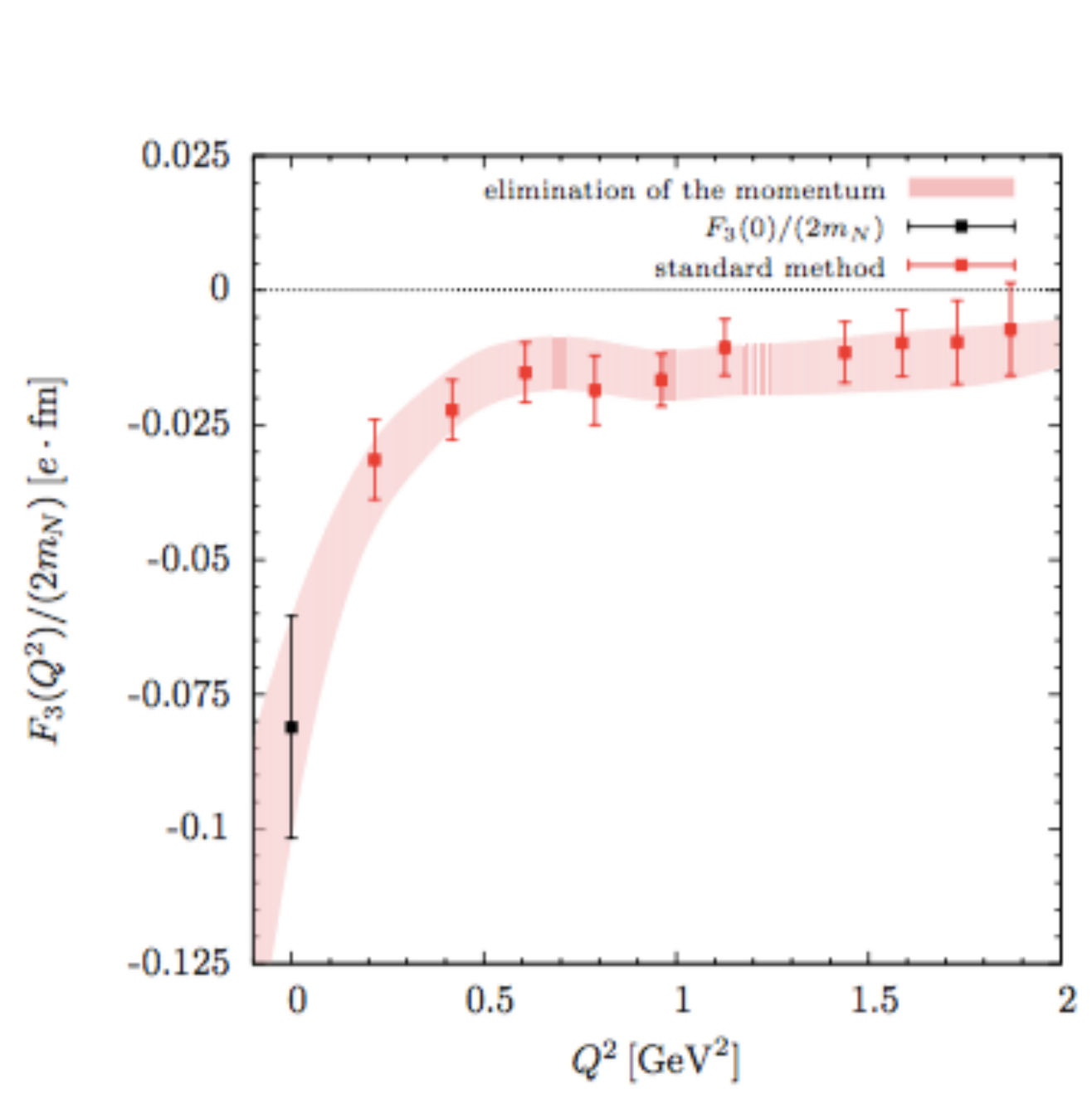}
\end{minipage}
\begin{minipage}{0.49\linewidth}
 \includegraphics[width=0.98\linewidth]{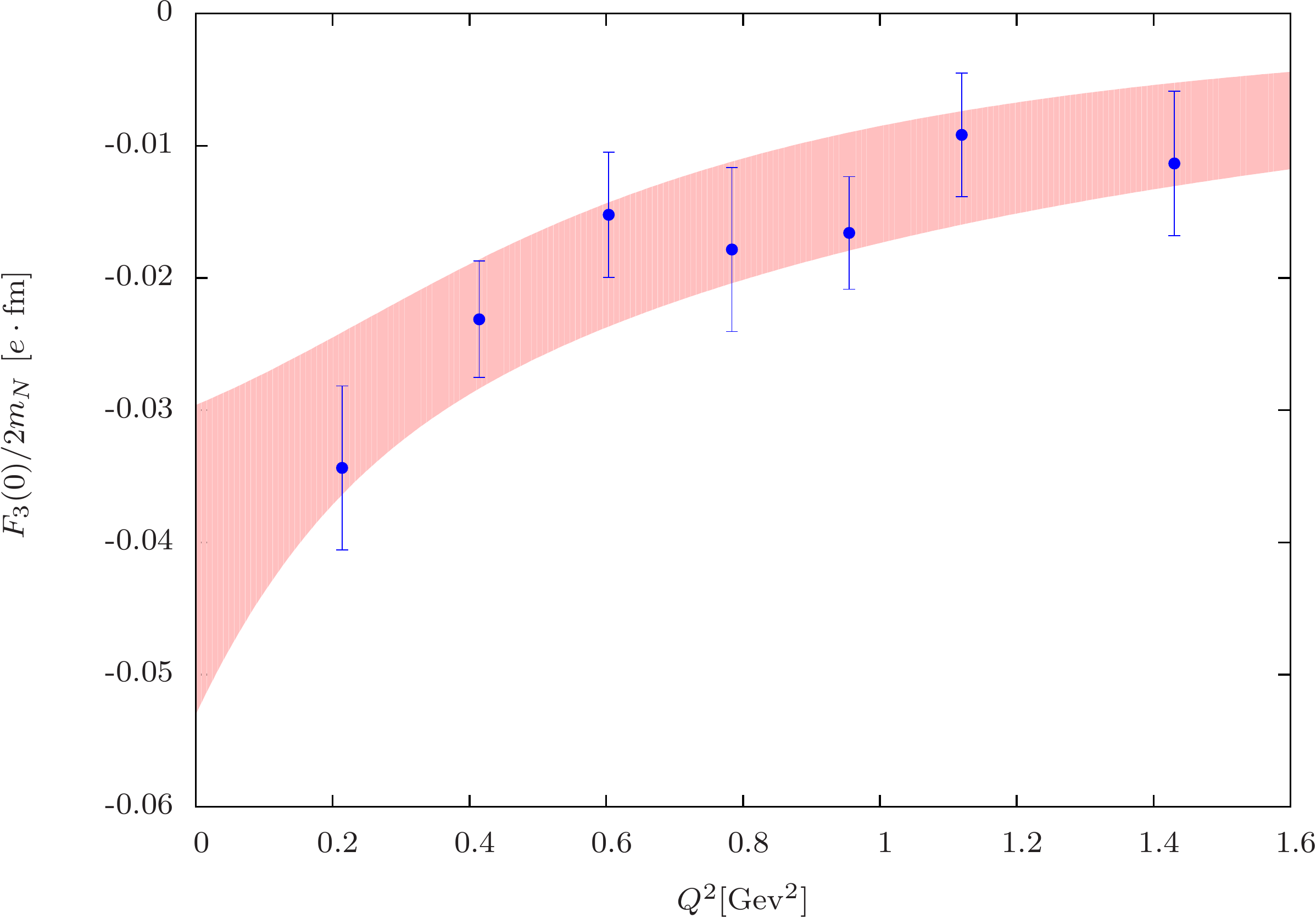}
   \end{minipage}
\caption{The CP-violating form factor $F_3(0)$ using $\mathcal{O}(4700)$ gauge configuration for the $B55$ ensemble. We use the improved gluonic definition for the topological charge  $Q_\mathrm{top}$ and the gradient flow to define $Q_{\rm top}$.}
\label{fig:nEDM B55}
\end{figure}
\begin{figure}[h]
\hspace*{2cm}\includegraphics[width=0.7\linewidth]{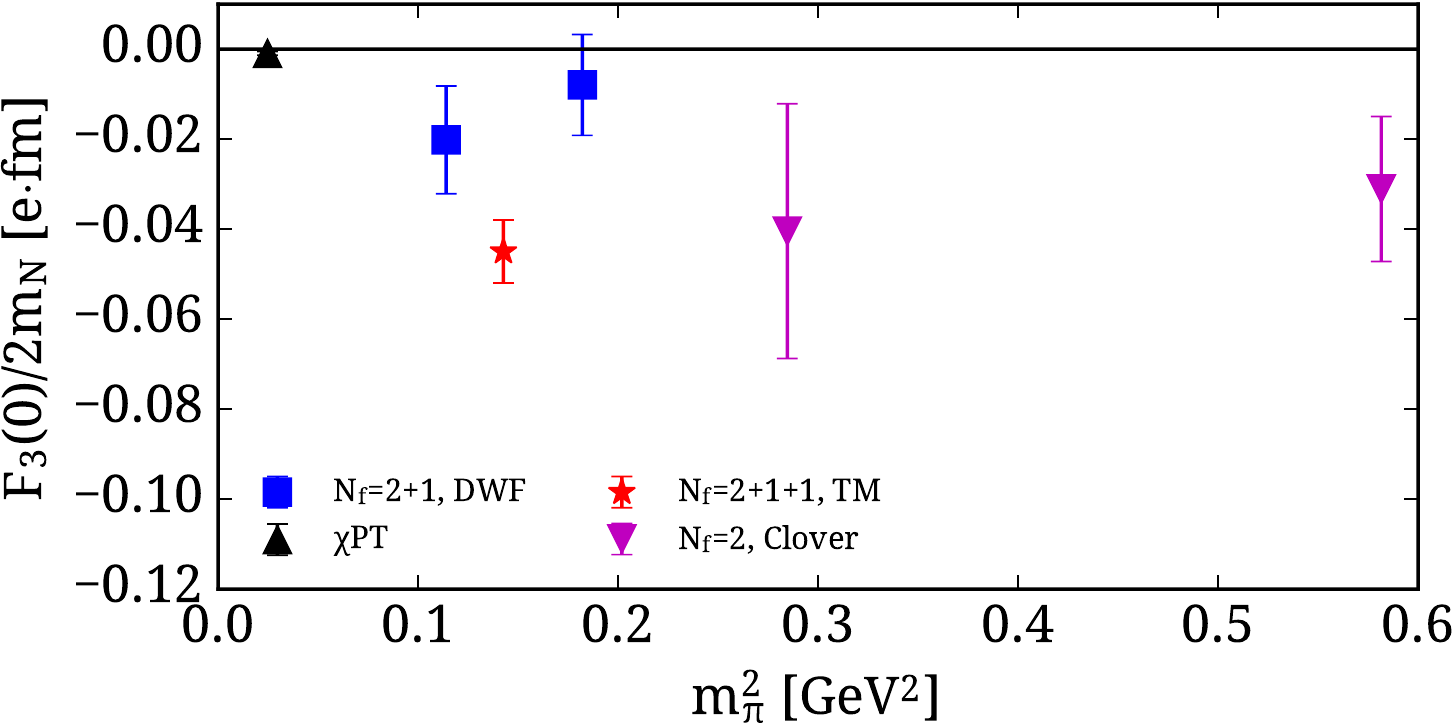}
\caption{$F_3(0)/(2 m_N)$
versus the pion mass squared ($m_\pi^2$). Results using $N_f{=}2{+}1{+}1$
twisted mass fermions are shown with a red asterisk. We also
show  results for $N_f{=}2{+}1$
domain wall fermions~\cite{Shintani:2014zra} at { $a \simeq 0.11$fm}
where  the $CP$-odd $F_3(Q^2)$ was evaluated and
$F_3(0)$ was determined by fitting its $Q^2$-dependence  (blue squares). Results obtained with $N_f{=}2$ Clover fermions at { $a \simeq 0.11$fm} using a
background electric field method are shown with
downward green triangles~\cite{Shintani:2008nt}. All errors shown are
statistical. A value determined in chiral perturbation theory { at one loop order} is
shown with the black  triangle~\cite{Ottnad:2009jw}.}
\label{fig:nEDM}
\end{figure}
 
 The three-point function is given by 
\beq
	C^{\theta,\mu}_{3pt}(t_s t, {\vec q},\Gamma_\nu)& =& \langle N(\vec{p_f},t_s) J_\mu^{\rm EM}(\vec q,t) \bar
N(\vec{p_i},0)e^{i\theta {\cal{Q}}}\rangle \nonumber\\ 
&=&\langle N(\vec{p_f},t_s) J_\mu^{\rm EM}(\vec q,t) \bar
N(\vec{p_i},0)\rangle + i\theta \langle N(\vec{p_f},t_s) J_\mu^{\rm EM}(\vec q,t) \bar
N(\vec{p_i},0) {\cal{Q}}\rangle \, + {\cal O}(\theta^2)
 	 \nonumber
\eeq
where the new element entering  is the correlation of the topological charge ${\cal {Q}}$ with the nucleon two-point and three-point functions.
The main steps involved in the computation of the nEDM is thus to use
the three-function to extract
  \be
	 \Pi_0^\theta \left(\vec{q},\Gamma_k \right) = \theta C \frac{i}{4m_N}\left[  {\alpha^1}  {{q_k}} F_1(Q^2) + \frac{{{q_k}} (E+3 m_N ) {\alpha^1} F_2(Q^2) }{2 m_N} + \frac{ {{q_k}} (E+m_N) F_3(Q^2)}{ 2m_N} \right] \,.
	\ee
where we the $\alpha^1$-parameter  as input. This is determined from ratios suitably projected of two-point functions at large $t$
 \be
     \frac{C_\mathrm{2pt}^\theta(t,\gamma_5)}{C_\mathrm{2pt}(t,1+\gamma_0)} \rightarrow 2i\alpha^1 \theta .
    \ee
Using
 a linear combination of $CP$-even ($\Pi_i\left(\vec{q},\Gamma_0\right)$, $\Pi_i\left(\vec{q},\Gamma_k\right)$) and $CP$-odd  ($\Pi_0^\theta \left(\vec{q},\Gamma_k \right)$)  one can isolate $F_3(Q^2)$ and then either use a fit Ansatz to extract $F_3(0)$ or  position space methods~\cite{Alexandrou:2015spa}.

 Results  on the $F_3$ form factor using TM fermions and the B55 ensemble are shown in Fig.~\ref{fig:nEDM B55}. We use a dipole fit to extrapolate to zero momentum transfer and  the momentum elimination in the plateau region technique to remove the kinematical momentum factor in front of $F_3(Q^2)$ enabling us to extract $F_3(0)$ directly~\cite{Alexandrou:2015spa, Alexandrou:2015ttm}.  We find a non-zero signal for the nEDM, with all definitions of ${\cal Q}$. The momentum  elimination method and the dipole fit yield results  compatible within the errors.
In Fig.~\ref{fig:nEDM} we show a comparison of lattice QCD results on $F_3(0)/(2m_N)$ obtained with dynamical simulations and treating the $\theta$-parameter  as a real parameter in the QCD Lagrangian keeping the comparison within a similar lattice methodology where lattice systematics are expected to be similar. We note that results obtained using formulations with an imaginary $\theta$
are in agreement with the value found using TM fermions.
The challenge to use this methodology to compute the nEDM at the physical point is the increased statistical fluctuations. Preliminary results show that unless one devises improved method to cut the noise such a calculation will require at least two order of magnitude more statistics.

\section{Conclusions}
Simulations at near physical parameters of QCD are yielding important results on benchmark quantities and  systematic studies of lattice artifacts are being pursued by a number of collaborations having gauge configurations simulated with physical value of the pion mass that are expected to lead to an understanding of the  small discrepancies between lattice QCD determinations and the experimental values.
 For example, lattice QCD results on the  nucleon axial charge, quark momentum fraction and intrinsic spin of the quark inside the nucleon for which there are experimental results show convergence to the experimental value. 
Taking into account the disconnected contributions is shown to be necessary for reproducing the experimental values and recent methods developed for the computation of disconnected contributions are  successfully applied to several matrix elements. The light and strange axial charges are examples where the  contribution of the disconnected quark loops is non-zero and needs to be include to obtain agreement with the experimental value. This well-established framework can thus be employed for predicting other quantities probing hadron structure such as   
 scalar and tensor charges,  tensor moments, and $\sigma$-terms, all of which have been computed at the physical point. 
 Exploration of new techniques to  evaluate hadron PDFs are presented with very promising results. Although the full renormalization has not been carried out yet this approach hold the promise of delivering  the complete parton distributions directly from QCD. Position space  methods are tested at heavier than physical pion mass yielding the anomalous magnetic moment of the nucleon. The same approach can be applied for the extraction of the nucleon radii and the CP-violating form factors $F_3(0)$ that determines the nEDM. A calculation at heavier than physical pion mass resulted in a non-zero value for the nEDM but noise reduction techniques are 
needed to evaluate this quantity at the physical point.
 We expect rapid progress in many of these aspects in the near future.

\section*{Acknowledgments}
I would like to thank my collaborators A. Abdel-Rehim, A. Athenodorou, K. Cichy,  M. Constantinou, K. Jansen, K. Hadjiyiannakou, Ch. Kallidonis, G. Koutsou, K. Ottnad, M. Petschlies, F. Steffens,  C. Wiese and A. Vaquero for their invaluable contributions, which made this presentation possible.
This work was supported by a grant from the Swiss National Supercomputing Centre (CSCS) under project ID s540 and in addition used computational resources  from
the John von Neumann-Institute for Computing on the Juropa system and
the BlueGene/Q system Juqueen  partly through PRACE allocation, which included Curie (CEA), Fermi (CINECA) and SuperMUC (LRZ).


\begin{thebibliography}{}


\bibitem{Frezzotti:2003ni} 
  R.~Frezzotti and G.~C.~Rossi,
  JHEP {\bf 0408}, 007 (2004)
  doi:10.1088/1126-6708/2004/08/007
  [hep-lat/0306014].
\bibitem{Frezzotti:2003xj} 
  R.~Frezzotti and G.~C.~Rossi,
  Nucl.\ Phys.\ Proc.\ Suppl.\  {\bf 128}, 193 (2004)
  doi:10.1016/S0920-5632(03)02477-0
  [hep-lat/0311008].
\bibitem{Aoki:2009ix} 
  S.~Aoki {\it et al.} [PACS-CS Collaboration],
  Phys.\ Rev.\ D {\bf 81}, 074503 (2010)
    [arXiv:0911.2561].
\bibitem{Ishikawa:2015rho} 
  K.-I.~Ishikawa {\it et al.},
  PoS LATTICE {\bf 2015}, 075 (2015)
  [arXiv:1511.09222].
\bibitem{Durr:2010aw} 
  S.~Durr {\it et al.},
  JHEP {\bf 1108}, 148 (2011)
    [arXiv:1011.2711].
\bibitem{Bazavov:2012xda} 
  A.~Bazavov {\it et al.} [MILC Collaboration],
  Phys.\ Rev.\ D {\bf 87}, no. 5, 054505 (2013)
    [arXiv:1212.4768].
\bibitem{Horsley:2013ayv} 
  R.~Horsley, Y.~Nakamura, A.~Nobile, P.~E.~L.~Rakow, G.~Schierholz and J.~M.~Zanotti,
  Phys.\ Lett.\ B {\bf 732}, 41 (2014)
    [arXiv:1302.2233];
  G.~S.~Bali {\it et al.},
  Phys.\ Rev.\ D {\bf 90}, no. 7, 074510 (2014)
  [arXiv:1408.6850].
\bibitem{Bruno:2014jqa} 
  M.~Bruno {\it et al.},
  JHEP {\bf 1502}, 043 (2015)
  [arXiv:1411.3982].
\bibitem{Boyle:2015hfa} 
  P.~A.~Boyle {\it et al.} [RBC/UKQCD Collaboration],
  JHEP {\bf 1506}, 164 (2015)
  [arXiv:1504.01692]; 
  T.~Blum {\it et al.} [RBC and UKQCD Collaborations],
  arXiv:1411.7017 [hep-lat].
\bibitem{Abdel-Rehim:2015pwa} 
  A.~Abdel-Rehim {\it et al.} [ETM Collaboration],
  arXiv:1507.05068; 
  R.~Baron {\it et al.},
  JHEP {\bf 1006}, 111 (2010)
   [arXiv:1004.5284].
\bibitem{Ji:2013dva} 
  X.~Ji,
  Phys.\ Rev.\ Lett.\  {\bf 110}, 262002 (2013)
  doi:10.1103/PhysRevLett.110.262002
  [arXiv:1305.1539 [hep-ph]].
\bibitem{Lin:2014zya} 
  H.~W.~Lin, J.~W.~Chen, S.~D.~Cohen and X.~Ji,
  Phys.\ Rev.\ D {\bf 91}, 054510 (2015)
  doi:10.1103/PhysRevD.91.054510
  [arXiv:1402.1462 [hep-ph]].
\bibitem{Alexandrou:2015rja} 
  C.~Alexandrou, K.~Cichy, V.~Drach, E.~Garcia-Ramos, K.~Hadjiyiannakou, K.~Jansen, F.~Steffens and C.~Wiese,
  Phys.\ Rev.\ D {\bf 92}, 014502 (2015)
  doi:10.1103/PhysRevD.92.014502
  [arXiv:1504.07455 [hep-lat]].
\bibitem{Maiani:1987by} 
  L.~Maiani, G.~Martinelli, M.~L.~Paciello and B.~Taglienti,
  Nucl.\ Phys.\ B {\bf 293}, 420 (1987).
  doi:10.1016/0550-3213(87)90078-2

\bibitem{Abdel-Rehim:2015owa} 
  A.~Abdel-Rehim {\it et al.},
  Phys.\ Rev.\ D {\bf 92}, no. 11, 114513 (2015)
  Erratum: [Phys.\ Rev.\ D {\bf 93}, no. 3, 039904 (2016)]
  doi:10.1103/PhysRevD.92.114513, 10.1103/PhysRevD.93.039904
  [arXiv:1507.04936 [hep-lat]].
\bibitem{Collins} S. Collins, PoS (Lattice2016) 009.
\bibitem{Anselmino:2013vqa} 
  M.~Anselmino, M.~Boglione, U.~D'Alesio, S.~Melis, F.~Murgia and A.~Prokudin,
  Phys.\ Rev.\ D {\bf 87}, 094019 (2013)
   [arXiv:1303.3822].
\bibitem{Radici:2015mwa} 
  M.~Radici, A.~Courtoy, A.~Bacchetta and M.~Guagnelli,
  JHEP {\bf 1505}, 123 (2015)
   [arXiv:1503.03495].
\bibitem{Alexandrou2016} C. Alexandrou {\it et al.},  PoS (Lattice2016) 153.
\bibitem{Abdel-Rehim:2016won} 
  A.~Abdel-Rehim {\it et al.} [ETM Collaboration],
  Phys.\ Rev.\ Lett.\  {\bf 116}, no. 25, 252001 (2016)
  doi:10.1103/PhysRevLett.116.252001
  [arXiv:1601.01624 [hep-lat]].
\bibitem{Pohl:2010zza} 
  R.~Pohl {\it et al.},
  Nature {\bf 466}, 213 (2010).
  doi:10.1038/nature09250
\bibitem{Bernauer:2013tpr} 
  J.~C.~Bernauer {\it et al.} [A1 Collaboration],
  Phys.\ Rev.\ C {\bf 90}, no. 1, 015206 (2014)
  doi:10.1103/PhysRevC.90.015206
  [arXiv:1307.6227 [nucl-ex]].
\bibitem{Griffioen:2015hta} 
  K.~Griffioen, C.~Carlson and S.~Maddox,
  Phys.\ Rev.\ C {\bf 93}, no. 6, 065207 (2016)
  doi:10.1103/PhysRevC.93.065207
  [arXiv:1509.06676 [nucl-ex]].
\bibitem{Lorenz:2012tm} 
  I.~T.~Lorenz, H.-W.~Hammer and U.~G.~Meissner,
  Eur.\ Phys.\ J.\ A {\bf 48}, 151 (2012)
  doi:10.1140/epja/i2012-12151-1
  [arXiv:1205.6628 [hep-ph]].
\bibitem{Higinbotham:2015rja} 
  D.~W.~Higinbotham, A.~A.~Kabir, V.~Lin, D.~Meekins, B.~Norum and B.~Sawatzky,
  Phys.\ Rev.\ C {\bf 93}, no. 5, 055207 (2016)
  doi:10.1103/PhysRevC.93.055207
  [arXiv:1510.01293 [nucl-ex]].
\bibitem{Hill}
R. Hill, these proceedings.
\bibitem{koutsou2016} C. Alexandrou {\it et al.} (ETMC) PoS(Lattice2016) 154.
\bibitem{Green:2014xba} 
  J.~R.~Green, J.~W.~Negele, A.~V.~Pochinsky, S.~N.~Syritsyn, M.~Engelhardt and S.~Krieg,
  Phys.\ Rev.\ D {\bf 90}, 074507 (2014)
  doi:10.1103/PhysRevD.90.074507
  [arXiv:1404.4029 [hep-lat]].

\bibitem{Jang}
T. Bhattacharya1, R. Gupta, Y.-Ch. Jang1, H.-W. Lin, B. Yoon, PoS (lattice2016)
\bibitem{Kuramashi}
Y. Kuramashi (PACS), PoS(Lattice2016).
\bibitem{Stathopoulos:2013aci} 
  A.~Stathopoulos, J.~Laeuchli and K.~Orginos,
  arXiv:1302.4018 [hep-lat].
\bibitem{Ahmed:2011vp} 
  Z.~Ahmed {\it et al.} [HAPPEX Collaboration],
  Phys.\ Rev.\ Lett.\  {\bf 108}, 102001 (2012)
  doi:10.1103/PhysRevLett.108.102001
  [arXiv:1107.0913 [nucl-ex]].
\bibitem{Green:2015wqa} 
  J.~Green {\it et al.},
  Phys.\ Rev.\ D {\bf 92}, no. 3, 031501 (2015)
  doi:10.1103/PhysRevD.92.031501
  [arXiv:1505.01803 [hep-lat]].
\bibitem{Sufian:2016pex} 
  R.~S.~Sufian, Y.~B.~Yang, A.~Alexandru, T.~Draper, K.~F.~Liu and J.~Liang,
  arXiv:1606.07075 [hep-ph].
\bibitem{Alexandrou:2016rbj} 
  C.~Alexandrou {\it et al.} [ETM Collaboration],
  Phys.\ Rev.\ D {\bf 94}, no. 7, 074508 (2016)
  doi:10.1103/PhysRevD.94.074508
  [arXiv:1605.07327 [hep-lat]].
\bibitem{Bhattacharya:2013ehc} 
  T.~Bhattacharya, S.~D.~Cohen, R.~Gupta, A.~Joseph, H.~W.~Lin and B.~Yoon,
  Phys.\ Rev.\ D {\bf 89}, no. 9, 094502 (2014)
  doi:10.1103/PhysRevD.89.094502
  [arXiv:1306.5435 [hep-lat]].
\bibitem{Capitani:2015sba} 
  S.~Capitani {\it et al.},
  Phys.\ Rev.\ D {\bf 92}, no. 5, 054511 (2015)
  doi:10.1103/PhysRevD.92.054511
  [arXiv:1504.04628 [hep-lat]].
\bibitem{Green2016} J. Green, M. Engelhardt, St. Krieg, J. Laeuchli, J. W. Negele, K. Orginos, A. Pochinsky, S. Syritsyn, PoS (Lattice2016).

\bibitem{Abdel-Rehim:2016pjw} 
  A.~Abdel-Rehim {\it et al.},
  PoS LATTICE {\bf 2016}, 155 (2016)
  [arXiv:1611.03802 [hep-lat]].

\bibitem{Alekhin:2016tkw} 
  S.~Alekhin, J.~Bluemlein, S.~O.~Moch and R.~Placakyte,
  PoS DIS {\bf 2016}, 016 (2016).
\bibitem{Blumlein:2010rn} 
  J.~Blumlein and H.~Bottcher,
  Nucl.\ Phys.\ B {\bf 841}, 205 (2010)
  doi:10.1016/j.nuclphysb.2010.08.005
  [arXiv:1005.3113 [hep-ph]].

\bibitem{Alexandrou:2013tfa} 
  C.~Alexandrou, V.~Drach, K.~Hadjiyiannakou, K.~Jansen, B.~Kostrzewa and C.~Wiese,
  PoS LATTICE {\bf 2013}, 289 (2014)
  [arXiv:1311.3174 [hep-lat]].
\bibitem{Alexandrou:2016ekb} 
  C.~Alexandrou, M.~Constantinou, K.~Hadjiyiannakou, K.~Jansen, H.~Panagopoulos and C.~Wiese,
  arXiv:1611.06901 [hep-lat].
\bibitem{Chen:2016utp} 
  J.~W.~Chen, S.~D.~Cohen, X.~Ji, H.~W.~Lin and J.~H.~Zhang,
  Nucl.\ Phys.\ B {\bf 911}, 246 (2016)
  doi:10.1016/j.nuclphysb.2016.07.033
  [arXiv:1603.06664 [hep-ph]].
\bibitem{Alexandrou:2016jqi} 
  C.~Alexandrou, K.~Cichy, M.~Constantinou, K.~Hadjiyiannakou, K.~Jansen, F.~Steffens and C.~Wiese,
  arXiv:1610.03689 [hep-lat].
03/PhysRevLett.97.131801;
\bibitem{Bali:2016lva} 
  G.~S.~Bali, B.~Lang, B.~U.~Musch and A.~Schäfer,
  Phys.\ Rev.\ D {\bf 93}, no. 9, 094515 (2016)
  doi:10.1103/PhysRevD.93.094515
  [arXiv:1602.05525 [hep-lat]].
\bibitem{Baker:2006ts} 
  C.~A.~Baker {\it et al.},
  Phys.\ Rev.\ Lett.\  {\bf 97}, 131801 (2006)
  doi:10.1103/PhysRevLett.97.131801
  [hep-ex/0602020].
\bibitem{Guo:2015tla} 
  F.-K.~Guo {\it et al.},
  Phys.\ Rev.\ Lett.\  {\bf 115}, no. 6, 062001 (2015)
  doi:10.1103/PhysRevLett.115.062001
  [arXiv:1502.02295 [hep-lat]].
%


\bibitem{Alexandrou:2015spa} 
  C.~Alexandrou, A.~Athenodorou, M.~Constantinou, K.~Hadjiyiannakou, K.~Jansen, G.~Koutsou, K.~Ottnad and M.~Petschlies,
  Phys.\ Rev.\ D {\bf 93}, no. 7, 074503 (2016)
  doi:10.1103/PhysRevD.93.074503
  [arXiv:1510.05823 [hep-lat]].
\bibitem{Shintani:2014zra} 
  E.~Shintani, T.~Blum, A.~Soni and T.~Izubuchi,
  {\it ``Neutron and proton EDM with $N_{f} = 2 + 1$ domain-wall fermion''}, PoS LATTICE {\bf 2013}, 298 (2014).
\bibitem{Shintani:2008nt} 
  E.~Shintani, S.~Aoki and Y.~Kuramashi, 
Phys.\ Rev.\ D {\bf 78}, 014503 (2008), [arXiv:0803.0797].
\bibitem{Ottnad:2009jw}
 K.~Ottnad, B.~Kubis, U.-G.~Meissner and F.-K.~Guo, 
Phys.\ Lett.\ B {\bf 687} 42 (2010), [arXiv:0911.3981].

\bibitem{Alexandrou:2015ttm} 
  C.~Alexandrou, A.~Athenodorou, M.~Constantinou, K.~Hadjiyiannakou, K.~Jansen, G.~Koutsou, K.~Ottnad and M.~Petschlies,
  PoS LATTICE {\bf 2015}, 131 (2016)
  [arXiv:1511.04942 [hep-lat]].

\end{thebibliography}
\end{document}